\documentclass[11pt]{article}
\usepackage{times,graphicx,theorem,epsfig,lscape,dcolumn,multirow,epic,float}
\usepackage{amsmath,amssymb}
\usepackage[normalem]{ulem}
\usepackage{natbib}
\usepackage[figuresright]{rotating}
\usepackage[english]{babel}
\usepackage{booktabs}
\usepackage{hhline}
\usepackage{bm}
\usepackage{enumerate}
\usepackage{verbatim}
\usepackage{dsfont}
\usepackage[colorlinks=true,linkcolor=black,citecolor=black]{hyperref}
\usepackage{threeparttable}
\usepackage{array}

\usepackage{comment}
\usepackage{pdflscape}
\usepackage{afterpage}
\usepackage{subcaption}
\usepackage[labelsep=quad,skip=-2pt]{caption}
\usepackage{setspace}

\theoremstyle{plain} \newtheorem{theorem}{{\sc Theorem}}
\theoremstyle{plain} 
\theoremstyle{plain} \newtheorem{remark}{{\sc Remark}}
\theoremstyle{plain} 
\theoremstyle{plain} 

\theoremstyle{plain} 

\theoremstyle{plain} \newtheorem{lemma}{{\sc Lemma}}

\theoremstyle{plain} 

\theoremstyle{plain}

\newcommand{\blinding}[2]{#1}   

\newcommand{\be}{\begin{eqnarray}}
\newcommand{\ee}{\end{eqnarray}}
\newcommand{\bee}{\begin{eqnarray*}}
\newcommand{\eee}{\end{eqnarray*}}
\newcommand{\bi}{\begin{enumerate}[(i)]}
\newcommand{\ei}{\end{enumerate}}

\usepackage{tikz}
\usetikzlibrary{shapes,decorations,arrows,calc,arrows.meta,fit,positioning}
\tikzset{
    -Latex,auto,node distance =1.5 cm and 1.5 cm,semithick,
    state/.style ={ellipse, draw, minimum width = 0.7 cm},
    point/.style = {circle, draw, inner sep=0.04cm,fill,node contents={}},
    bidirected/.style={Latex-Latex,dashed},
    el/.style = {inner sep=2pt, align=left, sloped}
}

\def\bSig\mathbf{\Sigma}

\newcommand{\eU}{\mathbb{U}}

\newenvironment{proof}{\paragraph{Proof:}}{\hfill$\square$}

\begin{document}

\begin{center}
\vspace*{-2.5cm}

{\Large Correcting for bias due to mismeasured exposure in mediation analysis with a survival outcome}

\medskip
\blinding{
Chao Cheng$^{1,2,*}$, Donna Spiegelman$^{1,2}$ and Fan Li$^{1,2}$
}{}

$^1$Department of Biostatistics, Yale University School of Public Health, New Haven, CT USA\\
$^2$ Center for Methods in Implementation and Prevention Science, Yale University, New Haven, CT USA\\
$*$c.cheng@yale.edu\\

\end{center}

\date{}

{\centerline{Abstract}
\noindent We investigate the impact of exposure measurement error on assessing mediation with a survival outcome modeled by Cox regression. We first derive the bias formulas of natural indirect and direct effects with a rare outcome and no exposure-mediator interaction. We then develop several calibration approaches to correct for the measurement error-induced bias, and generalize our methods to accommodate a common outcome and an exposure-mediator interaction. We apply the proposed methods to analyze Health Professionals Follow-up Study (1986--2016) and evaluate the extent to which reduced body mass index mediates the protective effect of physical activity on risk of cardiovascular diseases.

\vspace*{0.3cm}

\noindent {\sc Key words}: Bias analysis; measurement error correction; mediation proportion; natural indirect effect; regression calibration; time-to-event data  
}

\clearpage

\section{Introduction}
\label{sec:intro}

Mediation analysis is a popular statistical tool to disentangle the role of an intermediate variable, otherwise known as a mediator, in explaining the causal mechanism from an exposure to an outcome. It requires the decomposition of the total effect (TE) from the exposure on the outcome into a natural indirect effect (NIE) that works through the mediator and a natural direct effect (NDE) that works around the mediator (\citealp{baron1986moderator}). Oftentimes, the mediation proportion (MP), defined as the ratio of NIE and TE, is employed to measure the extent to which the mediator explains the underlying exposure-outcome relationship. The NIE, NDE, TE, and MP are four effect measures of primary interest in mediation analysis (\citealp{cheng2022difference}). Although existing mediation analysis methodologies primarily focus on a continuous or a binary outcome, recent advances also have been developed to accommodate survival or time-to-event outcomes subject to right censoring (e.g., \citealp{lange2011direct,vanderweele2011causal,wang2017causal}).


Mediation analysis is an increasingly popular tool for understanding causal mechanisms in observational epidemiologic studies, 
as exemplified by the recent empirical mediation analyses in the Health Professionals Follow-up Study \citep{petimar2019mediation,fu2020height}, the Nurses' Health Study \citep{lee2022professional}, and the Nurses' Health Study II \citep{harris2016endometriosis}. However, measurement error is a pervasive problem across such epidemiologic studies, and affects important variables such like physical activity, air pollution, smoking habits, and dietary intake \citep{camirand2022semiparametric,keogh2020stratos,morgenstern2021perspective,shaw2020stratos}. Ignoring such measurement error is likely to compromise the validity of mediation analyses. The problem of measurement error in mediation analysis with a continuous/binary outcome has been previously studied, and a literature review on methods that correct for the bias due to an error-prone exposure, mediator, or outcome can be found in \citet{cheng2022mediation}. These previous studies, however, have not fully addressed challenges due to measurement error in mediation analysis with a survival outcome. One exception is \cite{zhao2014covariate}, who developed regression calibration approaches to address estimation bias due to measurement error in the mediator, under the strong assumption that the measurement error process is known or can be accurately informed by prior information.

\begin{figure}[h]
\centering\includegraphics[width=0.5\textwidth]{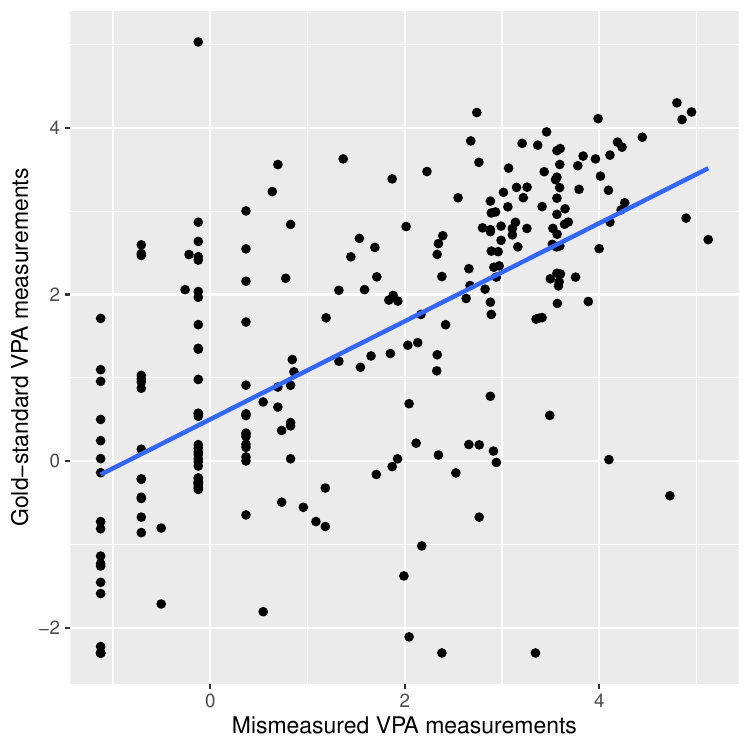}
\caption{A scatter plot of the logarithm of the vigorous physical activity (VPA) measurements (METs$\cdot$h$\cdot$wk$^{-1}$) in the HPFS validation study. The $y$-axis is the gold-standard measurements based on physical activity diaries and the $x$-axis is the error-prone measurements based on questionnaires. A linear regression line with $A=0.50+0.58A^*$ is added in blue color, where $A$ and $A^*$ are the gold-standard and error-prone VPA measurements, respectively.}
\label{fig:VGA}
\end{figure}

In this article, we investigate the implications of exposure measurement error in mediation analysis with a survival outcome, under a linear measurement error model for the mediator and a Cox proportional hazards regression for the survival outcome. Our methods are motivated by the need to address mismeasured exposure in evaluating the effect of vigorous physical activity through body mass index (BMI) on the risk of cardiovascular diseases in the Health Professionals Follow-up Study (HPFS) (\citealp{chomistek2012vigorous}). In HPFS, physical activity data was assessed through a self-administered physical activity questionnaire, whose accuracy was validated by physical activity diaries in the HPFS validation study. The measurement based on a physical activity diary is expensive but much more accurate, which is considered as gold-standard for assessing physical activity (\citealp{chasan1996reproducibility}). As shown in Figure \ref{fig:VGA}, the questionnaire-based measurements are notably different from the gold-standard diary-based measurements based on the HPFS validation study. The correlation between diary-based and questionnaire-based activity was estimated as 0.62 for vigorous activity, suggesting the questionnaire-based data used in the HPFS is error-prone and therefore may lead to bias for estimating the mediation effect measures.

Our contributions to the literature are two-folded. First, in the typical scenario with a rare outcome and no exposure-mediator interaction, we derive new asymptotic bias expressions of the uncorrected estimators for all mediation effect measures under a Berkson-like measurement error model (\citealp{carroll2006measurement}). This provides insights into the direction and magnitude of bias in mediation analysis due to exposure measurement error with a survival outcome.  Second, we propose bias-correction strategies to permit valid estimation and inference for the mediation effect measures with an error-prone continuous exposure, allowing for exposure-mediator interactions and accommodating both common as well as rare outcomes. Our work differs from \cite{zhao2014covariate} in two distinct ways: (i) \cite{zhao2014covariate} assumes that  parameters for characterizing the measurement error process are known or can be elicited from prior information. Relaxing that strong assumption, we leverage a validation study to estimate the measurement error process and account for the uncertainty in estimating the measurement error model parameters in inference about the mediation effect measures. (ii) \cite{zhao2014covariate} only considers mediation analysis with a rare outcome in the absence of exposure-mediator interaction in the outcome model. Our work relaxes these assumptions by allowing for exposure-mediator interactions and can be applied to either a rare or common outcome.

The paper is organized as follows. Section \ref{sec:preliminaries} reviews mediation analysis for a survival outcome without exposure measurement error. In Section \ref{sec:error}, we describe the assumed measurement error model and derive explicit bias expressions for the uncorrected estimators of the mediation effect measures. The bias-correction methods are developed in Section \ref{sec:methods} and \ref{sec:estimation_mediation}, followed by a simulation study in Section \ref{sec:sim} to evaluate their empirical performance. In Section \ref{sec:app}, we apply the methods to the HPFS. Section \ref{sec:disc} concludes.

\section{Preliminaries on mediation analysis with a survival outcome}
\label{sec:preliminaries}

\begin{figure}[ht]
\centering 
\large
\begin{tikzpicture}
    \node (1) at (0,0) {$A$};
    \node (2) [right = of 1] {$M$};
    \node (3) [right = of 2] {$\widetilde{T}$};
    \node (4) [above = of 1] {$\bm W$};
    \path (1) edge  (2);
    \path (2) edge (3);
    \path (4) edge (1);
    \path (4) edge (2);
    \path (4) edge (3);
    \path (1) edge[bend right = 50] (3);
\end{tikzpicture}
\caption{Mediation directed acyclic graph, where $\bm W$, $A$, $M$, $\widetilde{T}$ are the baseline covariates, the exposure, the mediator, the underlying survival outcome, respectively. Here, the $A \rightarrow M \rightarrow \widetilde{T}$ pathway denotes the natural indirect and the $A \rightarrow \widetilde{T}$ pathway denotes the natural direct effect.}
\label{fig1}
\end{figure}
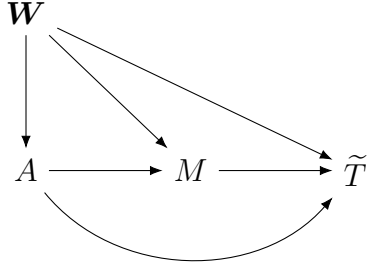

Let $A$ denote an exposure of interest, $M$ a mediator, $\widetilde{T}$ a time-to-event outcome, $\bm W$ a set of covariates. A diagram capturing the causal relationships among those variables is presented in Figure \ref{fig1}. {In large prospective cohort studies like HPFS, $\widetilde{T}$ is right-censored due to loss to follow-up or end of study. Therefore, we only observe $(T,\delta)$, where $\delta = \mathbb{I}(T \leq C)$ is the failure indicator and $T=\widetilde{T}\wedge C\wedge t^*$ is the observed failure time, $C$ is the censoring time and $t^*$ is the maximum follow-up time.} The observed data consists of $n_1$ copies of $\{T_i,\delta_i,M_i,A_i,\bm W_i\}$. 

Under the counterfactual outcomes framework, several definitions for the mediation effect measures for a survival outcome have been proposed, and can be based on the survival function, hazard ratio, or mean survival times (\citealp{lange2011direct,vanderweele2011causal}). Here we pursue the mediation effect measures on the log-hazard ratio scale as they are widely used in health science studies. To define the mediation effect measures, we let $M_a$ be the counterfactual mediator that would have been observed if the exposure $A$ has been set to level $a$. Similarly, let $\widetilde{T}_{am}$ be the counterfactual outcome that would be observed if $A$ and $M$ has been set to levels $a$ and $m$, respectively. Given these counterfactual variables, we can decompose the TE for contrasting $a^*$ to $a$ on the log-hazard ratio scale as
\begin{align*}
     &\underbrace{\log\{\lambda_{\widetilde T_{aM_a}}\!(t|\bm w)\} \!-\! \log\{\lambda_{\widetilde T_{a^*M_{a^*}}}\!(t|\bm w)\}}_{\text{TE}_{a,a^*|\bm w}(t)} \nonumber \\
    = &  \underbrace{\log\{\lambda_{\widetilde T_{aM_a}}(t|\bm w)\} \!-\! \log\{\lambda_{\widetilde T_{aM_{a^*}}}(t|\bm w)\}}_{\text{NIE}_{a,a^*|\bm w}(t)} 
     + \underbrace{\log\{\lambda_{\widetilde T_{aM_{a^*}}}(t|\bm w)\} \!-\! \log\{\lambda_{\widetilde T_{a^*M_{a^*}}}(t|\bm w)\}}_{\text{NDE}_{a,a^*|\bm w}(t)},
\end{align*}
where $\lambda_{\widetilde T_{aM_{a^*}}}(t|\bm w)$ is the hazard of the counterfactual outcome $\widetilde T_{aM_{a^*}}$ conditional on the covariates $\bm W=\bm w$. The NIE measures that, at time $t$, the extent to which the log-hazard of the outcome would change if the mediator were to change from the level that would have been observed under $a^*$ versus $a$ but the exposure is fixed at level $a$. The NDE measures how much the log-hazard of the outcome at time $t$ would change if the exposure is set to be level $a$ versus $a^*$ but the mediator is fixed at the value it would have taken under exposure level $a^*$. Given the TE and NIE, the MP is defined as $\text{MP}_{a,a^*|\bm w}(t)=\text{NIE}_{a,a^*|\bm w}(t)/\text{TE}_{a,a^*|\bm w}(t)$.  Notice that in the survival data context all of the mediation effect measures are functions of the follow-up time $t$. Our primary interest is to estimate $\tau_{a,a^*|\bm w}(t)$, where $\tau$ can be NIE, NDE, TE, or MP. We require several causal assumptions to identify the mediation effect measures, including consistency, positivity, composition and a set of no unmeasured confounding assumptions; details of these assumptions are given in Web Appendix A. In addition to the above standard assumptions for mediation analysis, we further assume that the censoring time is conditionally independent from the survival time given covariates such that $C \perp \widetilde T | \{A,M,\bm W\}$. 

To assess mediation, we consider the following semiparametric structural models for the mediator and outcome introduced in \citet{vanderweele2011causal}:
\begin{gather}
    M = \alpha_0 + \alpha_1 A + \bm \alpha_2^T \bm W + \epsilon_M, \label{m:mediator} \\
    \lambda_{\widetilde T} (t|A,M,\bm W) = \lambda_0 (t) e^{\beta_1 A + \beta_2 M + \beta_3 AM + \bm\beta_4^T \bm W}, \label{m:outcome}
\end{gather}
where $\epsilon_M \sim N(0,\sigma_M^2)$, $\lambda_{\widetilde T} (t|A,M,\bm W)$ is the hazard of the survival time conditional on levels of $[A,M,\bm W]$, $\lambda_0 (t)$ is the unspecified baseline hazard, and $\beta_3$ is the exposure-mediator interaction on the log-hazard ratio scale. Let $\bm\alpha=[\alpha_0,\alpha_1,\bm\alpha_2^T,\sigma_M^2]^T$, $\bm\beta=[\beta_1,\beta_2,\beta_3,\bm\beta_4^T]^T$, and $\bm\theta=\left[\bm\alpha^T,\bm\beta^T\right]^T$ be all the parameters in the mediator and outcome models. From \eqref{m:mediator} and \eqref{m:outcome} and under the aforementioned identifiability assumptions, \cite{vanderweele2011causal} showed that
\begin{equation}\label{exact_expressions}
\begin{aligned}
\text{NIE}_{a,a^*|\bm w}(t) & = \delta_{a,a|\bm w}\left(\Lambda_0(t),\bm\theta\right) - \delta_{a,a^*|\bm w}\left(\Lambda_0(t),\bm\theta\right), \\
\text{NDE}_{a,a^*|\bm w}(t) & = \delta_{a,a^*|\bm w}\left(\Lambda_0(t),\bm\theta\right) - \delta_{a^*,a^*|\bm w}\left(\Lambda_0(t),\bm\theta\right),
\end{aligned}
\end{equation}
where $\Lambda_0(t) = \int_{0}^t \lambda_0(u) du$ is the cumulative baseline hazard,
\begin{align*}
\delta_{a,a^*|\bm w}\left(\Lambda_0(t),\bm\theta\right) = &\log\left\{r_{a,a^*|\bm w}(t)\right\} + (\beta_1 a+ \bm \beta_4^T \bm w) + (\beta_2 + \beta_3 a)(\alpha_0 + \alpha_1 a^* + \bm\alpha_2^{T}\bm w) \\
&+ 0.5(\beta_2+\beta_3a)^2\sigma_M^2,
\end{align*}
and
\begingroup\makeatletter\def\f@size{11}\check@mathfonts
$$
r_{a,a^*|\bm w}(t) \!=\!\frac{ \int_m \exp\left(\!-\! \Lambda_0(t) e^{(\beta_2 + \beta_3 a)^2+\beta_1 a + \beta_2 m + \beta_3 am + \bm\beta_4^T \bm w} \right) \exp\left(-0.5(m\!-\!\alpha_0\!-\!\alpha_1a^*-\!\bm\alpha_2^T\bm w)^2\right) d m }{\int_m \exp\left(- \Lambda_0(t) e^{\beta_1 a + \beta_2 m + \beta_3 am + \bm\beta_4^T \bm w} \right) \exp\left(-0.5(m \!-\!\alpha_0\!-\!\alpha_1a^*\!-\!\bm\alpha_2^T\bm w)^2\right) d m}.
$$
\endgroup
Expressions for the $\text{TE}_{a,a^*|\bm w}(t)$ and $\text{MP}_{a,a^*|\bm w}(t)$ can be obtained by summing the NIE and NDE expressions and taking the ratio of the NIE and TE expressions, respectively. Therefore, once estimates of $\bm\theta$ and $\Lambda_0(t)$ are obtained, one can substitute them into \eqref{exact_expressions} to obtain the point estimates for mediation effect measures of interest. 

While the exact expressions \eqref{exact_expressions} are appealing to calculate the mediation effect measures from a theoretical point of view, it may require intensive computations because the expressions involve integrals that typically do not have a closed-form solution. In health science studies, the survival outcome is oftentimes rare (for example, the cumulative event rate is below 10\%, or $\Lambda_0(t) \approx 0$), in which case the mediation effect measure expressions can be much further simplified \citep{vanderweele2011causal}. Specifically, because $r_{a,a^*|\bm w}(t)\approx 1$ when $\Lambda_0(t) \approx 0$, we arrive at the following approximate expressions under the rare outcome assumption (also referred to as the rare disease assumption in epidemiologic case-control studies):
\begin{align*}
    \text{NIE}_{a,a^*|\bm w}(t) & \approx (\beta_2\alpha_1 + \beta_3\alpha_1 a)(a-a^*), \\
    \text{NDE}_{a,a^*|\bm w}(t) & \approx \left\{\beta_1+\beta_3(\alpha_0+\alpha_1 a^* + \bm\alpha_2^T \bm w +\beta_2\sigma_M^2)\right\}(a-a^*) + 0.5 \beta_3^2 \sigma_M^2 (a^2-{a^*}^2).
\end{align*}
which are free of the baseline hazard and therefore are constant over time $t$. Approximate expressions of TE and MP can be obtained accordingly. If we further assume no mediator-exposure interaction in the Cox outcome model ($\beta_3=0$), we obtain $\text{NIE}_{a,a^*|\bm w}(t)\approx \beta_2\alpha_1(a-a^*)$ and $\text{NDE}_{a,a^*|\bm w}(t)\approx \beta_1(a-a^*)$. In what follows, we primarily discuss the mediation effect measures based upon the approximate expressions under a rare outcome assumption. Since the approximate mediation effect measures are constant over $t$, we shall abbreviate $\tau_{a,a^*|\bm w}(t)$ to $\tau_{a,a^*|\bm w}$ without ambiguity. Extensions to compute the exact expressions under a common outcome assumption are provided in Section \ref{sec:common_outcome}.

\section{Implications of exposure measurement error on assessing mediation}
\label{sec:error}

{The exposure variable in health science research is usually measured with error due to technology or budget limitations. In the HPFS, vigorous physical activity (the exposure of interest in our application study) was assessed through self-administrative questionnaires. While it is known to be error-prone, using self-administrative questionnaires is a more cost-effective and practical option, comparing to the gold-standard measurement of physical activity through diaries.} In the presence of exposure measurement error, only a mismeasured version of exposure, $A^*$, is available so that we only observe $\{T_i,\delta_i,M_i,A_i^*,\bm W_i\}$, $i=1,\dots,n_1$. We denote this data sample as the main study to distinguish it from the validation study that will be introduced later. We assume the relationship between $A$ and $A^*$ can be characterized by the following Berkson-like measurement error model (\citealp{carroll2006measurement,rosner1990correction}):
\begin{equation}\label{m:error}
    A = \gamma_0 + \gamma_1 A^* + \bm\gamma_2^T \bm W + \epsilon_A,
\end{equation}
where $\epsilon_A$ is a mean-zero error term with a constant variance $\sigma_A^2$ and is independent of $A^*$ and $\bm W$. In two later occasions, we additionally assume that $\epsilon_A$ follows a normal distribution; these include the bias formula in Theorem \ref{thm1} and the ORC2 approach for bias-corrected estimation of the Cox outcome model parameters (Section \ref{sec:orc}). Beyond these two occasions, the normality assumption on $\epsilon_A$ is not required. In model \eqref{m:error}, we refer to $\gamma_1$ as the \textit{calibration coefficient} representing the association between $A$ and $A^*$ conditional on $\bm W$.  
Throughout, we employ the standard surrogacy assumption such that $P(\widetilde T |A,A^*,M,\bm W)=P(\widetilde T |A,M,\bm W)$ and $P(M|A,A^*,\bm W)=P(M|A,\bm W)$; that is, the error-prone measurement $A^*$ is not informative for the outcome and mediator given the true exposure $A$.

If the true exposure $A$ is replaced by the error-prone exposure $A^*$ in models \eqref{m:mediator} and \eqref{m:outcome}, then estimators of the regression parameters $\bm\theta$ may be biased; and hence estimates for the mediation effect measures will also be subject to bias. Let $\widehat{\tau}_{a,a^*|\bm w}^*$ be the uncorrected estimator of $\tau_{a,a^*|\bm w}$ when $A$ is not available in models \eqref{m:mediator} and \eqref{m:outcome} and  $A^*$ is used instead. The following result characterizes the bias of the uncorrected estimators under the rare outcome scenario and in the absence of exposure-mediator interaction ($\beta_3=0$ in model \eqref{m:outcome}).

\begin{theorem}\label{thm1}
Define $RelBias(\widehat\tau_{a,a^*|\bm w}^*)= \underset{n\rightarrow \infty}{\text{lim}}(\widehat\tau_{a,a^*|\bm w}^* - \tau_{a,a^*|\bm w})/{\tau_{a,a^*|\bm w}}$ as the relative bias of the uncorrected estimator $\widehat\tau_{a,a^*|\bm w}^*$ and
$$
\rho={\sigma_A^2}/(\sigma_A^2+\sigma_M^2/\alpha_1^2).
$$ 
Suppose the outcome is rare and $\beta_3=0$ in model \eqref{m:outcome}. Then, if the error term in model \eqref{m:error} is normally distributed such that $\epsilon_A \sim N(0,\sigma_A^2)$, the asymptotic relative bias for each uncorrected estimator of the mediation effect measures are given by 
\begin{align*}
RelBias(\widehat{\text{NIE}}_{a,a^*|\bm w}^*) & \approx \gamma_1-1 + \gamma_1\rho\left({1}/{\text{MP}_{a,a^*|\bm w}}-1\right)\\
RelBias(\widehat{\text{NDE}}_{a,a^*|\bm w}^*) & \approx  (\gamma_1-1)(1-\rho) - \rho, \\
RelBias(\widehat{\text{TE}}_{a,a^*|\bm w}^*) & \approx \gamma_1-1 \\
RelBias(\widehat{\text{MP}}_{a,a^*|\bm w}^*) & \approx \rho\left({1}/{\text{MP}_{a,a^*|\bm w}}-1\right).
\end{align*}
\end{theorem}

The proof of Theorem \ref{thm1} is presented in Web Appendix B. Theorem \ref{thm1} indicates that, as long as the outcome is rare with no exposure-mediator interaction, the relative bias of the uncorrected estimators only depends on two parameters, $\gamma_1$ and $\rho$, where $\gamma_1$ is the calibration coefficient and $\rho$, as is defined in Theorem \ref{thm1}, is a proportion that ranges from 0 to 1. In this paper, we refer to $\rho$ as the \textit{MP reliability index} as it drives the relative bias of the  uncorrected estimator of MP. Evidently, all uncorrected estimators will be unbiased when $\gamma_1=1$ and $\rho=0$, otherwise the uncorrected estimators tend to have a specific direction of bias. For example, the uncorrected NIE estimator will be biased away or toward the null when $\gamma_1$ is larger or smaller than 
$$
{1}\Big/\left(1+\rho(1/\text{MP}_{a,a^*|\bm w}^*-1)\right).
$$
The uncorrected NDE estimator will only have a positive relative bias when $\gamma_1> 1/(1-\rho)$. The direction of relative bias of the uncorrected TE estimator only depends on $\gamma_1$, with $\gamma_1>1$ and $<1$ corresponding to a positive and negative bias, respectively. The uncorrected MP estimator will always have a positive relative bias since $\rho \geq 0$, hence always exaggerating the true MP.

\begin{figure}[h]
\centering\includegraphics[width=0.6\textwidth]{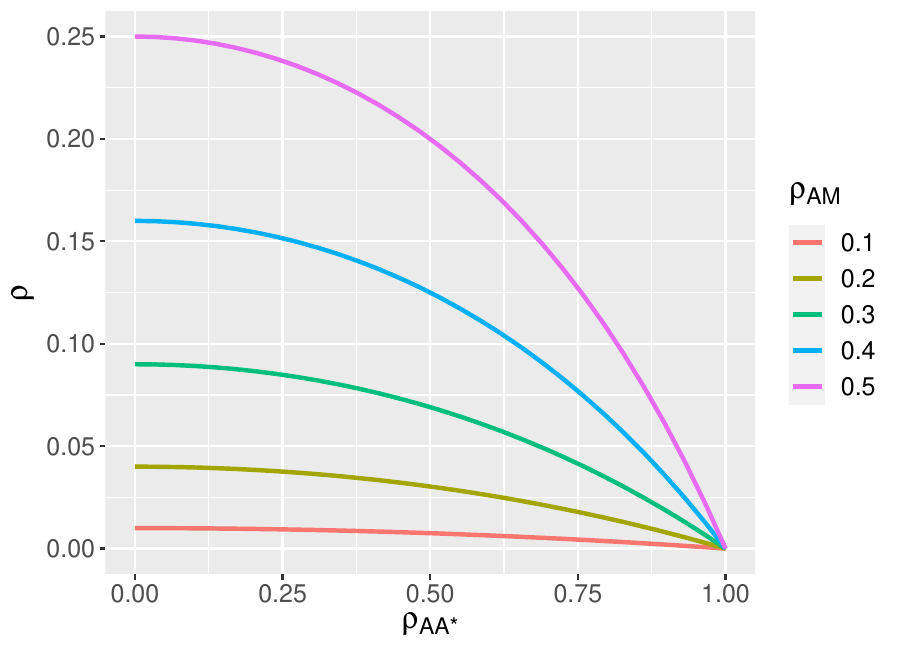}
\caption{A visualization of the MP reliability index ($\rho$) in terms of the correlation coefficient between exposure and mediator ($\rho_{AM}$) and the correlation coefficient between the exposure and its mismeasured counterpart ($\rho_{AA^*}$).}
\label{fig:reliability_index}
\end{figure}

Below we give some intuitions about the MP reliability index, $\rho$, in the absence of covariates. Let $\rho_{AA^*} = \text{Corr}(A,A^*)$ and $\rho_{AM}=\text{Corr}(A,M)$ be the correlation coefficients between $A$ and $A^*$ and between $A$ and $M$, respectively. When $\bm W=\emptyset$, We show in Web Appendix B.2 that 
$$
\rho = \frac{1-\rho_{AA^*}^2}{1/\rho_{AM}^2-\rho_{AA^*}^2},
$$ 
which is bounded within $\rho \in [0,\rho_{AM}^2]$ regardless of the value of $\rho_{AA^*}$. Therefore, $\rho$ will be close to 0 if either the exposure measurement error is small (i.e., $|\rho_{AA^*}|$ is large) or the exposure-mediator association is weak (i.e., $\rho_{AM}$ is small). Under either condition, the uncorrected NIE, NDE, and TE estimators will share the same asymptotic relative bias, $\gamma_1-1$, and the MP estimator will be approximately unbiased as $\rho\approx 0$. Figure \ref{fig:reliability_index} provides a visualization on the value of $\rho$ in terms of $\rho_{AA^*}$ and $\rho_{AM}$, and indicates that $\rho$ is below 0.1 when either (i) $\rho_{AM} \leq 0.3$ or (ii) $\rho_{AM} \leq 0.4$ and $\rho_{AA^*} \geq 0.65$. Therefore, the uncorrected MP estimator will often exhibit a relatively small bias when the exposure-mediator association is not strong alongside a small to moderate amount of exposure measurement error.

In Web Appendix B.3, we discuss an interesting observation that all bias formulas in Theorem \ref{thm1} happen to coincide with the exposure measurement error-induced bias in mediation analysis with a binary outcome in \citet{cheng2022mediation}, even although the mediation effect measures are defined differently and different estimators are considered. Specifically, the present work defines the mediation effect measures on the log-hazard ratio scale and performs mediation analysis based on a Cox outcome model and a linear mediator model. In contrast, \citet{cheng2022mediation} defined the mediation effect measures on the log-risk ratio scale and performed mediation analysis based on the difference method with two log-binomial outcome models (no mediator model is needed under their formulation). Despite these differences, the bias formulas due to exposure measurement error happen to be identical.

To draw analytic insights on the implications of exposure measurement error, Theorem \ref{thm1} is derived under the assumption with a rare outcome with no exposure-mediator interaction. In the presence of an exposure-mediator interaction, the bias will have a more complex form and can depend on all of regression parameters $\bm\theta$, and thus analytically intractable. Next, we develop bias correction approaches to ensure valid estimation for mediation effect measures with a mismeasured exposure, in the more general settings that permit an exposure-mediator interaction and even a common outcome.

\section{Estimation of the mediator and outcome model parameters}
\label{sec:methods}

We now propose methods to correct for the exposure measurement error in estimating mediation effect measures. First, this Section provides methods for bias-corrected estimation of the unknown parameters in the mediator and outcome models \eqref{m:mediator}--\eqref{m:outcome}. Then, one can substitute the bias-corrected parameter estimates from the mediator and outcome models into the approximate or exact expressions of the mediation effect measures to obtain the final estimates of mediation effect measures. 
In addition to the main study, $\{T_i,\delta_i,M_i,A_i^*,\bm W_i\}$, $i=1,\dots,n_1$, we require a validation study to model the measurement error process. Motivated by HPFS, we consider an external validation study such that the validation study consists of data $\{T_i,M_i,A_i^*,A_i,\bm W_i\}$, $i=n_1+1,\dots,n$,  as in previous work by \cite{liao2011survival}. Here, we include the observed failure time $T$ but not the failure indicator $\delta$ in the validation study, because $T$ along with $\{M,A^*,A,\bm W\}$ is the minimum set of variables required in validation study to perform all regression calibration methods developed in Section \ref{sec:methods}. Despite the availability of these data elements, it should be noted that data on $T$ in the validation study is not required to estimate the mediator model (Section \ref{sec:correction_mediator_model}) nor it is required to perform the ordinary regression calibration (Section \ref{sec:orc}); data on $T$ in the validation study is only required to perform the risk-set regression calibration (Section \ref{sec:rrc}).

To ensure that the parameters estimated in the validation study can be validly transported to the main study, we require the transportability assumption such that the joint distribution of $P(A, M,  T, C|A^*,\bm W)$ in the validation study is identical to its counterpart in the main study (\citealp{yi2015functional}). Remarks 1 and 2 in Web Appendix A provide further discussions about the transportability assumption and its connection to an ignorable selection into the validation study assumption \citep{spiegelman2001efficient}. {The transportability assumption is plausible in our data application, because (i) both the main and validition study in the HPFS uses the same way to assess the mismeasured version of exposure $A^*$ through self-administrated questionnaires and (ii) study participants in the validation study are draw from the same target population with the HPFS main study.}

The parameters in the measurement error model \eqref{m:error}, $\bm \gamma=[\gamma_0,\gamma_1,\bm\gamma_2^T,\sigma_A^2]^T$, can be estimated by regressing $A$ on $\{A^*,\bm W\}$ in the validation study; i.e., $\widehat{\bm \gamma}$ solves
\begin{equation}\label{e:error}
\sum_{i=1}^n (1-R_i)\left\{\begin{matrix}
(1,A_i^*,\bm W_i^T)^T (A_i-\gamma_0-\gamma_1 A_i^*-\bm\gamma_2^T \bm W_i) \\
\sigma_A^2 - (A_i-\gamma_0-\gamma_1 A_i^*-\bm\gamma_2^T \bm W_i)^2
\end{matrix}\right\} = \bm 0,
\end{equation}
where $R_i$ is an indicator with $R_i=0$ if the individual is in the validation study (i.e., $i > n_1$) and $R_i=1$ if the individual is in the main study (i.e., $i \leq n_1$). One can verify that $[\widehat\gamma_0,\widehat\gamma_1,\widehat{\bm\gamma}_2^T]$ is the ordinary least squares (OLS) estimator of the regression coefficients in model \eqref{m:error} and $\widehat{\sigma}_A^2$ is the empirical variance of the residuals of model \eqref{m:error} in the validation study. Notice that $\widehat{\bm\gamma}$ is consistent if \eqref{m:error} is correctly specified, regardless of whether $\epsilon_A$ is normally distributed. In what follows, we discuss estimation of  parameters in the mediator and outcome models, separately.

\subsection{Estimation of the mediator model parameters}\label{sec:correction_mediator_model}

The mediator model \eqref{m:mediator} involves variables $\{M_i,A_i,\bm W_i\}$, which are fully observed in the validation study. A convenient approach for estimating $\bm \alpha$ is to regress $M$ on $A$ and $\bm W$ in the validation study alone but this dispenses with any information from the main study. Below we pursue an approach to estimate the mediator model that efficiently uses the information from both studies. Combining the measurement error model \eqref{m:error} and mediator model \eqref{m:mediator}, we can show that the first two moments of the mediator given the mismeasured exposure and covariates are $E[M|A^*,\bm W]  = \alpha_0+\alpha_1 \mu_{A}(\bm\gamma) + \bm\alpha_2^T \bm W$ and $\text{Var}(M|A^*,\bm W)  = \sigma_M^2 + \alpha_1^2 \sigma_A^2$, 
where $\mu_{A}(\bm\gamma) = \gamma_0 + \gamma_1 A^* + \bm \gamma_2^T \bm W$ is the conditional mean of the $A|A^*,\bm W$ implied by model \eqref{m:error}. Comparing $E[M|A^*,\bm W]$ to $E[M|A,\bm W]$ in \eqref{m:mediator}, one can observe that fitting \eqref{m:mediator} with $A$ replaced by $\mu_{A}(\bm\gamma)$ leads to valid estimates of the regression coeffcients in the mediator model, $[\alpha_0,\alpha_1,\bm\alpha_2^T]$. This suggests the following estimating equations for $[\alpha_0,\alpha_1,\bm\alpha_2^T]$:
\begin{align}
    & \sum_{i=1}^n R_i\left(1,\mu_{A_i}(\widehat{\bm\gamma}),\bm W_i^T\right)^T(M_i\!-\!\alpha_0\!-\!\alpha_1 \mu_{A_i}(\widehat{\bm\gamma}) \!-\! \bm\alpha_2^T \bm W_i) \nonumber \\
    & \quad \quad + (1-R_i) \left(1,A_i,\bm W_i^T\right)^T(M_i\!-\!\alpha_0\!-\!\alpha_1 A_i \!-\! \bm\alpha_2^T \bm W_i) = \bm 0 \label{e:mediator1}
\end{align}
where $\mu_{A_i}(\widehat{\bm\gamma})=\widehat\gamma_0 + \widehat\gamma_1 A^* + \widehat{\bm \gamma}_2^T \bm W$ is obtained by fitting model \eqref{m:error} in the validation study. The estimating equation \eqref{e:mediator1} utilizes data from both the main and the validation studies. For observations in the validation study ($R_i=0$), equation \eqref{e:mediator1} uses the true score function for $[\alpha_0,\alpha_1,\bm\alpha_2^T]$; for observations in the main study ($R_i=1$), equation \eqref{e:mediator1} uses a pseudo score function that replaces $A_i$ in the original score by its predicted value $\mu_{A_i}(\widehat{\bm\gamma})$ based on the fitted measurement error model.

Similar to \eqref{e:mediator1}, we can construct the following estimating equation for $\sigma_M^2$:
\begin{align}
   & \sum_{i=1}^n R_i\Big(\sigma_M^2+\widehat\alpha_1^2\widehat\sigma_A^2-(M_i-\widehat\alpha_0-\widehat\alpha_1 \mu_{A}(\widehat{\bm\gamma}) - \widehat{\bm\alpha}_2^T \bm W_i)^2\Big) \nonumber \\
   & \quad\quad\quad+ (1-R_i) \Big(\sigma_M^2-(M_i-\widehat\alpha_0-\widehat\alpha_1 A_i - \widehat{\bm\alpha}_2^T \bm W_i)^2\Big) = 0, \label{e:mediator2}
\end{align}
where $[\widehat{\alpha}_0,\widehat\alpha_1,\widehat{\bm\alpha}_2^T]$ is obtained by \eqref{e:mediator1}. Equation \eqref{e:mediator2} has an explicit solution given by $$\widehat{\sigma}_M^2 = \frac{1}{n}\sum_{i=1}^n \left[M_i - \widehat{\alpha}_0 - \widehat{\alpha}_1 \Big(R_i\mu_{A}(\widehat{\bm\gamma})+(1-R_i)A_i
\Big)-\widehat{\bm\alpha}_2^T \bm W_i\right]^2 - \frac{n_1}{n}\widehat\alpha_1^2\widehat\sigma_A^2.$$  

\subsection{Estimation of the outcome model parameters via ORC}\label{sec:orc}

Based on the true hazard function \eqref{m:outcome},  the induced hazard function of $\widetilde T$ given the mismeasured exposure $A^*$ instead of $A$ is
\begin{align}
    \lambda_{\widetilde T}(t|A^*, M, \bm W) 
    & = \lambda_0(t)e^{\beta_2 M + \bm\beta_4^T \bm W }E[e^{ (\beta_1+\beta_3M)A }|A^*,M,\bm W, T \geq t]. \label{induced_hazard}
\end{align}
From \eqref{induced_hazard}, we see the critical quantity is $E[e^{ (\beta_1+\beta_3M)A }|A^*,M,\bm W, T \geq t]$, which does not have an explicit expression. The ordinary regression calibration (ORC) approach uses the rare outcome assumption to approximate this critical quantity. \citet{prentice1982covariate} has shown that  $E[e^{ (\beta_1+\beta_3M)A }|A^*,M,\bm W,  T \geq t] \approx E[e^{ (\beta_1+\beta_3M)A }|A^*,M,\bm W]$ if the the outcome is rare. Then, we can further use a first-order Taylor expansion to approximate $E[e^{(\beta_1+\beta_3M)A}|A^*,M,\bm W]$ by $e^{ (\beta_1+\beta_3M)E[A|A^*,M,\bm W]}$. The Taylor approximation is exact when $A|\{A^*,M,\bm W\}$ is normally distributed with a constant variance and in the absence of any exposure-mediator interaction. Otherwise, the approximation is valid when either (i) $A|\{A^*,M,\bm W\}$ is normal and $(\beta_1+\beta_3M)^2\text{Var}(A|A^*,M,\bm W)$ is small, or  (ii)  $|\beta_1|+|\beta_3|$ and $(\beta_1+\beta_3M)^2\text{Var}(A|A^*,M,\bm W)$ are small, or (iii) $|\beta_1|$ is small with no exposure-mediator interaction and $\text{Var}(A|A^*,M,\bm W)=\sigma^2$ is constant (see Web Appendix C.1 for a detail discussion about the approximation). Based on this approximation, we can rewrite the induced hazard  \eqref{induced_hazard} as the following Cox model 
\begin{equation}\label{cox:orc}
\lambda_{\widetilde T}(t|A^*, M, \bm W) \approx \lambda_0(t)e^{\beta_1 E[A|A^*,M,\bm W] + \beta_2 M + \beta_3 M E[A|A^*,M,\bm W] + \bm\beta_4^T \bm W }.
\end{equation}
This indicates that, with an appropriate estimator $\widehat{E}[A|A^*,M,\bm W]$, we can obtain $\widehat{\bm\beta}$ by simply implementing Cox model \eqref{m:outcome} based on the main study with $A$ replaced by $\widehat{E}[A|A^*,M,\bm W]$.

We provide two strategies to calculate $\widehat{E}[A|A^*,M,\bm W]$. First, observing that all of the variables used in $E[A|A^*,M,\bm W]$ are fully observed in the validation study, one can specify a model for $E[A|A^*,M,\bm W]$ based on the validation study directly. For example, we can first specify the following regression for $E[A|A^*,M,\bm W]$:
\begin{equation}\label{e_aamw1}
    E[A|A^*,M,\bm W] = \eta_0 + \eta_1 A^* + \eta_2 M + \bm\eta_3^T \bm W,
\end{equation}
with coefficients $\bm\eta=[\eta_0,\eta_1,\eta_2,\bm\eta_3]$. Next, we calculate the OLS estimator, $\widehat{\bm\eta}$, by solving
\begin{equation}\label{e:orc_error}
    \sum_{i=1}^n (1-R_i)\left(1,A_i^*,M_i,\bm W_i^T\right)^T\left(A_i-\eta_0-\eta_1A_i^*-\eta_2 M_i-\bm\eta_3^T\bm W_i\right)=\bm 0,
\end{equation}
and then obtain $\widehat{E}[A|A^*,\bm M,\bm W] = m_A^{(O1)}(\widehat{\bm\eta})$, where $m_A^{(O1)}(\widehat{\bm\eta})=\widehat\eta_0 + \widehat\eta_1 A^* + \widehat\eta_2 M + \widehat{\bm\eta}_3^T \bm W$ is the estimated regression function in \eqref{e_aamw1}. An alternative approach to calculate $\widehat{E}[A|A^*,\bm M,\bm W]$ is deriving its analytic form based on the measurement error model \eqref{m:error} and the mediator model \eqref{m:mediator}. This approach requires the distributional assumption of the error term in model \eqref{m:error}, $\epsilon_A$. When $\epsilon_A$ is normally distributed, we show that $E[A|A^*,M,\bm W]=m_{A}^{(O2)}({\bm\alpha},{\bm\eta})$, where
\begin{equation}\label{e_aamw2}
   m_{A}^{(O2)}({\bm\alpha},{\bm\gamma}) = \frac{\gamma_0\sigma_M^2-\alpha_0\alpha_1\sigma_A^2}{\alpha_1^2\sigma_A^2+\sigma_M^2} + \frac{\gamma_1\sigma_M^2}{\alpha_1^2\sigma_A^2+\sigma_M^2}A^* + \frac{\alpha_1\sigma_A^2}{\alpha_1^2\sigma_A^2+\sigma_M^2}M + \frac{\sigma_M^2\bm\gamma_2^T-\alpha_1\sigma_A^2\bm\alpha_2^T}{\alpha_1^2\sigma_A^2+\sigma_M^2}\bm W 
\end{equation}
by applying the Bayes formula on \eqref{m:error} and \eqref{m:mediator} (See Lemma 1 in the Supplementary Material for a proof). Therefore, the second strategy is to estimate $\widehat{E}[A|A^*,\bm M,\bm W]=m_{A}^{(O2)}(\widehat{\bm\alpha},\widehat{\bm\gamma})$ by replacing $(\bm\alpha,\bm\gamma)$ in \eqref{e_aamw2} with their corresponding point estimators.

We denote $\widehat{\bm\beta}^{(O1)}$ and $\widehat{\bm\beta}^{(O2)}$ as the ORC estimators of $\bm\beta$ by using the strategies \eqref{e_aamw1} and \eqref{e_aamw2}, respectively. Comparing both estimators, $\widehat{\bm\beta}^{(O1)}$ requires fitting an additional model for $E[A|A^*,\bm M, \bm W]$, whereas $\widehat{\bm\beta}^{(O2)}$ more effectively use the information from the measurement error model and mediator model to derive an analytic form of $E[A|A^*,\bm M, \bm W]$. Validity of $\widehat{\bm\beta}^{(O2)}$ depends on a correctly specified measurement error model \eqref{m:error} with a normal error term, where  $\widehat{\bm\beta}^{(O1)}$ only requires that the mean model \eqref{e_aamw1} is correctly specified.

\subsection{Estimation of the outcome model parameters via RRC}\label{sec:rrc}

Performance of the ORC may be compromised under departure from the rare outcome assumption. Following \cite{xie2001risk} and \cite{liao2011survival}, we further consider a risk set calibration (RRC) approach to estimate $\bm\beta$, which is based on an improved approximation to the induced hazard \eqref{induced_hazard} without the rare outcome assumption. Specifically, we consider a first order Taylor expansion to approximate the critical quantity $E[e^{ (\beta_1+\beta_3M)A }|A^*,M,\bm W,T \geq t]$ by $e^{ (\beta_1+\beta_3M)E[A|A^*,M,\bm W,T \geq t]}$ and then the induced hazard \eqref{induced_hazard} becomes
\begin{equation}\label{cox:rrc}
\lambda_{\widetilde T}(t|A^*, M, \bm W) \approx \lambda_0(t)e^{\beta_1 E[A|A^*,M,\bm W,T \geq t] + \beta_2 M + \beta_3 M E[A|A^*,M,\bm W,T \geq t] + \bm\beta_4^T \bm W }.
\end{equation}
This approximation is exact when $A|\{A^*,M,\bm W, T \geq t\}$ is normal, the variance $\text{Var}(A|A^*, M,\bm W, T \geq t)$ is only a function of $t$, and there is no exposure-mediator interaction ($\beta_3=0$). The approximation is also valid when (i) $A|\{A^*,M,\bm W,T\geq t\}$  is normal and $(\beta_1+\beta_3M)^2\text{Var}(A|A^*,M, \bm W,T\geq t)$ is small, 
or (ii)  $|\beta_1|+|\beta_3|$ and $(\beta_1+\beta_3M)^2\text{Var}(A|A^*,M,\bm W,T\geq t)$ are small, or (iii) $|\beta_1|$ is small with no exposure-mediator interaction and $\text{Var}(A|A^*,M,\bm W,T\geq t)$ only depends on $t$ (see Web Appendix C.2 for a detailed discussion of the conditions for the approximation). We then use the following regression model for $E[A|A^*,M,\bm W,T\geq t]$:
\begin{equation}\label{e_aamwt}
 E[A|A^*,M,\bm W,T\geq t] = \xi_0(t) + \xi_1(t) A^* + \xi_2(t) M + \bm\xi_3^T(t) \bm W,
\end{equation}
where $\bm\xi(t)=[\xi_0(t),\xi_1(t),\xi_2(t),\bm\xi_0^T(t)]^T$ are time-dependent coefficients. The basic idea underlying the RRC is to estimate model \eqref{e_aamwt} from the validation study at each main study failure time point and then running the Cox model \eqref{m:outcome} among the main study samples with $A_i$ replaced by its predicted value from \eqref{e_aamwt}. However, this may be computationally burdensome when the number of unique failure times  in the main study is large. {More importantly, because a precise measurement of the exposure is often expensive in health sciences research, the validation study is usually too small to accurately estimate model \eqref{e_aamwt} at each main study failure time. For example, the validation study in HPFS includes a total of 238 participants, which is much smaller compared to the main study sample size ($n_1=43,573$).
Given the small validation study size, we propose to group the risk sets to stabilize the estimation results and facilitate computation, in a similar spirit to \cite{zhao2014covariate} and \cite{liao2018survival}.}

The proposed RRC estimation procedure is as follows: Step 1, divide the main study follow-up duration $[0,t^*]$ in to $K$ intervals, $[t_1,t_2)$, $[t_2,t_3)$, $\dots$, $[t_{K},t^*]$, where $t_1=0$. Step 2, find the $K$ risk sets in the validation study, $R_v(t_k)$, $k=1,\dots,K$, such that $R_v(t_k)$ includes all individuals with $(1\!-\!R_i)Y_i(t_k) = 1$, where $Y_i(t) = \mathbb{I}(T_i \geq t)$ is the at-risk indicator. Step 3, for each $t_k$, obtain $\widehat{\bm \xi}(t_k)$ by regressing $A$ on $(A^*,M,\bm W)$ among individuals in the risk set $R_v(t_k)$; i.e., obtain $\widehat{\bm \xi}(t_k)$ by solving the estimating equation,  \begin{equation}\label{e:rrc_error}
        \sum_{i=1}^n  (1\!-\!R_i)Y_i(t_k)\left(1,
    A_i^*,
    M_i,
    \bm W_i^T\right)^T\Big(A_i - \xi_0(t_k) - \xi_1(t_k) A_i^* - \xi_2(t_k) M_i - \bm\xi_3^T(t_k) \bm W_i\Big) = \bm 0.
\end{equation}
Step 4, let $m_{A_i}^{(R)}(\widehat{\bm\xi},t) = \widehat \xi_0(\widetilde t) + \widehat \xi_1(\widetilde t) A_i^* + \widehat \xi_2(\widetilde t) M_i + \widehat{\bm \xi}_3^T(\widetilde t) \bm W_i$, where $\widetilde t$ is the most recent available $t_k$ that is at or prior to $t$. Step 5, run the Cox model \eqref{m:outcome} in the main study with $A_i$ replaced by the time-varying variable $m_{A_i}^{(R)}(\widehat{\bm\xi},t)$ to get the point estimator of $\bm\beta$, denoted by $\widehat{\bm \beta}^{(R)}$. 

The choice of the split points $t_k$, $k=1,\dots,K$, can be flexible. For example, following \cite{zhao2014covariate}, one can equally divide the time scale into $K$ intervals such that $t_k = \frac{(k-1)t_v^*}{K}$, where $t_v^*$ is the maximum observed failure time in the validation study.  Notice that $\widehat{\bm\beta}^{(R)}$ coincides with $\widehat{\bm\beta}^{(O1)}$ for $K=1$ since there are no split points between $[0,t^*]$. Performance of RRC with different number of $K$ is evaluated in Section \ref{sec:sim}. To determine the optimal number of risk sets ($K$), one can also consider a range of candidate values of $K$, e.g., $K\in\{1,\dots,10\}$, and then use a cross-validation approach in the validation study samples to select the one with smallest mean squared error or mean absolute error.

\subsection{Asymptotic properties and variance estimation}\label{sec:asmp}

We have proposed one estimator for the mediator model, $\widehat{\bm\alpha}$, and three estimators for the Cox outcome model, the ORC1 ($\widehat{\bm\beta}^{(O1)}$), the ORC2 ($\widehat{\bm\beta}^{(O2)}$), and the RRC ($\widehat{\bm\beta}^{(R)}$). Notably, these three estimators of $\bm\beta$ can all be cast as the solution to the following estimating equation induced from model \eqref{m:outcome}, but with different choice of $m_{A_i}(t)$: 
 \begin{equation}\label{e:outcome}
    \sum_{i=1}^n R_i\int_0^{t^*} \left\{\bm Z_i(t)-\frac{\sum_{j=1}^{n_1}  Y_j(t) \exp\{\bm\beta^T\bm Z_j(t)\}\bm Z_j(t)}{\sum_{j=1}^{n_1}  Y_j(t) \exp\{\bm\beta^T\bm Z_j(t)\}} \right\}d N_i(t) = \bm 0,
\end{equation}
where $\bm Z_i(t) = \left[m_{A_i}(t),M_i,m_{A_i}(t)\times M_i,\bm W_i^T\right]^T$ is the design matrix, $N_i(t)=\mathbb{I}(T_i\leq t, \delta_i=1)$ is the counting process, and $Y_i(t) = \mathbb{I}(T_i\geq t)$ is the at-risk process. Here, $m_{A_i}(t)$ is the imputed value of $A_i$ in the Cox outcome model, which is set as  $m_{A_i}^{(O1)}(\widehat{\bm\eta})$, $m_{A_i}^{(O2)}(\widehat{\bm\alpha},\widehat{\bm\gamma})$, and $m_{A_i}^{(R)}(\widehat{\bm\xi},t)$ by the ORC1, ORC2, and RRC methods, respectively. Notice that $m_{A_i}(t) \equiv m_{A_i}$ is a time-invariant variable for the ORC1 and ORC2 methods, but is a time-dependent variable for the RRC method.
To summarize, we have three estimators for $\bm\theta$ characterized by their outcome model estimation approach, the ORC1, $\widehat{\bm\theta}^{(O1)}=\left[\widehat{\bm\alpha}^T,\widehat{\bm\beta}^{(O1)^T}\right]^T$, the ORC2, $\widehat{\bm\theta}^{(O2)}=\left[\widehat{\bm\alpha}^T,\widehat{\bm\beta}^{(O2)^T}\right]^T$, and the RRC, $\widehat{\bm\theta}^{(R)}=\left[\widehat{\bm\alpha}^T,\widehat{\bm\beta}^{(R)^T}\right]^T$. 

Theorem \ref{thm:orc1} states that  $\widehat{\bm\theta}^{(O1)}$ is approximately consistent and asymptotically normal, and the proof is given in Web Appendix D.
\begin{theorem}\label{thm:orc1}
Under regularity conditions in Web Appendix D.1, $\widehat{\bm\theta}^{(O1)} \overset{p}{\to} \bm\theta_0^{(O1)}$, where $\bm\theta_0^{(O1)}=\left[\bm\alpha_0^T, {\bm\beta_0^{(O1)}}^T\right]^T$, $\bm\alpha_0$ is the true value of $\bm\alpha$ in the mediator model \eqref{m:mediator}, and $\bm\beta_0^{(O1)}$ is the true value of $\bm\beta$ in the approximate hazard model \eqref{cox:orc} with $E[A|A^*,M,\bm W]$ defined in \eqref{e_aamw1}. In addition, we have $\sqrt{n}\left(\widehat{\bm\theta}^{(O1)}-\bm\theta_0^{(O1)}\right) \overset{D}{\to} N\left(\bm 0,\bm V^{(O1)}\right)$, where $\bm V^{(O1)}$ is provided in Web Appendix D.2. 
\end{theorem}
This result shows that $\widehat{\bm\alpha}$ converges to the true value of $\bm\alpha$ but  $\widehat{\bm\beta}^{(O1)}$ converges to $\bm\beta_0^{(O1)}$ that is close to
 but generally different from $\bm\beta$ in the true Cox model \eqref{m:outcome}. Also, Theorem \ref{thm:orc1} suggests the following sandwich variance estimator of $\bm V^{(O1)}$:
$$
\widehat{\bm V}^{(O1)} = \left[\widehat{\bm I}_{\bm\theta}^{(O1)}\right]^{-1} \widehat{\bm M}_{\bm\theta}^{(O1)} \left[\widehat{\bm I}_{\bm\theta}^{(O1)}\right]^{-T},
$$ 
where $
\widehat{\bm M}_{\bm\theta}^{(O1)} \!=\! \frac{1}{n}\displaystyle\sum_{i=1}^n \left\{\eU_{\bm\theta,i}^{(O1)}(\widehat{\bm\theta}^{(O1)},\widehat{\bm\gamma},\widehat{\bm\eta}) - \widehat{\bm I}_{\bm\theta,\bm\gamma}^{(O1)} \widehat{\bm I}_{\bm\gamma}^{-1}\eU_{\bm\gamma,i}(\widehat{\bm\gamma}) - \widehat{\bm I}_{\bm\theta,\bm\eta}^{(O1)} \widehat{\bm I}_{\bm\eta}^{-1}\eU_{\bm\eta,i}(\widehat{\bm\eta}) \right\}^{\otimes 2}
$ and $\widehat{\bm I}_{\bm\theta}^{(O1)}=\frac{1}{n}\frac{\partial}{\partial \bm\theta}\eU_{\bm\theta}^{(O1)}(\widehat{\bm\theta}^{(O1)},\widehat{\bm\gamma},\widehat{\bm\eta})$ and $\bm v^{\otimes 2} = \bm v\bm v^T$ for a vector $\bm v$. Here, we use $\eU_{\bm\omega}(\bm\omega,\bm\omega^*)$ and $\eU_{\bm\omega,i}(\bm\omega,\bm\omega^*)$ to denote the estimating equation and its $i$-th component of some unknown parameter $\bm\omega$, which are functions of $\bm\omega$ and may also depend on some nuisance parameters $\bm\omega^*$. Therefore, $\eU_{\bm\theta}^{(O1)}(\bm\theta,\bm\gamma,\bm\eta)=\left[\begin{matrix}\eU_{\bm\alpha}(\bm\alpha,\bm\gamma) \\ \eU_{\bm\beta}^{(O1)}(\bm\beta,\bm\eta) \end{matrix}\right]$, where $\eU_{\bm\alpha}(\bm\alpha,\bm\gamma)$ is obtained by stacking the left side of \eqref{e:mediator1} and \eqref{e:mediator2}, $\eU_{\bm\beta}^{(O1)}(\bm\beta,\bm\eta)$ is the left side of \eqref{e:outcome} with $m_{A_i}=m_{A_i}^{(O1)}(\bm\eta)$, $\eU_{\bm\gamma}(\bm\gamma)$ is the left side of \eqref{e:error},  $\eU_{\bm\eta}(\bm\eta)$ is the left side of \eqref{e:orc_error}, $\widehat{\bm I}_{\bm\theta,\bm\gamma}^{(O1)}=\frac{1}{n}\frac{\partial}{\partial \bm\gamma}\eU_{\bm\theta}^{(O1)}(\widehat{\bm\theta}^{(O1)},\widehat{\bm\gamma},\widehat{\bm\eta})$, $\widehat{\bm I}_{\bm\theta,\bm\eta}^{(O1)}=\frac{1}{n}\frac{\partial}{\partial \bm\eta}\eU_{\bm\theta}^{(O1)}(\widehat{\bm\theta}^{(O1)},\widehat{\bm\gamma},\widehat{\bm\eta})$, $\widehat{\bm I}_{\bm\gamma}=\frac{1}{n}\frac{\partial}{\partial \bm\gamma}\eU_{\bm\gamma}(\widehat{\bm\gamma})$, and $\widehat{\bm I}_{\bm\eta}=\frac{1}{n}\frac{\partial}{\partial \bm\eta}\eU_{\bm\eta}(\widehat{\bm\eta})$.

Similarly, Theorem \ref{thm:orc2} and \ref{thm:rrc} state the asymptotic properties of  $\widehat{\bm\theta}^{(O2)}$ and $\widehat{\bm\theta}^{(R)}$, and their proofs are given in Web Appendices D.3 and E, respectively.
\begin{theorem}\label{thm:orc2}
Under regularity conditions outlined in Web Appendix D.3, $\widehat{\bm\theta}^{(O2)} \overset{p}{\to} \bm\theta_0^{(O2)}$, where $\bm\theta_0^{(O2)}=\left[\bm\alpha_0^T, {\bm\beta_0^{(O2)}}^T\right]^T$ and $\bm\beta_0^{(O2)}$ is the true value of $\bm\beta$ in the approximate hazard model \eqref{cox:orc} with $E[A|A^*,M,\bm W]$ defined in \eqref{e_aamw2}. In addition, we have that $\sqrt{n}\left(\widehat{\bm\theta}^{(O2)}-\bm\theta_0^{(O2)}\right) \overset{D}{\to} N\left(\bm 0,\bm V^{(O2)}\right)$, where $\bm V^{(O2)}$ is provided in Web Appendix D.3. 
\end{theorem}

\begin{theorem}\label{thm:rrc}
Under regularity conditions outlined in Web Appendix E, $\widehat{\bm\theta}^{(R)} \overset{p}{\to} \bm\theta_0^{(R)}$, where $\bm\theta_0^{(R)}=\left[\bm\alpha_0^T, {\bm\beta_0^{(R)}}^T\right]^T$ and $\bm\beta_0^{(R)}$ is the true value of $\bm\beta$ in the approximate hazard model \eqref{cox:rrc} with $E[A|A^*,M,\bm W, T\geq t]$ defined in \eqref{e_aamwt}. In addition, we have that $\sqrt{n}\left(\widehat{\bm\theta}^{(R)}-\bm\theta_0^{(R)}\right) \overset{D}{\to} N\left(\bm 0,\bm V^{(R)}\right)$, where $\bm V^{(R)}$ is provided in Web Appendix E. 
\end{theorem}

We also derive consistent estimators of $\bm V^{(O2)}$ and $\bm V^{(R)}$ in Web Appendices D.3 and E. The proposed variance estimators for $\bm V^{(O1)}$, $\bm V^{(O2)}$ and $\bm V^{(R)}$ will be evaluated in the subsequent simulations (Section \ref{sec:sim}) and applied to the analysis of HFPS (Section \ref{sec:app}).

\section{Estimation of the mediation effect measures}
\label{sec:estimation_mediation}

\subsection{Estimation based on the approximate expressions with a rare outcome}\label{sec:estimation_approx_expression}

We first consider the approximate mediation effect measure expressions under a rare outcome assumption. In Section \ref{sec:methods}, three point estimators of $\widehat{\bm\theta}^{(m)}$ and their corresponding variance estimator $\widehat{\bm V}^{(m)}$ were developed, where $m$ can be $O1$, $O2$, and $R$ for the ORC1, ORC2, and RRC approaches, respectively. Then, one can substitute $\widehat{\bm\theta}^{(m)}$ into the expressions for the approximate mediation effect measures to obtain their corresponding estimators, $\widehat{\tau}_{a,a^*|\bm w}^{(m)}$. The asymptotic variance of $\widehat{\tau}_{a,a^*|\bm w}^{(m)}$ can be estimated by the multivariate delta method such that $\widehat{\text{Var}}\left(\widehat{\tau}_{a,a^*|\bm w}^{(m)}\right) = \frac{1}{n} \left[\widehat{\bm{I}}_{\tau,\bm\theta}^{(m)}\right] \widehat{\bm V}^{(m)}\left[\widehat{\bm{I}}_{\tau,\bm\theta}^{(m)}\right]^T$, where $\widehat{\bm{I}}_{\tau,\bm\theta}^{(m)}=\frac{\partial \tau_{a,a^*|\bm w}^{(m)}}{\partial \bm\theta}\big|_{\bm\theta=\widehat{\bm \theta}^{(m)}}$ and its explicit expressions are given in Appendix F. Finally, a Wald-type 95\% confidence interval can constructed as  $\Big(\widehat{\tau}_{a,a^*|\bm w}^{(m)}-1.96\times\sqrt{\widehat{\text{Var}}(\widehat{\tau}_{a,a^*|\bm w}^{(m)})},  \widehat{\tau}_{a,a^*|\bm w}^{(m)}+1.96\times\sqrt{\widehat{\text{Var}}(\widehat{\tau}_{a,a^*|\bm w}^{(m)})}\Big)$. 

Alternatively, we can also employ the non-parametric bootstrap to obtain the variance and interval estimates by re-sampling the data for $B$ iterations (for example, $B=1000$). For each iteration, we re-sample individuals with replacement in the main and validation studies, separately. Then, we re-estimate $\widehat{\tau}_{a,a^*|\bm w}^{(m)}$ across each  bootstrap dataset. The empirical variance of $\widehat{\tau}_{a,a^*|\bm w}^{(m)}$ among the the $B$ bootstrap datasets can be a consistent estimator of $\text{Var}\left(\widehat{\tau}_{a,a^*|\bm w}^{(m)}\right)$, from which the confidence interval can be constructed. Alternatively, the 2.5\% and 97.5\% percentiles of empirical distribution of $\widehat{\tau}_{a,a^*|\bm w}^{(m)}$ among the $B$ bootstrap datasets can determine its 95\% confidence interval. Comparing to estimators based on the asymptotic variance, the bootstrap is easier to implement but it can be computationally intensive when the sample size is  large.

\subsection{Estimation based on the exact expressions with a common outcome}
\label{sec:common_outcome}

The approximate mediation effect measure expressions used in the previous sections are derived under a rare outcome assumption such that the cumulative baseline hazard $\Lambda_0(t)$ in \eqref{m:outcome} is small. 
{In health science studies, assuming a rare outcome when the outcome is in fact common might bias the analysis results even in the absence of measurement error. In the HPFS, for example, the cumulative event rate of cardiovascular diseases is 14.5\% during 30 years of follow-up, which is slightly above the usual 10\% threshold for a rare/common outcome in health sciences research (\citealp{mcnutt2003estimating}). Below, we proceed with assessing mediation without invoking a rare outcome assumption.}

If the outcome is common, the baseline hazard function $\Lambda_0(t)$ in the outcome model \eqref{m:outcome} must be estimated, since the exact mediation effect measures are functions of both $\bm\theta$ and $\Lambda_0(t)$. Corresponding to each $\widehat{\bm \beta}^{(m)}$ ($m$ can be $O1$, $O2$, $R$), we consider the Breslow estimator for $\Lambda_0(t)$ based on the main study data: 
$$
\widehat{\Lambda}_0^{(m)}(t) = \sum_{i=1}^{n} R_i \int_0^{t^*} \frac{1}{ \sum_{j=1}^{n} R_j Y_j(t) \exp\left(\widehat{\bm\beta}^{(m)^T} \bm Z_j(t)  \right)} d N_i(t),
$$
where $\bm Z_i(t) = \left[m_{A_i}(t),M_i,m_{A_i}(t)\times M_i,\bm W_i^T\right]^T$ is the design matrix. Once we obtain $\widehat{\bm\theta}^{(m)}$ and $\widehat{\Lambda}_0^{(m)}(t)$, we plug them into \eqref{exact_expressions} and obtain $\widehat{\tau}_{a,a^*|\bm w}^{(m)}(t)$. The integrals in \eqref{exact_expressions} can be approximated by numerical integration (for example, the Gauss–Hermite Quadrature approach in \cite{liu1994note}). In principle, a plug-in estimator of the asymptotic variance can be developed to construct the standard error and confidence interval, but its implementation is generally  
cumbersome due to complexity of the exact mediation effect expressions. 
Instead, we recommend the nonparametric bootstrap approach (as in Section \ref{sec:estimation_approx_expression}) to estimate the standard error and confidence interval.

To summarize, we have presented six bias-corrected estimators of the mediation effect measures, with three approaches for estimating mediator and outcome model parameters (ORC1, ORC2, and RRC) along with two sets of expressions of mediation effect measures (approximate versus exact). The validity of each estimator requires a set of assumptions for identifying the mediation effect measures as well as additional assumptions for characterizing the measurement error process. A summary of the requisite assumptions for each estimator is provided in Table \ref{tab:assumptions}. From a theoretical perspective, it should be noted that only the RRC estimator based on the exact mediation effect expressions is valid to accommodate a common outcome, whereas the other methods may be biased in the presence of a common outcome due to either using ORC1 or ORC2 to estimate the Cox outcome model coefficients or using the approximate mediation effect expressions.

\begin{table}[htbp]
\centering
\begin{threeparttable}
\caption{A summary of the assumptions used in each approach for correcting exposure measurement error-induced bias in mediation analysis with a survival outcome. A check mark indicates that this approach uses this assumption.\label{tab:assumptions}}
\renewcommand{\arraystretch}{0.5}
\begin{tabular}{cc ccc ccc}
\hline
\hline
Groups$^{1}$ & Assumptions & \multicolumn{3}{c}{Approximate Expression} & \multicolumn{3}{c}{Exact Expression} \\
\cmidrule(lr){3-5} \cmidrule(lr){6-8}
       &                      &  ORC1   &  ORC2   &   RRC & ORC1 & ORC2 & RRC \\ \hline
 I& Identification assumptions$^{2}$  & \checkmark   & \checkmark    & \checkmark & \checkmark   & \checkmark    & \checkmark   \\ 
& Independent censoring$^{3}$ & \checkmark   & \checkmark    & \checkmark & \checkmark   & \checkmark    & \checkmark   \\
\hline
 II & Surrogacy assumption$^{4}$        &   \checkmark   &  \checkmark    & \checkmark & \checkmark   &  \checkmark    & \checkmark    \\ 
& Transportability assumption$^{5}$ &   \checkmark   &   \checkmark   & \checkmark & \checkmark   &   \checkmark   & \checkmark    \\ 
& Rare outcome assumption$^{6}$     & \checkmark    & \checkmark    &  \checkmark & \checkmark    & \checkmark    &    \\ 
\hline
 III & Measurement error model \eqref{m:error}$^{7}$ & \checkmark & \checkmark & \checkmark & \checkmark & \checkmark & \checkmark \\ 
& Measurement error model \eqref{e_aamw1}$^{8}$ & \checkmark & \checkmark & & \checkmark & \checkmark &  \\
& Measurement error model \eqref{e_aamwt}$^{8}$ &  &  & \checkmark & &  & \checkmark \\
\hline\hline
\end{tabular}
{\footnotesize
\begin{tablenotes}
\item[$^1$] We divide the assumptions into three groups: Group I includes identification assumptions for survival mediation effects in the absence of measurement error, Group II includes assumptions characterizing for the exposure measurement error process, and Group III includes the modelling assumptions of the specified  measurement error models.
\item[$^2$] The identification assumptions are used to identify the mediation measure effects in the absence of measurement error, which include the consistency, positivity, composition, and four no unmeasured confounding assumptions. More details are given in Web Appendix A.
\item[$^3$] Also known as the covariate-dependent censoring; it assumes $T\perp C|A,M,\bm W$. 
\item[$^4$] The surrogacy assumption, also known as nondifferential measurement error assumption, assumes that the error-prone measurement $A^*$ is not informative for the outcome and mediator given the true exposure $A$.
\item[$^5$] It requires $P(T,C,M,A|A^*,W,R) = P(T,C,M,A|A^*,W)$, where $R$ is the indicator for being in the main study. This assumption ensures that regression parameters in the measurement error models \eqref{m:error}, \eqref{e_aamw1}, and \eqref{e_aamwt} are identical for individuals in the main and validation studies. See Remarks 1 and 2 in Web Appendix A for a further discussion about the transportability assumptions.
\item[$^6$] The rare outcome assumption is used in ORC1 and ORC2 approaches to correct for bias of the mediator and outcome model parameter estimators, and is also used in the approximate expressions to calculate the mediation effect measures.
\item[$^7$] All ORC1, ORC2, and RRC assumes measurement error model \eqref{m:error} for correcting the bias in the mediator model \eqref{m:mediator}. ORC1 and RRC only assume the conditional mean and variance in  measurement error model \eqref{m:error} are correctly specified, whereas the ORC2 additionally requires the the error term in measurement error model \eqref{m:error} is normally distributed.
\item[$^8$] Only the conditional mean in measurement error models \eqref{e_aamw1} and \eqref{e_aamwt} are required to be correctly specified, and no additional distributional assumptions are required in \eqref{e_aamw1} and \eqref{e_aamwt}. 
\end{tablenotes}}
\end{threeparttable}
\end{table}

\section{Simulation Study}
\label{sec:sim}

\subsection{Simulation setup and main results}
\label{sec:sim_1}

We studied the empirical performance of the proposed methods via simulation studies. We considered a main study with $n_1=10,000$ and a validation study with $n_2=250$ and 1000 to represent a moderate and large validation study size. 
We considered a continuous baseline covariate $W$ and generated the mediator $M$, true exposure $A$, and
$W$ from a multivariate normal distribution with mean $\bm 0$ and variance-covariance matrix $0.5\times\left[\begin{matrix} 1 & 0.2 & 0.2 \\ \cdot & 1 & 0.2 \\ \cdot & \cdot & 1 \end{matrix}\right]$ such that the marginal variance for each variable is 0.5 and the mutual correlation between any two of the three variables are 0.2. Then, given the true exposure $A$, the mismeasured exposure $A^*$ was generated by $N(\rho_{AA^*} A, 0.5(1-\rho_{AA^*}^2))$  to ensure that the correlation coefficient between $A$ and $A^*$ is $\rho_{AA^*}$ and that $A^*$, $A$ have the same marginal distribution. We chose $\rho_{AA^*} \in \{0.4,0.6,0.8\}$ to reflect a large, medium, and small measurement error. The true failure time was generated by the Cox model $\lambda_{\widetilde T}(t|A,M,\bm W) = \lambda_0(t) e^{\beta_1 A + \beta_2 M + \beta_3 AM + \beta_4 W}$, where $\beta_3=\beta_4=\log(1.1)$ and $(\beta_1,\beta_2)$ were selected such that TE and MP, defined for $A$ in change from 0 to 1 conditional on $W=0$ under a rare outcome assumption, are $\log(1.5)$ and 30\%, respectively. Similar to \cite{liao2011survival}, the baseline hazard was set to be of the Weibull form $\lambda_0(t) = \theta v (vt)^{\theta-1}$ with $\theta=6$ and $v$ was selected to achieve a relative rare cumulative event rate at 5\% at the maximum follow-up time with $t^*=50$. The censoring time $C$ was assumed to be exponentially distributed with the rate parameter set at 0.01. The observed failure time $T_i$ was set to $\text{min}\{\widetilde T_i,C_i,t^*\}$. According to the approximate expressions of the mediation effect measures, the true values of the NIE, NDE, and MP on the log-hazard ratio scale were 0.121, 0.283, and 30\%, respectively.

For each scenario, we conducted 2,000 simulations and compared the bias, empirical standard error, and 95\% confidence interval coverage rate among the ORC1 ($\widehat{\tau}^{(O1)}$), ORC2 ($\widehat{\tau}^{(O2)}$), RRC ($\widehat{\tau}^{(R)}$), the uncorrected approach ($\widehat{\tau}^{(U)}$), and a gold standard approach ($\widehat{\tau}^{(G)}$). The uncorrected approach ignored exposure measurement error by treating $A^*$ as the true exposure. The gold standard approach had access to $A$ in the main study and was expected to provide high-quality estimates since it is free of measurement error. The RRC grouped the risk sets by equally dividing the time scale into $K=4$ intervals to ensure that the results are numerically stable.

The simulation results are summarized in Table \ref{tab:sim_main}. The uncorrected method provided notable bias with substantially attenuated coverage rate for estimating NIE and NDE. However, the uncorrected MP estimator exhibited small bias because, as implied from Theorem \ref{thm1}, the uncorrected MP estimator should be nearly unbiased when the MP reliability index $\rho$ is small (here, $\rho=0.03,0.02,0.01$ for $\rho_{AA^*}=0.4,0.6,0.8$, respectively) and the exposure-mediator interaction effect is weak ($\beta_3\approx 0.09$ in the simulation). Overall, all of the three proposed correction approaches, ORC1, ORC2, and RRC, performed well with minimal bias across different levels of validation study sample size and exposure measurement error. The interval estimators based on estimated asymptotic variance formula also provided close to nominal coverage rates, although the coverage rate for MP was slightly attenuated when the exposure measurement error is large ($\rho_{AA^*}=0.4$). The bootstrap confidence interval always exhibited  nominal coverage rates among all the levels of exposure measurement error considered. Among the three proposed estimators, ORC2 had the smallest Monte Carlo standard error, followed by the ORC1 and then the RRC. However, the differences in the Monte Carlo standard errors were usually small when the exposure measurement error was moderate or small ($\rho_{AA^*}=0.6$ or $0.8$). 

\begin{table}[htbp]
\caption{Simulation results for a rare outcome with a 5\% cumulative event rate at the maximum follow-up time $t^*=50$. Here, $\widehat{\tau}^{(U)}$, $\widehat{\tau}^{(G)}$, $\widehat{\tau}^{(O1)}$, $\widehat{\tau}^{(O2)}$, and $\widehat{\tau}^{(R)}$ denote the estimators given by the uncorrected, gold-standard, ORC1, ORC2, RRC approach, respectively. Upper panel: performance of the NIE estimators; middle panel: performance of the NDE estimators; lower panel: performance of the MP estimators.}
\label{tab:sim_main}
\scalebox{0.75}[0.75]{
\begin{threeparttable}
\renewcommand{\arraystretch}{0.7}
\setlength\tabcolsep{4pt}
\begin{tabular}{rl rrrrr rrrrr }
  \hline
& & \multicolumn{5}{c}{Percent Bias (SE$\times$100)} &   \multicolumn{5}{c}{Empirical Coverage Rate v1/v2} \\
\cmidrule(lr){3-7} \cmidrule(lr){8-12}
$n_2$ &   $\rho_{AA^*}$  & 
 $\widehat{\text{NIE}}^{(U)}$ &  $\widehat{\text{NIE}}^{(G)}$ & $\widehat{\text{NIE}}^{(O1)}$ & $\widehat{\text{NIE}}^{(O2)}$ & $\widehat{\text{NIE}}^{(R)}$ &
 $\widehat{\text{NIE}}^{(U)}$ &  $\widehat{\text{NIE}}^{(G)}$ & $\widehat{\text{NIE}}^{(O1)}$ & $\widehat{\text{NIE}}^{(O2)}$ & $\widehat{\text{NIE}}^{(R)}$ \\ 
  \hline
250 & 0.4 & $-$62.2 (1.1) & $-$0.3 (2.1) & $-$2.8 (4.5) & $-$2.3 (4.3) & $-$3.4 (4.5) & 0.1/0.1 & 93.8/93.8 & 92.7/95.0 & 92.5/95.0 & 92.8/95.5 \\ 
   & 0.6 & $-$41.9 (1.4) & $-$0.7 (2.0) & $-$1.4 (3.2) & $-$1.2 (3.2) & $-$1.6 (3.2) & 7.2/13.3 & 95.0/95.8 & 94.5/95.4 & 94.3/95.4 & 94.5/95.3 \\ 
   & 0.8 & $-$21.4 (1.7) & $-$1.3 (2.0) & $-$1.3 (2.5) & $-$1.2 (2.5) & $-$1.3 (2.5) & 65.8/71.6 & 95.3/95.5 & 95.5/95.4 & 95.2/95.5 & 95.1/95.2 \\ 
  1000 & 0.4 & $-$61.7 (1.1) & $-$0.4 (2.1) & $-$1.2 (3.9) & $-$1.5 (3.9) & $-$1.1 (3.9) & 0.0/0.1 & 95.2/95.1 & 95.4/95.1 & 95.3/95.0 & 95.6/95.3 \\ 
   & 0.6 & $-$41.6 (1.4) & 0.2 (2.0) & $-$0.4 (3.0) & $-$0.4 (3.0) & $-$0.2 (3.0) & 8.2/14.6 & 94.8/94.8 & 95.9/95.2 & 95.7/95.0 & 95.8/95.1 \\ 
   & 0.8 & $-$21.5 (1.7) & $-$0.9 (2.0) & $-$1.0 (2.4) & $-$1.1 (2.4) & $-$1.0 (2.4) & 64.5/71.0 & 95.3/95.8 & 95.5/94.9 & 95.5/94.8 & 95.5/94.7 \\ 
   \hline
   \hline
 & & \multicolumn{5}{c}{Percent Bias (SE$\times$100)} &   \multicolumn{5}{c}{Empirical Coverage Rate v1/v2} \\
\cmidrule(lr){3-7} \cmidrule(lr){8-12}
$n_2$ &   $\rho_{AA^*}$  & 
 $\widehat{\text{NDE}}^{(U)}$ &  $\widehat{\text{NDE}}^{(G)}$ & $\widehat{\text{NDE}}^{(O1)}$ & $\widehat{\text{NDE}}^{(O2)}$ & $\widehat{\text{NDE}}^{(R)}$ &
 $\widehat{\text{NDE}}^{(U)}$ &  $\widehat{\text{NDE}}^{(G)}$ & $\widehat{\text{NDE}}^{(O1)}$ & $\widehat{\text{NDE}}^{(O2)}$ & $\widehat{\text{NDE}}^{(R)}$  \\ 
  \hline
  250 & 0.4 & $-$61.7 (7.6) & 1.6 (7.9) & 1.1 (21.7) & 1.5 (21.8) & 1.5 (22.1) & 36.6/45.0 & 94.6/94.9 & 94.6/95.8 & 94.5/96.0 & 94.6/96.0 \\ 
   & 0.6 & $-$41.9 (7.8) & $-$0.0 (7.9) & 2.3 (14.1) & 2.7 (14.1) & 1.7 (14.4) & 65.2/70.8 & 95.0/95.5 & 94.0/95.0 & 94.0/94.8 & 93.8/94.3 \\ 
   & 0.8 & $-$22.6 (7.7) & $-$0.4 (8.0) & $-$1.0 (10.0) & $-$1.1 (10.0) & $-$1.1 (10.1) & 87.8/89.0 & 94.0/94.1 & 94.9/94.7 & 95.0/94.7 & 95.0/94.7 \\ 
  1000 & 0.4 & $-$62.4 (7.8) & 0.3 (7.9) & 0.0 (20.9) & 0.1 (20.9) & 0.8 (21.1) & 36.1/44.4 & 95.2/95.8 & 94.2/94.4 & 94.3/94.5 & 94.3/94.5 \\ 
   & 0.6 & $-$43.0 (7.9) & 0.2 (8.0) & $-$0.8 (13.8) & $-$0.7 (13.8) & $-$0.8 (13.8) & 65.0/70.7 & 94.2/94.8 & 95.0/95.2 & 95.0/95.0 & 94.8/95.2 \\ 
   & 0.8 & $-$20.6 (7.6) & 1.1 (8.0) & 2.1 (9.8) & 2.1 (9.8) & 1.8 (9.8) & 88.5/90.2 & 94.8/94.9 & 95.7/95.8 & 95.8/95.7 & 95.5/95.7 \\ 
   \hline
   \hline
    & & \multicolumn{5}{c}{Percent Bias (SE$\times$100)} &   \multicolumn{5}{c}{Empirical Coverage Rate v1/v2} \\
\cmidrule(lr){3-7} \cmidrule(lr){8-12}
$n_2$ &   $\rho_{AA^*}$  & 
 $\widehat{\text{MP}}^{(U)}$ &  $\widehat{\text{MP}}^{(G)}$ & $\widehat{\text{MP}}^{(O1)}$ & $\widehat{\text{MP}}^{(O2)}$ & $\widehat{\text{MP}}^{(R)}$ &
 $\widehat{\text{MP}}^{(U)}$ &  $\widehat{\text{MP}}^{(G)}$ & $\widehat{\text{MP}}^{(O1)}$ & $\widehat{\text{MP}}^{(O2)}$ & $\widehat{\text{MP}}^{(R)}$  \\ 
  \hline
    250 & 0.4 & $-$2.2 (449.3) & $-$1.3 (7.9) & $-$6.2 (465.0) & $-$5.3 (186.6) & $-$5.2 (817.9) & 90.1/97.6 & 93.8/94.8 & 89.9/97.0 & 89.7/97.2 & 89.3/97.2 \\ 
   & 0.6 & 1.0 (28.3) & $-$0.8 (8.1) & $-$1.4 (27.6) & $-$1.0 (27.2) & $-$0.9 (32.6) & 92.9/97.2 & 94.5/96.4 & 92.3/96.8 & 92.4/96.8 & 91.8/96.6 \\ 
   & 0.8 & 1.2 (10.9) & 0.3 (8.4) & $-$0.1 (11.2) & $-$0.0 (11.2) & 0.5 (11.2) & 93.8/95.9 & 94.2/94.7 & 94.0/95.6 & 94.0/95.8 & 93.9/95.6 \\ 
  1000 & 0.4 & $-$0.9 (506.7) & $-$0.6 (8.2) & $-$4.4 (196.0) & $-$4.0 (153.9) & $-$3.7 (805.6) & 89.8/97.0 & 94.3/95.7 & 90.1/96.9 & 90.0/97.0 & 90.0/97.0 \\ 
   & 0.6 & 1.8 (95.2) & 0.0 (8.3) & $-$0.5 (129.0) & $-$0.5 (97.0) & $-$0.3 (185.6) & 92.2/97.1 & 94.0/94.9 & 92.6/96.7 & 92.2/96.8 & 92.5/97.0 \\ 
   & 0.8 & $-$0.8 (11.1) & $-$1.4 (8.3) & $-$2.0 (11.4) & $-$2.2 (11.4) & $-$2.0 (11.3) & 93.9/96.6 & 94.0/95.7 & 93.8/96.0 & 93.8/96.2 & 93.8/96.2 \\ 
 \hline
   \hline
\end{tabular}
\begin{tablenotes}
\item[1] The data generation process is given in Section \ref{sec:sim}, where the mediator follows a linear regression \eqref{m:mediator} and outcome follows the Cox model \eqref{m:outcome}, with a linear measurement error process characterized in \eqref{m:error}. $n_2$ is the validation study sample size and $\rho_{AA^*}$ is the correlation coefficient between the true exposure $A$ and its mismeasured value $A^*$. The percent bias is defined as the ratio of bias to the true value over 2,000 replications, i.e., $mean(\frac{\widehat{\tau}-\tau}{\tau}) \times 100\%$. The standard error (SE) is defined as the square root of empirical variance of mediation effect measure estimates from the 2,000 replications. The empirical coverage rate v1 is obtained based on a Wald-type 95\% confidence interval using the derived asymptotic variance formula. The empirical coverage rate v2 is obtained based on a non-parametric bootstrap confidence interval.  
\end{tablenotes}
\end{threeparttable}}
\end{table}

We also explored the impact of increasing the number of split points ($K$) for risk set formation on the RRC estimator in Table \ref{tab:sim_rrc}, which showed the empirical performance of the RRC with equally-spaced split points and $K$ ranging from 2 to 8. The RRC performed well and the results were robust to the choice of $K$. 

\begin{table}[htbp]
\caption{Simulation results of the RRC approach with different numbers of split points ($K$) for a rare outcome with a 5\% cumulative event rate at the maximum follow-up time $t^*=50$. Here, $\widehat{\tau}^{(R8)}$, $\widehat{\tau}^{(R5)}$, $\widehat{\tau}^{(R4)}$, $\widehat{\tau}^{(R3)}$, $\widehat{\tau}^{(R2)}$ denote the RRC estimators with equally-spaced split points for $K=$8, 5, 4, 3, and 2, respectively. Upper panel: performance of the NIE estimators; middle panel: performance of the NDE estimators; lower panel: performance of the MP estimators.}
\label{tab:sim_rrc}
\scalebox{0.7}[0.7]{
\begin{threeparttable}
\renewcommand{\arraystretch}{0.7}
\setlength\tabcolsep{4pt}
\begin{tabular}{rl ccccc ccccc }
  \hline
& & \multicolumn{5}{c}{Percent Bias (SE$\times$100)} &   \multicolumn{5}{c}{Empirical Coverage Rate v1/v2} \\
\cmidrule(lr){3-7} \cmidrule(lr){8-12}
$n_2$ &   $\rho_{AA^*}$  & 
 $\widehat{\text{NIE}}^{(R8)}$ &  $\widehat{\text{NIE}}^{(R5)}$ & $\widehat{\text{NIE}}^{(R4)}$ & $\widehat{\text{NIE}}^{(R3)}$ & $\widehat{\text{NIE}}^{(R2)}$ &
 $\widehat{\text{NIE}}^{(R8)}$ &  $\widehat{\text{NIE}}^{(R5)}$ & $\widehat{\text{NIE}}^{(R4)}$ & $\widehat{\text{NIE}}^{(R3)}$ & $\widehat{\text{NIE}}^{(R2)}$ \\ 
  \hline
250 & 0.4 & $-$1.4 (4.5) & $-$1.4 (4.5) & $-$1.3 (4.5) & $-$1.3 (4.5) & $-$0.9 (4.5) & 93.2/96.1 & 93.2/96.0 & 93.5/95.9 & 93.4/96.2 & 93.5/95.8 \\ 
   & 0.6 & $-$0.5 (3.2) & $-$0.5 (3.2) & $-$0.7 (3.2) & $-$0.6 (3.2) & $-$0.6 (3.2) & 94.2/95.7 & 94.0/95.6 & 94.3/95.8 & 94.3/95.6 & 94.0/95.5 \\ 
   & 0.8 & $-$0.6 (2.5) & $-$0.6 (2.5) & $-$0.7 (2.5) & $-$0.6 (2.5) & $-$0.6 (2.5) & 95.2/95.5 & 95.0/95.5 & 95.0/95.5 & 95.0/95.5 & 94.9/95.4 \\ 
  1000 & 0.4 & $-$2.5 (3.8) & $-$2.5 (3.8) & $-$2.4 (3.8) & $-$2.4 (3.8) & $-$2.5 (3.8) & 95.5/94.7 & 95.5/95.0 & 95.7/94.8 & 95.6/94.7 & 95.6/94.4 \\ 
   & 0.6 & 0.7 (3.0) & 0.6 (3.0) & 0.5 (3.0) & 0.5 (3.0) & 0.4 (3.0) & 95.5/94.8 & 95.6/94.6 & 95.5/94.5 & 95.5/94.5 & 95.5/94.7 \\ 
   & 0.8 & $-$0.8 (2.4) & $-$0.7 (2.4) & $-$0.9 (2.4) & $-$0.7 (2.4) & $-$0.9 (2.4) & 95.8/95.6 & 95.7/95.6 & 95.6/95.7 & 95.7/95.6 & 95.7/95.7 \\ 
   \hline
   \hline
 & & \multicolumn{5}{c}{Percent Bias (SE$\times$100)} &   \multicolumn{5}{c}{Empirical Coverage Rate v1/v2} \\
\cmidrule(lr){3-7} \cmidrule(lr){8-12}
$n_2$ &   $\rho_{AA^*}$  & 
 $\widehat{\text{NDE}}^{(R8)}$ &  $\widehat{\text{NDE}}^{(R5)}$ & $\widehat{\text{NDE}}^{(R4)}$ & $\widehat{\text{NDE}}^{(R3)}$ & $\widehat{\text{NDE}}^{(R2)}$ &
 $\widehat{\text{NDE}}^{(R8)}$ &  $\widehat{\text{NDE}}^{(R5)}$ & $\widehat{\text{NDE}}^{(R4)}$ & $\widehat{\text{NDE}}^{(R3)}$ & $\widehat{\text{NDE}}^{(R2)}$  \\ 
  \hline
   250 & 0.4 & 1.3 (21.7) & 0.9 (21.8) & 0.3 (21.6) & $-$0.3 (21.7) & 0.5 (21.6) & 94.3/95.4 & 94.7/95.5 & 94.4/95.8 & 94.7/95.5 & 94.5/95.7 \\ 
   & 0.6 & 2.0 (14.4) & 2.1 (14.4) & 2.0 (14.4) & 2.0 (14.4) & 1.8 (14.4) & 94.2/94.3 & 94.1/94.5 & 94.2/94.3 & 94.0/94.5 & 94.3/94.3 \\ 
   & 0.8 & $-$1.1 (10.0) & $-$0.7 (10.0) & $-$0.8 (10.0) & $-$0.5 (10.0) & $-$0.4 (10.0) & 94.7/95.4 & 94.5/95.3 & 94.5/95.4 & 94.5/95.5 & 94.5/95.5 \\ 
  1000 & 0.4 & 1.4 (20.7) & 1.1 (20.6) & 0.8 (20.6) & 1.4 (20.6) & 1.4 (20.6) & 94.5/95.0 & 94.4/95.0 & 94.2/95.0 & 94.5/94.9 & 94.2/94.8 \\ 
   & 0.6 & 2.1 (13.4) & 2.2 (13.4) & 2.6 (13.4) & 2.3 (13.4) & 2.7 (13.4) & 95.1/95.4 & 95.0/95.2 & 95.0/95.2 & 95.0/95.3 & 95.0/95.2 \\ 
   & 0.8 & 1.3 (9.8) & 1.6 (9.8) & 1.4 (9.8) & 1.5 (9.8) & 1.3 (9.8) & 95.3/95.2 & 95.3/95.2 & 95.5/95.2 & 95.5/95.1 & 95.2/95.1 \\ 
   \hline
   \hline
    & & \multicolumn{5}{c}{Percent Bias (SE$\times$100)} &   \multicolumn{5}{c}{Empirical Coverage Rate v1/v2} \\
\cmidrule(lr){3-7} \cmidrule(lr){8-12}
$n_2$ &   $\rho_{AA^*}$  & 
 $\widehat{\text{MP}}^{(R8)}$ &  $\widehat{\text{MP}}^{(R5)}$ & $\widehat{\text{MP}}^{(R4)}$ & $\widehat{\text{MP}}^{(R3)}$ & $\widehat{\text{MP}}^{(R2)}$ &
 $\widehat{\text{MP}}^{(R8)}$ &  $\widehat{\text{MP}}^{(R5)}$ & $\widehat{\text{MP}}^{(R4)}$ & $\widehat{\text{MP}}^{(R3)}$ & $\widehat{\text{MP}}^{(R2)}$  \\ 
  \hline
   250 & 0.4 & $-$3.9 (867.5) & $-$3.8 (292.8) & $-$4.3 (1613.6) & $-$3.6 (994.9) & $-$4.2 (3964.4) & 91.0/97.9 & 90.8/98.0 & 90.8/97.8 & 91.0/97.8 & 90.8/97.9 \\ 
   & 0.6 & $-$1.4 (4341.6) & $-$1.7 (61.6) & $-$1.3 (87.2) & $-$1.3 (256.9) & $-$1.4 (109.7) & 92.0/97.0 & 91.5/96.9 & 91.5/96.8 & 91.6/96.9 & 91.5/96.5 \\ 
   & 0.8 & $-$0.3 (14.8) & $-$0.2 (14.7) & $-$0.4 (14.4) & $-$0.4 (14.5) & $-$0.1 (14.5) & 94.4/96.3 & 94.3/96.2 & 94.5/96.2 & 94.3/96.3 & 94.4/96.1 \\ 
  1000 & 0.4 & $-$3.8 (419.6) & $-$3.6 (4076.0) & $-$3.6 (296.9) & $-$3.7 (161.7) & $-$3.5 (589.4) & 89.8/96.8 & 89.8/96.9 & 89.7/96.8 & 89.8/96.8 & 89.7/97.0 \\ 
   & 0.6 & $-$1.0 (28.9) & $-$0.9 (31.5) & $-$1.1 (27.1) & $-$1.0 (33.3) & $-$1.1 (28.5) & 92.8/97.3 & 92.8/97.2 & 92.7/97.2 & 92.8/97.3 & 92.8/97.2 \\ 
   & 0.8 & $-$1.7 (11.5) & $-$1.7 (11.5) & $-$1.8 (11.5) & $-$1.8 (11.5) & $-$2.0 (11.5) & 94.7/96.0 & 94.7/96.0 & 94.7/95.9 & 94.7/96.0 & 94.5/95.9 \\ 
 \hline
   \hline
\end{tabular}
\begin{tablenotes}
\item[1] The data generation process is given in Section \ref{sec:sim}, where the mediator follows a linear regression \eqref{m:mediator} and outcome follows the Cox model \eqref{m:outcome}, with a linear measurement error process characterized in \eqref{m:error}. $n_2$ is the validation study sample size and $\rho_{AA^*}$ is the correlation coefficient between the true exposure $A$ and its mismeasured value $A^*$. The percent bias is defined as the ratio of bias to the true value over 2,000 replications, i.e., $mean(\frac{\widehat{\tau}-\tau}{\tau}) \times 100\%$. The standard error (SE) is defined as the square root of empirical variance of mediation effect measure estimates from the 2,000 replications. The empirical coverage rate v1 is obtained based on a Wald-type 95\% confidence interval using the derived asymptotic variance formula. The empirical coverage rate v2 is obtained based on a non-parametric bootstrap confidence interval.  
\end{tablenotes}
\end{threeparttable}}
\end{table}

\subsection{When the outcome is common}

We conducted additional simulations under the common outcome scenario, where the data generation process was identical to Section \ref{sec:sim_1} except that the parameter, $v$, in the baseline hazard $\lambda_0(t) = \theta v (vt)^{\theta-1}$, was specified to achieve a cumulative event rate of around 50\%. We evaluated the performance based on the exact mediation effect measures \eqref{exact_expressions} at different times $t$, ranging from 10 to 45. The true values of the mediation effect measures were evaluated based on the exact expressions in Section \ref{sec:preliminaries}, and the integrals were approximated using the Gauss–Hermite quadrature. From Web Table 1, satisfactory performance was observed among all proposed bias-correction approaches. The ORC approaches still provided robust performance under this common outcome scenario, which agreed with previous research in \cite{xie2001risk} that the ORC approach is insensitive to the rare outcome assumption as long as the exposure effect on Cox model is small ($\beta_1=0.25$ in our simulation). We further assessed the performance of the proposed methods with a relatively larger $\beta_1$ in Web Table 2 ($\beta_1=1$) and Web Table 3 ($\beta_1=1.5$), while retaining all other aspects of the data generation process. In those scenarios, ORC1 and ORC2 led to excessive bias with attenuated coverage as the exposure effect increases. In contrast, the RRC approach had small bias and maintained closer to normal coverage rates.

In the previous simulation study with a common outcome, we set $\theta$ to 6 in $\lambda_0(t)=\theta v (vt)^{\theta-1}$, resulting in an increasing baseline hazard function over time. Next, we investigated robustness of the proposed methods to a constant baseline hazard function. We set $\theta=1$ and fix $v = 0.012$ so that the cumulative event rate is around 50\% until the maximum follow-up time $t^*=50$. We also considered a moderate exposure effect in the Cox outcome model ($\beta_1=0.25$), and two relatively large exposure effects at $\beta_1=1$ and $\beta_1=1.5$. The simulation results were provided in Web Tables 4--6 for $\beta_1=0.25$, $1$, and 1.5, respectively. To summarize, the ORC1, ORC2, and RRC approaches always displayed minimal bias and desirable coverage with a moderate exposure effect size ($\beta_1=0.25$). As the exposure effect increased, the percent bias among all estimators increased with attenuated coverage rate, and the percent bias of RRC approach was smaller than these of the ORC1 and ORC2.

\subsection{When the true measurement error process deviates from normality}

As shown in Table \ref{tab:assumptions}, the ORC2 approach assumes the error term $\epsilon_A$ in the measurement error model \eqref{m:error} is normally distributed, whereas ORC1 and RRC do not require distributional assumptions on the measurement error process. We conducted additional simulations to examine the sensitivity of performance of ORC2 under departures from the normality assumption. The data generation process for this additional simulation is detailed in Web Appendix G.1. Briefly, we consider two scenarios to generate $\epsilon_A$ based on a skew normal distribution (\citealp{fernandez1998bayesian}) and a uniform distribution, while holding other aspects of the data generation process to be the same as Section \ref{sec:sim_1}. Specifically, when generating $\epsilon_A$ based on a skew normal distribution, we set the skew parameter, $\xi$, at 5, leading to a coefficient of skewness at approximately 1.  Web Figure 1 provided a graph comparing the skew normal and uniform distribution to the normal distribution.

Web Tables 7--8 presented the simulation results when $\epsilon_A$ follows a skew normal or a uniform distribution, respectively, and validation study sample size is set to 250. The ORC1, ORC2, and RRC approaches exhibited little percent bias with the desired  coverage rates. Notably, despite its requirement for the normality assumption in $\epsilon_A$ to derive $m_A^{(O2)}(\bm\alpha,\bm\gamma)$, the ORC2 approach had robust performance and its percent bias fell below 7\% across all mediation effect measures.

\subsection{When the mediator or outcome model is misspecified}\label{sec:mis-M-O}

In this section, we further investigated performance of the proposed estimators under violations of two key assumptions in the mediator and outcome models: (i) the normality assumption in the mediator model \eqref{m:mediator} and (ii) the proportional hazards assumption in the Cox outcome model  \eqref{m:outcome}. Notice that the expressions for the mediation effect measures developed in Section \ref{sec:preliminaries} will no longer be valid because they were derived under the postulated mediator and outcome models \eqref{m:mediator}--\eqref{m:outcome}, which make both of these assumptions. 
Accurate expressions of $\text{NIE}_{a,a^*|\bm w}(t)$, $\text{NDE}_{a,a^*|\bm w}(t)$, and $\text{MP}_{a,a^*|\bm w}(t)$ that account for violations of (i) or (ii) are provided in Web Appendix G.2 based on results in \cite{vanderweele2011causal}. 

We first considered the scenario when the normality assumption in the mediator model is violated. The data generation process was detailed in Web Appendix G.2, where we followed the process in Section \ref{sec:sim_1} except that we now generated $\epsilon_M$ from a skew normal or uniform distribution. We calculated the true values of the mediation effect measures based on the expressions in Web Appendix G.2 as functions over time $t$. We evaluated robustness of the proposed ORC1, ORC2, and RRC estimators based on the approximate expressions of mediation effect measures in Section \ref{sec:preliminaries}, which assumed normality in $\epsilon_M$. Web Tables 9--10 presented the simulation results when $\epsilon_M$ followed a skew normal and a uniform distribution, the validation study sample size was set to 250 and $\rho_{AA^*}$ was set to 0.6. The results suggested that all proposed estimators exhibited small bias with close-to-nominal coverage rates, and the percent bias was controlled below 4\% among all proposed estimators. 

Next, we considered the scenario when the proportional hazards assumption was violated. The data generation process followed Section \ref{sec:sim_1} with $n_2=250$ and $\rho_{AA^*}=0.6$ except that the survival outcome was now generated based on a proportional odds model (see Web Appendix G.2 for more details). We investigated robustness of the proposed methods based on the approximate expressions in Section \ref{sec:preliminaries} that incorrectly assumed proportional hazards. We used the coefficients in the proportional odds model to calculate the true values of NIE, NDE, and MP that were determined by this model (see Web Appendix G.2). As shown in Web Table 11, ORC1, ORC2, and RRC maintained robust performance with relatively small bias and nominal coverage, although the percent bias of NIE and MP estimates were elevated to around 5\%.

\section{Applications to the Health Professionals Follow-up Study}
\label{sec:app}

\subsection{Main analysis under the rare outcome assumption}

We apply our proposed methods to analyze the Health Professionals Follow-up Study (HPFS), an ongoing prospective cohort study of the causes of cancer and heart disease. The HPFS began enrollment in 1986 and included 51,929 male health professionals aged 40 to 75 years at baseline. Every two years post baseline, study participants completed a follow-up questionnaire reporting information about disease, diet, medications, and lifestyle conditions. Further details on HPFS can be found in \cite{choi2005obesity}.

Previous work suggested that vigorous-intensity physical activity (VPA) was associated with lowering the risk of cardiovascular diseases (CVD) in HPFS (\citealp{chomistek2012vigorous}).  The goal of this analysis is to explore the role of body mass index (BMI) in mediating the effect of VPA on CVD. 
To improve the normality, we considered the exposure as the log-transformation of the VPA (MET$\cdot$h$\cdot$wk$^{-1}$), measured in the year of 1986. The mediator is BMI ($\text{kg}/\text{m}^2$), also evaluated in the year of 1986. The time to CVD incidence was considered as the outcome, which was subject to right censoring. Here, we used age as the time scale, as a typical choice in analyses of prospective cohort studies of chronic disease. After excluding participants with a history of CVD at baseline and who did not report their physical activity and BMI data at baseline, we included 43,547 participants in the main study. A total of 1,040,567 person-years were observed in the main study, where 6,308 participants developed CVD between 1986 to 2016. In our mediation analysis, we adjusted for baseline characteristics in \cite{chomistek2012vigorous} that are potential confounders of the exposure-mediator relationship, exposure-outcome relationship, and mediator-outcome relationship. Full details of the baseline characteristics used in the analysis are shown in Table \ref{tab:hpfs}.

\begin{figure}[htbp]
\centering\includegraphics[width=0.6\textwidth]{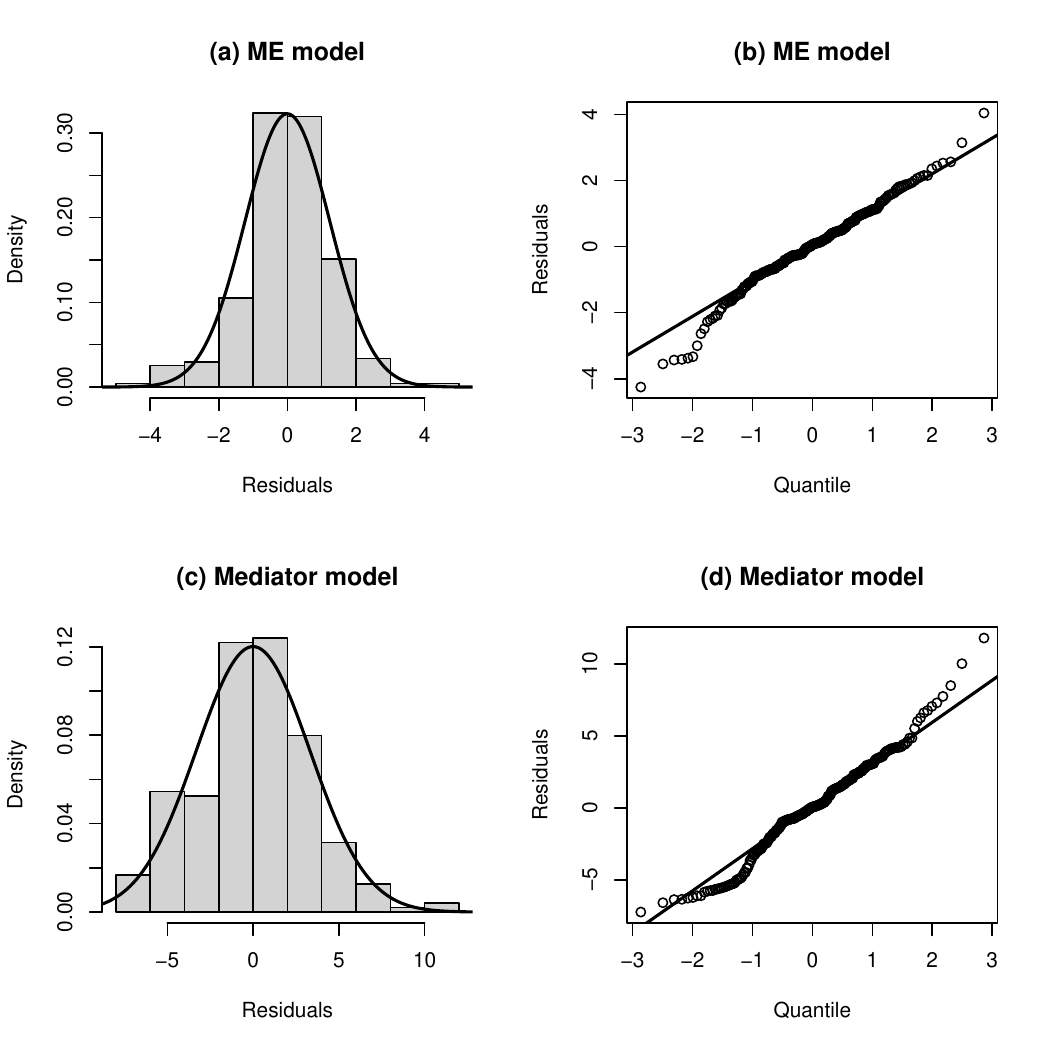}
\caption{Normality diagnostic plots for the residuals in the measurement error model \eqref{m:error} and mediator model \eqref{m:mediator} in the HPFS. Panel (a): histogram of residuals after fitting the measurement error model \eqref{m:error}; Panel (b): QQ plot of residuals after fitting the measurement error model \eqref{m:error}; Panel (c): histogram of residuals after fitting the mediator model \eqref{m:mediator}; Panel (d): QQ plot of residuals after fitting the mediator model \eqref{m:mediator}.}
\label{fig:hist}
\end{figure}

In the HPFS, physical activity were assessed via physical activity questionnaires. As previously reported, the questionnaire measurements of physical activity are known to be error-prone, and a more reliable assessment of VPA is through physical activity diaries. In the HPFS validation study, physical activity diaries were observed in 238 person-years among 238 study participants (\citealp{chasan1996reproducibility}). As a preliminary step, we carried out an empirical assessment of the normality assumption of the error term in the measurement error model \eqref{m:error} and the mediator model \eqref{m:mediator}, where the first is required in ORC2 approach and the second serves as a key assumption for derivation of the expressions of the mediation effect measures. We fit the measurement error model and mediator model in the HPFS validation study and the diagnostic plots for the normality of their residuals are presented in Figure \ref{fig:hist}. The diagnostic plots did not suggest strong evidence against normality in the measurement error model, but the lower quantiles of the residuals in the mediator model appeared to slightly deviate from normality. We thus further conducted a Kolmogorov–Smirnov test for a given reference normality assumption based on the residuals in the fitted measurement error model and mediator model, and obtained p-values as 0.23 and 0.13, respectively. These results provided no evidence against normality for both mediator and outcome models at the 0.05 level.  
We first investigated the potential bias of the uncorrected estimator. Based on the HPFS validation study, we observed that $\widehat{\gamma}_1$ and $\widehat{\rho}$, the two key quantities to determine the bias in Theorem \ref{thm1}, were $0.58$ and $0.01$, respectively. We therefore expect that the uncorrected MP estimator may be relatively less biased since $\widehat{\rho}$ is small, but the uncorrected estimator for other three mediation effect measures will be biased towards the null, if exposure-mediator interaction effect is weak in the Cox outcome model. 

We considered the mediation effect measures on the log-hazard ratio scale for a change in physical activity  from the lower quartile to the upper quartile of the diary-based METs$\cdot$h$\cdot$wk$^{-1}$, conditional on the median levels of the confounding variables. Table \ref{tab:hpfs} (Panel A) presents the estimated mediation effect measures from ORC1, ORC2, RRC, and the uncorrected approaches, based on the approximate mediation effect measure expressions with allowing for an exposure-mediator interaction. The RRC estimator used two equally spaced split points to form the risk sets in the validation study. When $K\geq 3$, the sample size in the last risk set was smaller than 42, which was insufficient for reaching stable estimates in the measurement error model \eqref{e_aamwt} given the large number of confounders included in the analysis. All methods suggested a negative NIE, NDE, and TE, indicating that VGA leads to a lower risk of cardiovascular diseases. However, the NIE and NDE estimates of the uncorrected method were attenuated towards 0 as compared to the proposed methods, due to exposure measurement error ($\widehat{\rho}_{AA^*}=0.62$). The uncorrected approach provided a similar MP estimate to the measurement error corrected approaches,  all indicating that about 40\% of the total effect from the VGA to CVD was due to BMI. 
The exposure-mediator interaction term was not statistically significant under any of the bias correction strategies (e.g., $\widehat{\beta}_3^{(R)}=0.001$ (standard error $=0.002$) with p-value $=0.38$). Therefore, we also explored additional analyses without  the exposure-mediator interaction (Panel B in Table \ref{tab:hpfs}). We found that point estimates without considering this interaction were close to their counterparts with an interaction, because the exposure-mediator interaction effect is statistically insignificant and is much smaller compared to the main effects of exposure and mediator  (e.g., RRC gives $\widehat{\beta}_1^{(R)}=-0.062$ (standard error $=0.057$), $\widehat{\beta}_2^{(R)}=0.045$ (standard error $=0.003$), and $\widehat{\beta}_3^{(R)}=0.001$ (standard error $=0.002$); ORC1 and ORC2 also give similar coefficient estimates).

\begin{table}[htbp]
\centering
\caption{Mediation analysis of physical activity on cardiovasular diseases incidence by body mass index, with and without considering the exposure-mediator interaction, in HPFS ($n_1=43,547$, $n_2=238$), 1986--2016. The mediation effect measures were estimated based on the approximate expressions and were defined on the a log-hazard ratio scale from increasing physical activity from the lower quartile to the upper quartile of the diary-based METs$\cdot$h$\cdot$wk$^{-1}$. The results are conditional on the medians of all confounding variables.  \label{tab:hpfs}}
\begin{threeparttable}
\renewcommand{\arraystretch}{0.7}
\begin{tabular}{ccccc}
\hline
Method & NIE & NDE & TE & MP \\
\hline
\multicolumn{5}{l}{\textbf{Panel A: with exposure-mediator interaction}}\\
  Uncorrected & $-$0.025 (0.003) & $-$0.039 (0.015) & $-$0.064 (0.015) & 0.384 (0.094) \\ 
   ORC1 & $-$0.049 (0.006) & $-$0.066 (0.026) & $-$0.115 (0.025) & 0.429 (0.106) \\ 
   ORC2 & $-$0.043 (0.005) & $-$0.067 (0.025) & $-$0.110 (0.025) & 0.390 (0.095) \\ 
   RRC & $-$0.043 (0.005) & $-$0.068 (0.026) & $-$0.110 (0.026) & 0.388 (0.096) \\ 
\hline
\multicolumn{5}{l}{\textbf{Panel B: without exposure-mediator interaction}}\\
 Uncorrected & $-$0.023 (0.002) & $-$0.039 (0.015) & $-$0.062 (0.015) & 0.367 (0.090) \\ 
 ORC1 & $-$0.038 (0.003) & $-$0.069 (0.026) & $-$0.108 (0.025) & 0.355 (0.092) \\ 
 ORC2 & $-$0.040 (0.003) & $-$0.068 (0.025) & $-$0.108 (0.025) & 0.369 (0.089) \\ 
 RRC & $-$0.040 (0.003) & $-$0.068 (0.026) & $-$0.108 (0.026) & 0.369 (0.090) \\ 
\hline
\end{tabular}
\begin{tablenotes}
\item[1] Numbers in the brackets are the standard errors of the point estimates given by the plug-in estimator of the asymptotic variance. Each analysis considered the following baseline variables as potential confounders: alcohol intake (yes/no), smoking status (past only/never/current), aspirin use (yes/no), parental history of myocardial infarction at or before 60 years old (yes/no), parental history of cancer at or before 60 years old (yes/no), intake of polyunsaturated fat, trans fat, omega-3 fatty acids, and fiber, as well as diabetes (yes/no), hypertension (yes/no), and hypercholesterolemia (yes/no). 
\end{tablenotes}
\end{threeparttable}
\end{table}

\subsection{Accounting for a common outcome with the exact mediation effect expressions}

\begin{figure}[htbp]
\centering\includegraphics[width=0.7\textwidth]{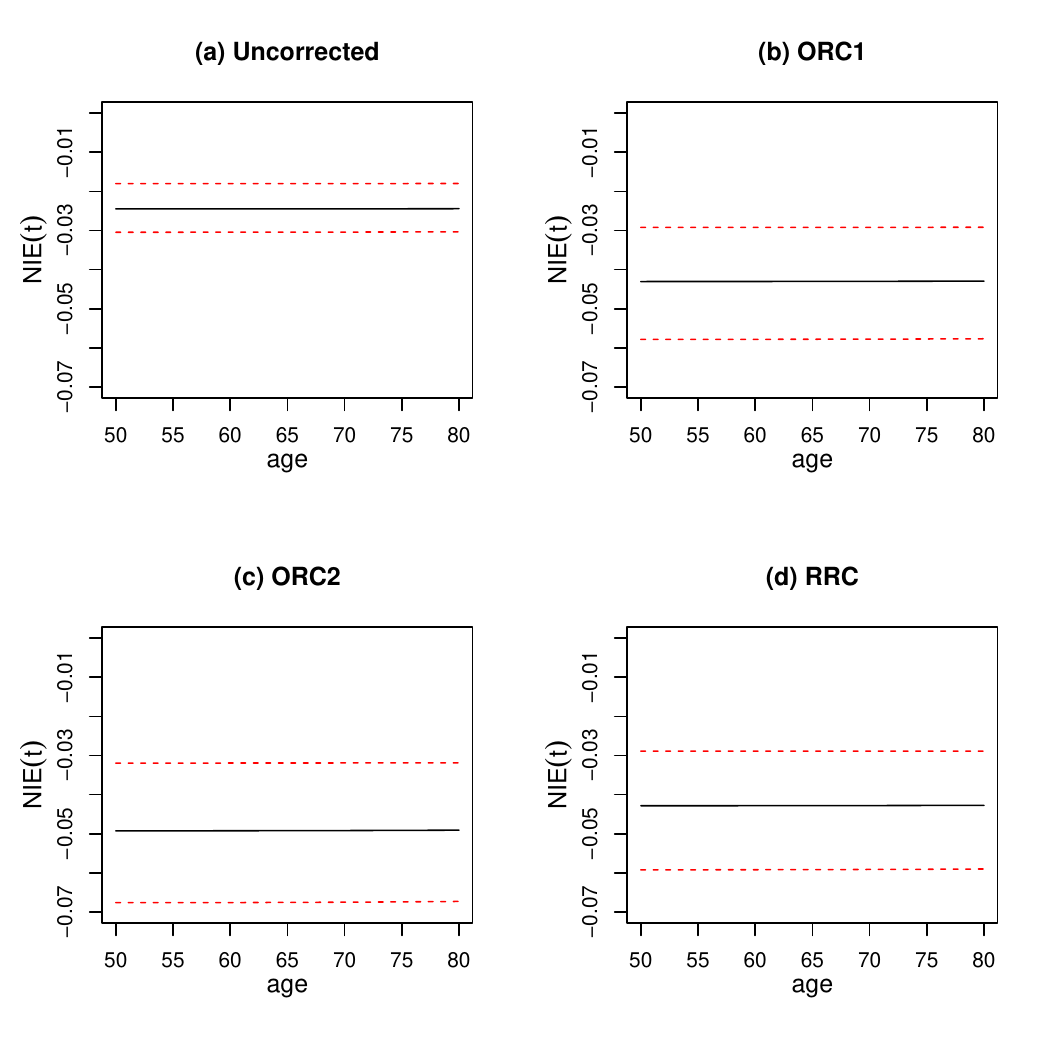}
\caption{NIE estimates based on the exact mediation effect measure expressions for the HPFS, 1986--2016. The black solid line is the point estimates and the red dotted lines are the 95\% confidence interval obtained via  bootstrapping.}
\label{fig:nie}
\end{figure}

The cumulative event rate for the CVD is 14.5\% in the HPFS, which is slightly higher than the suggestive threshold (10\%) for a rare outcome.  Therefore, the mediation effect measures may be time-varying. We repeated the mediation analysis in Table \ref{tab:hpfs}  (Panel A) using the exact mediation effect measure expressions without assuming a rare outcome. Results of $\widehat{\text{NIE}}_{a,a^*|\bm w}(t)$ are provided in Figure \ref{fig:nie}. We observed that the estimates given by all approaches were nearly constant with respect to the time ($t$, here, age in years) and were almost identical to the NIE estimates given in Table \ref{tab:hpfs} (Panel A). Similar results has been observed for the NDE estimates (See Web Figure 2). Therefore, the approximate mediation effect measures we derived served as an adequate approximation in HPFS.  Based on exact mediation effect measure expressions, we also compared the RRC estimates on NIE, NDE, and MP both with and without incorporating the exposure-mediator interactions; the point and interval estimates were quantitatively similar to the results based on the approximate mediation effect measure expressions (see Web Table 12).

\subsection{Sensitivity analysis for departures from the proportional hazards assumption}

A key assumption of our primary analysis with Cox model is proportional hazards. 
We carried out a Grambsch-Therneau test \citep{grambsch1994proportional} for the proportional hazards assumption for the fitted Cox model based on the RRC approach. Although the results suggested no violation of the proportional hazards assumptions for the exposure (p-value $=0.88$) and the exposure-mediator interaction (p-value $=0.80$), they indicated a potential violation of the proportional hazards assumption for the mediator (p-value $<0.001$). The test also indicated non-proportional hazards with respect to five baseline covariates (p-value $<0.05$), including  smoking status, parental history of myocardial infarction, intake of polyunsaturated fat, diabetes, and hypercholesterolemia. Notice that the original analysis of HPFS by \cite{chomistek2012vigorous} employed a Cox model with time-varying covariates updated at each 2-year follow-up interval and found no evidence of a violation of the proportional hazards assumption. In contrast, our analysis only addressed baseline covariates which may explain potential non-proportional hazards in certain covariates.

Although our simulations in Section \ref{sec:mis-M-O} indicated robustness of our proposed methods under certain types of departure from proportional hazards, we carried out additional sensitivity analyses that relaxed the proportional hazards assumption with time-dependent coefficients across different time intervals in the Cox model \citep{lehr2007parsimonious}. Specifically, we split the entire follow-up time $[0,t^*]$ into $J\geq 2$ intervals, with 
$\mathcal T_1=[0,t_2)$, $\mathcal T_2 = [t_2,t_3)$, $\dots$, $\mathcal T_J=[t_J,t^*]$, and specified the following Cox outcome model with piece-wise constant coefficients:
\begin{equation}\label{m:cox_piecewise_constant}
\lambda_{\widetilde T|A,M,\bm W}(t|A,M,\bm W) = \lambda_0(t)\exp\left(\sum_{j=1}^J \mathbb{I}(t\in \mathcal T_j)\left\{\beta_{1,j}A+\beta_{2,j}M+\beta_{3,j}AM+\bm\beta_{4,j}^T \bm W\right\}\right),
\end{equation}
where $\bm\beta_{\mathcal T_j} = [\beta_{1,j},\beta_{2,j},\beta_{3,j},\bm\beta_{4,j}^T]^T$ is the coefficients in the $j$-th interval. Although we presented the most general model, one can enforce a coefficient to be a constant value over the different time intervals if needed, and the subsequent mediation analysis result will become a special case of our general development below. Under this piecewise Cox outcome model specification, the expressions of the mediation effect measures were re-derived in Web Appendix H and represented by $\text{NIE}_{a,a^*|\bm w}(t) = \widetilde \delta_{a,a|\bm w}\left(\Lambda_0(t),\bm\theta\right) - \widetilde \delta_{a,a^*|\bm w}\left(\Lambda_0(t),\bm\theta\right)$ and $\text{NDE}_{a,a^*|\bm w}(t) = \widetilde \delta_{a,a^*|\bm w}\left(\Lambda_0(t),\bm\theta\right) - \widetilde \delta_{a^*,a^*|\bm w}\left(\Lambda_0(t),\bm\theta\right)$, where $\widetilde \delta_{a,a^*|\bm w}\left(\Lambda_0(t),\bm\theta\right)$ is a function of $\Lambda_0(t)$ (the cumulative baseline hazard in \eqref{m:cox_piecewise_constant}) and $\bm \theta = [\bm\alpha^T,\bm\beta_{\mathcal T_1}^T,\dots,\bm\beta_{\mathcal T_J}^T]^T$ (coefficients in models \eqref{m:mediator} and \eqref{m:cox_piecewise_constant}). The ORC1, ORC2, and RRC approaches can then be naturally extended to accommodate model \eqref{m:cox_piecewise_constant}, by running the Cox model \eqref{m:cox_piecewise_constant} with $A_i$ in the main study replaced by $m_{A_i}(t)$ (as in Section \ref{sec:asmp}). Specific implementation details for Cox model \eqref{m:cox_piecewise_constant} were given in Web Appendix H.

In the application, we considered the Cox model \eqref{m:cox_piecewise_constant} with piece-wise constant coefficients, with 5 disjoint intervals on $[40,54)$, $[54,67)$, $[67,80)$, $[80,93)$, $[93,107]$ years of age. These intervals evenly partition the observed age range for time to CVD incidence, from 40 to 107 years, based on range observed in the HPFS study. To avoid over-fitting, we only specified coefficients as piecewise constant when their main effects were significant and the p-value from the Grambsch-Therneau test in the original Cox model is below 0.05, which include the mediator, smoking status, parental history of myocardial infarction, intake of polyunsaturated fat, diabetes, and hypercholesterolemia. 
Web Table 13 summarized the mediation results based on the Cox model with piece-wise constant coefficients and the RRC approach. Comparing to the parallel results under the proportional hazards assumption (Web Table 12), we observed that the NDE is robust to these observed departures from proportional hazards, but the NIE was somewhat more sensitive under non-proportional hazards.  Specifically, the original RRC estimator under the proportional hazards assumption suggested $\widehat{\text{NIE}}^{(R)}(t)\approx -0.043$ and $\widehat{\text{NDE}}^{(R)}(t)\approx -0.068$ with $\widehat{\text{MP}}^{(R)}(t)\approx 0.388$, and all these mediation effect measures were approximately constant over time. After accounting for potential non-proportional hazards in the Cox model, the RRC estimator indicated that $\widehat{\text{NDE}}^{(R)}(t)$ fluctuated between $-$0.066 and $-$0.065 and $\widehat{\text{NIE}}^{(R)}(t)$ varied between $-$0.074 and $-$0.029 with mediation proportion varying between 0.305 to 0.534. Throughout, all the updated mediation effect estimates remained statistically significant as the 95\% bootstrap confidence intervals excluded the null.

\section{Discussion}
\label{sec:disc}

Motivated by the need to account for exposure measurement error in analysis of the HPFS, this article developed methods for addressing exposure measurement error-inducted bias in mediation analysis of a censored survival outcome. Under a main study/external validation study design, we proposed three bias-correction approaches for the mediator and outcome model parameters (ORC1, ORC2, RRC) along with two sets of expressions of mediation effect measures (the approximate expressions under a rare outcome and the exact expressions under both rare/common outcome), leading to a total of six bias-corrected estimators. With a rare outcome, all proposed estimators are effective for correcting the exposure measurement error-induced bias, as demonstrated by extensive simulations. With a common outcome, only the RRC approach with the exact expressions of the mediation effect is theoretically valid. Interestingly, our simulations also showed that ORC1 and ORC2 also maintain satisfactory bias and coverage metrics with a common outcome at 50\% cumulative event rate, as long as the exact expressions of mediation measures are used. However, using the approximate expressions of mediation effect may induce bias in the common outcome scenario, regardless of whether ORC1, ORC2, and RRC have been used to estimate the mediator and outcome models. Although 10\% is a typical threshold for a rare outcome in epidemiological studies, there is little previous work that investigates the performance of the approximate mediation effect measures in survival analysis under a rare disease assumption when the outcome is common, and our current work fills in that gap. As in our application to the HPFS, when the outcome prevalence is common (e.g., the cumulative event rate exceeds 10\%), we recommend using the exact mediation effect measures, which vary with time, to check the robustness of the estimates based on the approximate expressions. If the mediation effect estimates are observed to be nearly constant over time based on the exact expression, it empirically supports the use of approximation under the rare outcome assumption. Otherwise, it would be more informative to report the exact mediation effect estimates defined in \eqref{exact_expressions} and to present the potentially different results at pre-specified time points.

Although our motivating example aims to assess mediation with a survival outcome and exposure measurement error, the implications from this work extend beyond the context of mediation analysis. First, to the best of our knowledge, our simulation is the first study to present a direct comparison of the finite-sample performance of ORC and RRC methods under a main study/external validation study design. While previous work conducted simulations on regression calibration methods for measurement errors in the Cox model, they either only evaluated the ORC or RRC approach (\citealp{prentice1982covariate,liao2011survival,liao2018survival}) or considered the main study/reliability study design or the scenario with classical additive error and known variance components (\citealp{zhao2014covariate,xie2001risk}). Because the mediation effect measures we focused are essentially functions of the regression parameters, our simulation results can also inform the bias-correction properties for general regression analysis. For example, our results suggests that the ORC is relatively insensitive to violation of the common outcome assumption as long as the exposure effect in the outcome model is not large (with $\beta_1<\sqrt{0.5}$ under a unit exposure variance), supporting its use with Cox analysis with mismeasured covariates in more general regression settings (providing the estimated coefficients are not large). Second, we have created the \texttt{survmed} R package and a short vignette that are available via the GitHub link \url{https://github.com/chaochengstat/survmed} to facilitate the implementation of the proposed ORC and RRC methods. The R package can be applied to settings beyond the mediation contexts, including but not limited to (i) performing standard survival mediation analysis without exposure measurement error (this is equivalent to implementing an uncorrected estimator) and (ii) performing bias-corrected coefficient estimation in Cox regression models under a main study/validation study design. In the Supplementary Material, we provide example syntax of such applications with a simulated data example.

Our development is subject to two limitations that will be addressed in future work. First, following existing literature on measurement error correction, we considered a linear measurement error model to describe the exposure measurement error process. This assumption may be inadequate if the underlying true measurement error process is more complex. In future work, we plan to consider more flexible regression methods (for example, spline-based or other nonparametric approaches) to estimate the measurement error process, and compare its performance to the current methods under misspecification of the linear measurement error model. Second, building on \citet{vanderweele2011causal}, our development assumes a Cox outcome model to describe the outcome process and requires proportional hazards. This assumption is important both conceptually and operationally. Conceptually, as discussed in \citet{martinussen2020subtleties} and \citet{fay2024causal}, the causal interpretation of hazard-based contrasts generally requires proportional hazards, and this implication carries to the interpretation of our mediation effect measures. Operationally, violation of the proportional hazards assumption may necessitate more complicated mediation effect expressions even in the absence of exposure measurement error. To address potential non-proportional hazards in the analysis of HPFS, we have further considered a Cox outcome model with piece-wise constant coefficients, which allows for non-proportional hazards across different disjoint intervals in the time axis (the technical details are presented in Web Appendix H). Investigation of finite-sample performance of the proposed measurement error correction methods under this more complicated Cox model, as well as the development of bias-corrected mediation methods for alternative effect measures (based on survival probabilities and restricted mean survival times) represent important directions for future work.

\section*{Data Availability}
The code of the simulation studies is provided in  Supplementary Material. The HPFS used in Section \ref{sec:app} is not publicly available due to ethical restrictions. Data requests may be directed to the first author.

\section*{Funding}
Research in this article was supported by the National Institutes of Health (NIH) grants DP1ES025459 and R01CA279175, and the Patient-Centered Outcomes Research Institute\textsuperscript{\textregistered} award ME-2023C1-31350. The statements presented in this article are solely the responsibility of the authors and do not necessarily represent the official views of NIH, PCORI\textsuperscript{\textregistered}, their Board of Governors or Methodology Committee.

\section*{Conflict of Interests}
None to declare.

\section*{Supplementary Material}
The supplementary material includes web appendices, figures, and tables unshown in the manuscript and reproducible R code for the simulation studies in Section \ref{sec:sim}. An R package for implementing the proposed method is available at \url{https://github.com/chaochengstat/survmed}.

\bibliographystyle{jasa3}
\bibliography{refs}

\begin{thebibliography}{41}
\newcommand{\enquote}[1]{``#1''}
\expandafter\ifx\csname natexlab\endcsname\relax\def\natexlab#1{#1}\fi
\expandafter\ifx\csname url\endcsname\relax
  \def\url#1{{\tt #1}}\fi
\expandafter\ifx\csname urlprefix\endcsname\relax\def\urlprefix{URL }\fi

\bibitem[\protect\citeauthoryear{Andersen and Gill}{Andersen and
  Gill}{1982}]{andersen1982cox}
Andersen, P.~K. and Gill, R.~D. (1982), \enquote{Cox's regression model for
  counting processes: a large sample study,} {\em The Annals of Statistics\/},
  1100--1120.

\bibitem[\protect\citeauthoryear{Baron and Kenny}{Baron and
  Kenny}{1986}]{baron1986moderator}
Baron, R.~M. and Kenny, D.~A. (1986), \enquote{The moderator--mediator variable
  distinction in social psychological research: Conceptual, strategic, and
  statistical considerations.} {\em Journal of Personality and Social
  Psychology\/}, 51, 1173.

\bibitem[\protect\citeauthoryear{Camirand~Lemyre, Carroll, and
  Delaigle}{Camirand~Lemyre et~al.}{2022}]{camirand2022semiparametric}
Camirand~Lemyre, F., Carroll, R.~J., and Delaigle, A. (2022),
  \enquote{Semiparametric estimation of the distribution of episodically
  consumed foods measured with error,} {\em Journal of the American Statistical
  Association\/}, 117, 469--481.

\bibitem[\protect\citeauthoryear{Carroll, Ruppert, Stefanski, and
  Crainiceanu}{Carroll et~al.}{2006}]{carroll2006measurement}
Carroll, R.~J., Ruppert, D., Stefanski, L.~A., and Crainiceanu, C.~M. (2006),
  {\em Measurement error in nonlinear models: a modern perspective\/}, Chapman
  and Hall/CRC.

\bibitem[\protect\citeauthoryear{Chasan-Taber, Rimm, Stampfer, Spiegelman, and
  et~al}{Chasan-Taber et~al.}{1996}]{chasan1996reproducibility}
Chasan-Taber, S., Rimm, E.~B., Stampfer, M.~J., Spiegelman, D., and et~al
  (1996), \enquote{Reproducibility and validity of a self-administered physical
  activity questionnaire for male health professionals,} {\em Epidemiology\/},
  81--86.

\bibitem[\protect\citeauthoryear{Cheng, Spiegelman, and Li}{Cheng
  et~al.}{2022}]{cheng2022difference}
Cheng, C., Spiegelman, D., and Li, F. (2022), \enquote{Is the product method
  more efficient than the difference method for assessing mediation?} {\em
  American Journal of Epidemiology\/}.

\bibitem[\protect\citeauthoryear{Cheng, Spiegelman, and Li}{Cheng
  et~al.}{2023}]{cheng2022mediation}
--- (2023), \enquote{Mediation analysis in the presence of continuous exposure
  measurement error,} {\em Statistics in Medicine\/}.

\bibitem[\protect\citeauthoryear{Choi, Atkinson, Karlson, and Curhan}{Choi
  et~al.}{2005}]{choi2005obesity}
Choi, H.~K., Atkinson, K., Karlson, E.~W., and Curhan, G. (2005),
  \enquote{Obesity, weight change, hypertension, diuretic use, and risk of gout
  in men: the health professionals follow-up study,} {\em Archives of Internal
  Medicine\/}, 165, 742--748.

\bibitem[\protect\citeauthoryear{Chomistek, Cook, Flint, and Rimm}{Chomistek
  et~al.}{2012}]{chomistek2012vigorous}
Chomistek, A.~K., Cook, N.~R., Flint, A.~J., and Rimm, E.~B. (2012),
  \enquote{Vigorous-intensity leisure-time physical activity and risk of major
  chronic disease in men,} {\em Medicine and Science in Sports and Exercise\/},
  44, 1898.

\bibitem[\protect\citeauthoryear{Fay and Li}{Fay and Li}{2024}]{fay2024causal}
Fay, M.~P. and Li, F. (2024), \enquote{Causal interpretation of the hazard
  ratio in randomized clinical trials,} {\em Clinical Trials\/},
  17407745241243308.

\bibitem[\protect\citeauthoryear{Fern{\'a}ndez and Steel}{Fern{\'a}ndez and
  Steel}{1998}]{fernandez1998bayesian}
Fern{\'a}ndez, C. and Steel, M.~F. (1998), \enquote{On Bayesian modeling of fat
  tails and skewness,} {\em Journal of the American Statistical Association\/},
  93, 359--371.

\bibitem[\protect\citeauthoryear{Fu, Song, Li, Han, Adami, Giovannucci, and
  Mucci}{Fu et~al.}{2020}]{fu2020height}
Fu, B., Song, M., Li, X., Han, J., Adami, H., Giovannucci, E., and Mucci, L.
  (2020), \enquote{Height as a mediator of sex differences in cancer risk,}
  {\em Annals of Oncology\/}, 31, 634--640.

\bibitem[\protect\citeauthoryear{Grambsch and Therneau}{Grambsch and
  Therneau}{1994}]{grambsch1994proportional}
Grambsch, P.~M. and Therneau, T.~M. (1994), \enquote{Proportional hazards tests
  and diagnostics based on weighted residuals,} {\em Biometrika\/}, 81,
  515--526.

\bibitem[\protect\citeauthoryear{Harris, Costenbader, Mu, Kvaskoff, Malspeis,
  Karlson, and Missmer}{Harris et~al.}{2016}]{harris2016endometriosis}
Harris, H.~R., Costenbader, K.~H., Mu, F., Kvaskoff, M., Malspeis, S., Karlson,
  E.~W., and Missmer, S.~A. (2016), \enquote{Endometriosis and the risks of
  systemic lupus erythematosus and rheumatoid arthritis in the Nurses’ Health
  Study II,} {\em Annals of the Rheumatic Diseases\/}, 75, 1279--1284.

\bibitem[\protect\citeauthoryear{Keogh, Shaw, Gustafson, Carroll, Deffner,
  Dodd, K{\"u}chenhoff, Tooze, Wallace, Kipnis et~al.}{Keogh
  et~al.}{2020}]{keogh2020stratos}
Keogh, R.~H., Shaw, P.~A., Gustafson, P., Carroll, R.~J., Deffner, V., Dodd,
  K.~W., K{\"u}chenhoff, H., Tooze, J.~A., Wallace, M.~P., Kipnis, V., et~al.
  (2020), \enquote{STRATOS guidance document on measurement error and
  misclassification of variables in observational epidemiology: part 1—basic
  theory and simple methods of adjustment,} {\em Statistics in Medicine\/}, 39,
  2197--2231.

\bibitem[\protect\citeauthoryear{Lange and Hansen}{Lange and
  Hansen}{2011}]{lange2011direct}
Lange, T. and Hansen, J.~V. (2011), \enquote{Direct and indirect effects in a
  survival context,} {\em Epidemiology\/}, 575--581.

\bibitem[\protect\citeauthoryear{Lee, Wehrlen, Ding, and Ross}{Lee
  et~al.}{2022}]{lee2022professional}
Lee, L.~J., Wehrlen, L., Ding, Y., and Ross, A. (2022), \enquote{Professional
  quality of life, sleep disturbance and health among nurses: A mediation
  analysis,} {\em Nursing Open\/}, 9, 2771--2780.

\bibitem[\protect\citeauthoryear{Lehr and Schemper}{Lehr and
  Schemper}{2007}]{lehr2007parsimonious}
Lehr, S. and Schemper, M. (2007), \enquote{Parsimonious analysis of
  time-dependent effects in the Cox model,} {\em Statistics in Medicine\/}, 26,
  2686--2698.

\bibitem[\protect\citeauthoryear{Liao, Zhou, Wang, Hart, Laden, and
  Spiegelman}{Liao et~al.}{2018}]{liao2018survival}
Liao, X., Zhou, X., Wang, M., Hart, J.~E., Laden, F., and Spiegelman, D.
  (2018), \enquote{Survival analysis with functions of mismeasured covariate
  histories: the case of chronic air pollution exposure in relation to
  mortality in the nurses’ health study,} {\em Journal of the Royal
  Statistical Society: Series C (Applied Statistics)\/}, 67, 307--327.

\bibitem[\protect\citeauthoryear{Liao, Zucker, Li, and Spiegelman}{Liao
  et~al.}{2011}]{liao2011survival}
Liao, X., Zucker, D., Li, Y., and Spiegelman, D. (2011), \enquote{Survival
  analysis with error-prone time-varying covariates: A risk set calibration
  approach,} {\em Biometrics\/}, 67, 50--58.

\bibitem[\protect\citeauthoryear{Lin and Wei}{Lin and
  Wei}{1989}]{lin1989robust}
Lin, D.~Y. and Wei, L.-J. (1989), \enquote{The robust inference for the Cox
  proportional hazards model,} {\em Journal of the American Statistical
  Association\/}, 84, 1074--1078.

\bibitem[\protect\citeauthoryear{Liu and Pierce}{Liu and
  Pierce}{1994}]{liu1994note}
Liu, Q. and Pierce, D.~A. (1994), \enquote{A note on Gauss—Hermite
  quadrature,} {\em Biometrika\/}, 81, 624--629.

\bibitem[\protect\citeauthoryear{Martinussen, Vansteelandt, and
  Andersen}{Martinussen et~al.}{2020}]{martinussen2020subtleties}
Martinussen, T., Vansteelandt, S., and Andersen, P.~K. (2020),
  \enquote{Subtleties in the interpretation of hazard contrasts,} {\em Lifetime
  Data Analysis\/}, 26, 833--855.

\bibitem[\protect\citeauthoryear{McNutt, Wu, Xue, and Hafner}{McNutt
  et~al.}{2003}]{mcnutt2003estimating}
McNutt, L.-A., Wu, C., Xue, X., and Hafner, J.~P. (2003), \enquote{Estimating
  the relative risk in cohort studies and clinical trials of common outcomes,}
  {\em American Journal of Epidemiology\/}, 157, 940--943.

\bibitem[\protect\citeauthoryear{Morgenstern, Rosella, Costa, de~Souza, and
  Anderson}{Morgenstern et~al.}{2021}]{morgenstern2021perspective}
Morgenstern, J.~D., Rosella, L.~C., Costa, A.~P., de~Souza, R.~J., and
  Anderson, L.~N. (2021), \enquote{Perspective: Big data and machine learning
  could help advance nutritional epidemiology,} {\em Advances in Nutrition\/},
  12, 621--631.

\bibitem[\protect\citeauthoryear{Pearl}{Pearl}{2001}]{pearl2001direct}
Pearl, J. (2001), \enquote{Direct and indirect effects,} in {\em Proceedings of
  the Seventeenth Conference on Uncertainty and Artificial Intelligence,
  2001\/}, Morgan Kaufman, 411--420.

\bibitem[\protect\citeauthoryear{Petimar, Tabung, Valeri, Rosner, Chan,
  Smith-Warner, and Giovannucci}{Petimar et~al.}{2019}]{petimar2019mediation}
Petimar, J., Tabung, F.~K., Valeri, L., Rosner, B., Chan, A.~T., Smith-Warner,
  S.~A., and Giovannucci, E.~L. (2019), \enquote{Mediation of associations
  between adiposity and colorectal cancer risk by inflammatory and metabolic
  biomarkers,} {\em International Journal of Cancer\/}, 144, 2945--2953.

\bibitem[\protect\citeauthoryear{Prentice}{Prentice}{1982}]{prentice1982covariate}
Prentice, R.~L. (1982), \enquote{Covariate measurement errors and parameter
  estimation in a failure time regression model,} {\em Biometrika\/}, 69,
  331--342.

\bibitem[\protect\citeauthoryear{Rosner, Spiegelman, and Willett}{Rosner
  et~al.}{1990}]{rosner1990correction}
Rosner, B., Spiegelman, D., and Willett, W.~C. (1990), \enquote{Correction of
  logistic regression relative risk estimates and confidence intervals for
  measurement error: the case of multiple covariates measured with error,} {\em
  American Journal of Epidemiology\/}, 132, 734--745.

\bibitem[\protect\citeauthoryear{Shaw, Gustafson, Carroll, Deffner, Dodd,
  Keogh, Kipnis, Tooze, Wallace, K{\"u}chenhoff et~al.}{Shaw
  et~al.}{2020}]{shaw2020stratos}
Shaw, P.~A., Gustafson, P., Carroll, R.~J., Deffner, V., Dodd, K.~W., Keogh,
  R.~H., Kipnis, V., Tooze, J.~A., Wallace, M.~P., K{\"u}chenhoff, H., et~al.
  (2020), \enquote{STRATOS guidance document on measurement error and
  misclassification of variables in observational epidemiology: part 2—more
  complex methods of adjustment and advanced topics,} {\em Statistics in
  Medicine\/}, 39, 2232--2263.

\bibitem[\protect\citeauthoryear{Spiegelman, Carroll, and Kipnis}{Spiegelman
  et~al.}{2001}]{spiegelman2001efficient}
Spiegelman, D., Carroll, R.~J., and Kipnis, V. (2001), \enquote{Efficient
  regression calibration for logistic regression in main study/internal
  validation study designs with an imperfect reference instrument,} {\em
  Statistics in Medicine\/}, 20, 139--160.

\bibitem[\protect\citeauthoryear{Struthers and Kalbfleisch}{Struthers and
  Kalbfleisch}{1986}]{struthers1986misspecified}
Struthers, C.~A. and Kalbfleisch, J.~D. (1986), \enquote{Misspecified
  proportional hazard models,} {\em Biometrika\/}, 73, 363--369.

\bibitem[\protect\citeauthoryear{Therneau, Crowson, and Atkinson}{Therneau
  et~al.}{2017}]{therneau2017using}
Therneau, T., Crowson, C., and Atkinson, E. (2017), \enquote{Using time
  dependent covariates and time dependent coefficients in the cox model,} {\em
  Survival Vignettes\/}, 2, 1--25.

\bibitem[\protect\citeauthoryear{Van~der Vaart}{Van~der
  Vaart}{2000}]{van2000asymptotic}
Van~der Vaart, A.~W. (2000), {\em Asymptotic statistics\/}, volume~3, Cambridge
  university press.

\bibitem[\protect\citeauthoryear{VanderWeele}{VanderWeele}{2011}]{vanderweele2011causal}
VanderWeele, T.~J. (2011), \enquote{Causal mediation analysis with survival
  data,} {\em Epidemiology (Cambridge, Mass.)\/}, 22, 582.

\bibitem[\protect\citeauthoryear{VanderWeele and Vansteelandt}{VanderWeele and
  Vansteelandt}{2009}]{vanderweele2009conceptual}
VanderWeele, T.~J. and Vansteelandt, S. (2009), \enquote{Conceptual issues
  concerning mediation, interventions and composition,} {\em Statistics and its
  Interface\/}, 2, 457--468.

\bibitem[\protect\citeauthoryear{Wang and Albert}{Wang and
  Albert}{2017}]{wang2017causal}
Wang, W. and Albert, J.~M. (2017), \enquote{Causal mediation analysis for the
  Cox proportional hazards model with a smooth baseline hazard estimator,} {\em
  Journal of the Royal Statistical Society. Series C, Applied statistics\/},
  66, 741.

\bibitem[\protect\citeauthoryear{Xie, Wang, and Prentice}{Xie
  et~al.}{2001}]{xie2001risk}
Xie, S.~X., Wang, C., and Prentice, R.~L. (2001), \enquote{A risk set
  calibration method for failure time regression by using a covariate
  reliability sample,} {\em Journal of the Royal Statistical Society: Series B
  (Statistical Methodology)\/}, 63, 855--870.

\bibitem[\protect\citeauthoryear{Yi, Ma, Spiegelman, and Carroll}{Yi
  et~al.}{2015}]{yi2015functional}
Yi, G.~Y., Ma, Y., Spiegelman, D., and Carroll, R.~J. (2015),
  \enquote{Functional and structural methods with mixed measurement error and
  misclassification in covariates,} {\em Journal of the American Statistical
  Association\/}, 110, 681--696.

\bibitem[\protect\citeauthoryear{Yi, Yan, Liao, and Spiegelman}{Yi
  et~al.}{2018}]{yi2018parametric}
Yi, G.~Y., Yan, Y., Liao, X., and Spiegelman, D. (2018), \enquote{Parametric
  regression analysis with covariate misclassification in main study/validation
  study designs,} {\em The International Journal of Biostatistics\/}, 15,
  20170002.

\bibitem[\protect\citeauthoryear{Zhao and Prentice}{Zhao and
  Prentice}{2014}]{zhao2014covariate}
Zhao, S. and Prentice, R.~L. (2014), \enquote{Covariate measurement error
  correction methods in mediation analysis with failure time data,} {\em
  Biometrics\/}, 70, 835--844.

\end{thebibliography}

\clearpage

\section*{\centering Supplementary Material to ``Correcting for bias due to mismeasured exposure in mediation analysis with a survival outcome"}

\renewcommand{\theequation}{s\arabic{equation}}
\setcounter{equation}{0}
\setcounter{section}{0}
\setcounter{figure}{0}
\setcounter{table}{0}

\singlespacing

\section*{Appendix A: Further explanations of the identifiability assumptions for mediation analysis with a survival outcome}

We make the following assumptions based on the counterfactual outcomes framework \citep{pearl2001direct} to identify the mediation effect measures with a survival outcome. The first is the consistency assumption, which assumes that each individual's potential outcome ($\widetilde{T}_{a}$) and potential mediator ($M_a$) under his or her observed exposure are the outcome and mediator that will actually be observed for that individual under the corresponding treatment level, i.e., $\widetilde{T}_{a}=\widetilde{T}$ and $M_a = M$ if $A=a$ is observed. In the mediation analysis, the consistency assumption also requires that $\widetilde{T}_{am}=T$ if $A=a$ and $M=m$ are observed. As discussed in \citet{vanderweele2009conceptual}, the consistency assumption is usually plausible when the exposure and mediator is well-defined and there is no interference among individuals in the study. The second assumption, positivity, requires that $P(A=a|\bm W)>0$ and $P(M=m|A=a,\bm W)>0$ for any values of $a$ and $m$ in their support. The third assumption is the composition assumption, which assumes that an individual's potential outcome under the exposure $A=a$ is equal to the potential outcome when $A=a$ is observed and the mediator is set to its value corresponding to $A=a$, i.e., $Y_{aM_a}=Y_a$. The fourth assumption consists of four sub-assumptions regarding the confounders, including (i) $\widetilde{T}_{am} \perp A | \bm W$, (ii) $\widetilde{T}_{am} \perp M | A, \bm W$, (iii) $M_a \perp A | \bm W$, and (iv) $\widetilde{T}_{am} \perp M_{a^*} | \bm W$, where $A \perp B | C$ means that $A$ and $B$ are independent conditional on $C$. Intuitively, assumption (i) states that there are no unmeasured confounders for the association between the outcome and exposure, assumption (ii) states that there are no unmeasured confounders for the association between the outcome and mediator, and assumption (iii) states that there are no unmeasured confounders for the association between the mediator and exposure.  Assumption (iv) is sometimes referred to as the cross-world independence assumption, which states that there are no exposure-induced covariates that confound the mediator-outcome relationship. Assumptions (i)--(iv) are typically considered plausible when investigators measure all pre-exposure covariates that confound the exposure-mediator-outcome mechanisms and that there are no exposure-induced confounders for the mediator-outcome relationship \citep{vanderweele2009conceptual}. 


\section*{Appendix B: Bias of the uncorrected estimators for the mediation effect measures with exposure measurement error}

\subsection*{Appendix B.1: Proof of Theorem 1}

When the outcome is rare and in the absence of an exposure-mediator interaction effect ($\beta_3=0$), the mediation effect measure expressions simplify to
\begin{align*}
    \text{NIE}_{a,a^*|\bm w} & \approx \beta_2\alpha_1(a-a^*), \\
    \text{NDE}_{a,a^*|\bm w} & \approx \beta_1(a-a^*), \\
    \text{TE}_{a,a^*|\bm w} & \approx (\beta_1+\beta_2\alpha_1)(a-a^*),\\
    \text{MP}_{a,a^*|\bm w} & \approx \frac{\beta_2\alpha_1}{\beta_1+\beta_2\alpha_1}.
\end{align*}
One can verify that each one of the above expressions depends at most on three regression coefficients, $\alpha_1$ in the mediator model and $\beta_1$ and $\beta_2$ in the Cox outcome model. 
To derive the asymptotic bias of the uncorrected mediation effect measure estimators, our first step is to understand the limiting values of the uncorrected estimator of these three regression coefficients, $(\alpha_1,\beta_1,\beta_2)$.

\subsubsection*{Step 1. Deriving the limiting values of uncorrected regression coefficient estimators}

We first provide the following lemma.
\begin{lemma}\label{lemma:bayes}
If the error term in measurement error model (4) is normally distributed with $\epsilon_A \sim N(0,\sigma_A^2)$, then we have
\begin{equation*}
A|\{A^*,M,\bm W\} \sim N\left(E[A|A^*,M,\bm W], \frac{\sigma_A^2\sigma_M^2}{\alpha_1^2\sigma_M^2+\sigma_A^2}\right),
\end{equation*}
where
$$
E[A|A^*,M,\bm W] =  \frac{\gamma_0\sigma_M^2-\alpha_0\alpha_1\sigma_A^2}{\alpha_1^2\sigma_A^2+\sigma_M^2} + \frac{\gamma_1\sigma_M^2}{\alpha_1^2\sigma_A^2+\sigma_M^2}A^* + \frac{\alpha_1\sigma_A^2}{\alpha_1^2\sigma_A^2+\sigma_M^2}M + \frac{\sigma_M^2\bm\gamma_2^T-\alpha_1\sigma_A^2\bm\alpha_2^T}{\alpha_1^2\sigma_A^2+\sigma_M^2}\bm W.
$$
\end{lemma}
\begin{proof}
By the Bayes' theorem, one can show the distribution of $A|\{A^*,M,\bm W\}$ is
\begin{align*}
    P(A|A^*,M,\bm W) \propto & P(M|A,A^*,\bm W)P(A|A^*,\bm W) \\
    & \quad \text{(by the surrogacy assumption in Section 3 of the manuscript, we have)}\\
    \propto & P(M|A,\bm W)P(A|A^*,\bm W) \\
    \propto & \exp\left\{-\frac{(M-\alpha_0-\alpha_1A-\bm\alpha_2^T\bm W)^2}{2\sigma_M^2}\right\} \exp\left\{-\frac{(A-\gamma_0-\gamma_1A^*-\bm\gamma_2^T\bm W)^2}{2\sigma_{A}^2}\right\} \\
    \propto & \exp\left\{ \frac{\left(A-\frac{\alpha_1(M-\alpha_0-\bm\alpha_2^T \bm W)\sigma_A^2 + (\gamma_0+\gamma_1A^* + \bm \gamma_2^T \bm W)\sigma_M^2    }{ \alpha_1^2\sigma_A^2 + \sigma_M^2 }\right)^2}{2{\sigma_A^2\sigma_M^2}/\left(\alpha_1^2\sigma_M^2+\sigma_A^2\right)}\right\}.
\end{align*}
Thus,
$$
A|\{A^*,M,\bm W\} \sim N\left(\frac{\alpha_1(M-\alpha_0-\bm\alpha_2^T \bm W)\sigma_A^2 + (\gamma_0+\gamma_1A^* + \bm \gamma_2^T \bm W)\sigma_M^2    }{ \alpha_1^2\sigma_A^2 + \sigma_M^2 }, \frac{\sigma_A^2\sigma_M^2}{\alpha_1^2\sigma_M^2+\sigma_A^2}\right).
$$
Then, after some algebra, one can verify that $E[A|A^*,M,\bm W]$ has the form presented in Lemma \ref{lemma:bayes}.
\end{proof}

Recall that the mean structure of the true mediator model is
$$
E[M|A,\bm W] = \alpha_0 + \alpha_1 A + \bm \alpha_2^{T} \bm W,
$$
when each variable is measured without error. With a mismeasured exposure, if we mistakenly treat $A^*$ as the true exposure in the mediator model, the mean structure of the observed mediator model is
$$
E[M|A^*,\bm W] = \alpha_0^* + \alpha_1^* A^* + \bm \alpha_2^{*^T} \bm W.
$$
Denote the unknown regression coefficients in the true and observed mediator model as $\widetilde{\bm \alpha}=[\alpha_0,\alpha_1,\bm \alpha_2^{T}]^T$ and $\widetilde{\bm \alpha}^*=[\alpha_0^*,\alpha_1^*,\bm \alpha_2^{*^T}]^T$ and their corresponding estimators as $\widehat{\widetilde{\bm \alpha}}$ and $\widehat{\widetilde{\bm \alpha}}^*$, respectively. Based on the measurement error model (4), we have that
$$
\begin{aligned}
E[M|A^*,\bm W] & = E\left[E[M|A,A^*,\bm W]|A^*, \bm W\right]\\
& = E\left[\alpha_0 + \alpha_1 A + \bm \alpha_2^T \bm W |A^*, \bm W\right]\\
& = \alpha_0 + \bm \alpha_2^T \bm W +  \alpha_1 E[A|A^*,\bm W] \\
& = \alpha_0 + \bm \alpha_2^T \bm W +  \alpha_1( \gamma_0 + \gamma_1 A^* + \bm \gamma_2^T \bm W) \\
& = \alpha_0 + \alpha_1 \gamma_0 + \alpha_1 \gamma_1 A^* + (\bm \alpha_2^T+\alpha_1 \bm \gamma_2^T) \bm W 
\end{aligned},
$$
Comparing the above coefficients with those in observed mediator regression model, we have $\widehat \alpha_0^* \overset{p}{\to} \alpha_0 + \alpha_1 \gamma_0$, $\widehat \alpha_1^* \overset{p}{\to} \alpha_1 \gamma_1$, $\widehat{\bm \alpha}_2^* \overset{p}{\to} \bm \alpha_2+\alpha_1 \bm \gamma_2$, where `$\overset{p}{\to}$' denotes the convergence in probability.

Next, we derive the limiting value of the regression coefficients in the Cox outcome model. In the absence of exposure-mediator interaction, the true Cox outcome model and the observed Cox outcome model with the true exposure in replace by $A^*$ are
\begin{align*}
\lambda_{\widetilde T} (t|A,M,\bm W) & = \lambda_0 (t) e^{\beta_1 A + \beta_2 M  + \bm\beta_4^T \bm W}, \\
\lambda_{\widetilde T} (t|A^*,M,\bm W) & = \lambda_0^* (t) e^{\beta_1^* A^* + \beta_2^* M + \bm\beta_4^{*^T} \bm W},
\end{align*}
respectively. Denote the unknown regression coefficients in the true and observed outcome model as $\bm \beta=[\beta_1,\beta_2,\bm \beta_4^{T}]^T$ and $\bm \beta^*=[\beta_1^*,\beta_2^*,\bm \beta_4^{*^T}]^T$ and their corresponding estimators as $\widehat{\bm \beta}$ and $\widehat{\bm \beta}^*$, respectively. To obtain the relationship between the observed Cox model $\lambda_{\widetilde T} (t|A^*,M,\bm W)$ and the true Cox model $\lambda_{\widetilde T} (t|A,M,\bm W)$, we must solve
\begin{align*}
    \lambda_{\widetilde T} (t|A^*,M,\bm W) = &  E\left[\lambda_{\widetilde T} (t|A,M,\bm W)\Big|A^*,M,\bm W,T\geq t\right]  \\
    = & E\left[\lambda_0 (t) e^{\beta_1 A + \beta_2 M  + \bm\beta_4^T \bm W}\Big|A^*,M,\bm W,T\geq t\right] \\
    = & \lambda_0 (t) e^{\beta_2 M  + \bm\beta_4^T \bm W} E\left[\exp(\beta_1 A)|A^*,M,\bm W,T\geq t\right].
\end{align*}
Following \cite{prentice1982covariate}, $T \geq t$ can be dropped out from the expectation in the previous formula when the event is rare. Then, we have 
\begin{align*}
    & \lambda_{\widetilde T} (t|A^*,M,\bm W) \\
    \approx & \lambda_0 (t) e^{\beta_2 M  + \bm\beta_4^T \bm W} E\left[\exp(\beta_1 A)|A^*,M,\bm W\right] \\
    & \quad\quad\quad (\text{from Lemma \ref{lemma:bayes}, we have})\\
    = & \lambda_0 (t) e^{\beta_2 M  + \bm\beta_4^T \bm W} e^{\beta_1E[A|A^*,M,\bm W]+0.5\beta_1^2\text{Var}[A|A^*,M,\bm W]} \\
    = & \lambda_0 (t)e^{\frac{0.5\beta_1^2\sigma_A^2\sigma_M^2}{\alpha_1^2\sigma_M^2+\sigma_A^2}+\frac{\gamma_0\sigma_M^2-\alpha_0\alpha_1\sigma_A^2}{\alpha_1^2\sigma_A^2+\sigma_M^2}} e^{\frac{\beta_1\gamma_1\sigma_M^2}{\alpha_1^2\sigma_A^2+\sigma_M^2}A + \left(\beta_2+\frac{\beta_1\alpha_1\sigma_A^2}{\alpha_1^2\sigma_A^2+\sigma_M^2}\right) M  + \left(\bm\beta_4^T+\frac{\beta_1\sigma_M^2\bm\gamma_2^T-\beta_1\alpha_1\sigma_A^2\bm\alpha_2^T}{\alpha_1^2\sigma_A^2+\sigma_M^2}\right) \bm W}. 
\end{align*}
Comparing the above coefficients with those in observed Cox outcome regression, we have $\widehat \beta_1^* \overset{p}{\to} \displaystyle\frac{\beta_1\gamma_1\sigma_M^2}{\alpha_1^2\sigma_A^2+\sigma_M^2}$, $\widehat \beta_2^* \overset{p}{\to} \beta_2+\displaystyle\frac{\beta_1\alpha_1\sigma_A^2}{\alpha_1^2\sigma_A^2+\sigma_M^2}$, and $\widehat{\bm \beta}_4^* \overset{p}{\to} \bm\beta_4+\displaystyle\frac{\beta_1\sigma_M^2\bm\gamma_2-\beta_1\alpha_1\sigma_A^2\bm\alpha_2}{\alpha_1^2\sigma_A^2+\sigma_M^2}$.

\subsubsection*{Step 2: Deriving the asymptotic relative bias for the uncorrected mediation effect measure estimators}

Here, the asymptotic relative bias ($RelBias$) is defined as the ratio of the asymptotic bias and the true value of the mediation effect measure such that $RelBias(\widehat \tau_{a,a^*|\bm w}^*) = \underset{n\rightarrow \infty }{\text{lim}} \displaystyle\frac{\widehat \tau_{a,a^*|\bm w}^* - \tau_{a,a^*|\bm w}}{\tau_{a,a^*|\bm w}}$. From Step 1, we obtain that $\widehat \alpha_1^* \overset{p}{\to} \alpha_1 \gamma_1$, $\widehat \beta_1^* \overset{p}{\to} \displaystyle\frac{\beta_1\gamma_1\sigma_M^2}{\alpha_1^2\sigma_A^2+\sigma_M^2}$, and $\widehat \beta_2^* \overset{p}{\to} \beta_2+\displaystyle\frac{\beta_1\alpha_1\sigma_A^2}{\alpha_1^2\sigma_A^2+\sigma_M^2}$.

The asymptotic relative bias of $\widehat{\text{NIE}}_{a,a^*|\bm w}^*$ is thus
\begin{align*}
RelBias(\widehat{\text{NIE}}_{a,a^*|\bm w}^*) =& \underset{n\rightarrow \infty }{\text{lim}} \frac{\widehat{\text{NIE}}_{a,a^*|\bm w}^* - \text{NIE}_{a,a^*|\bm w}}{\text{NIE}_{a,a^*|\bm w}} 
\\
\approx & \frac{\left(\beta_2+\frac{\beta_1\alpha_1\sigma_A^2}{\alpha_1^2\sigma_A^2+\sigma_M^2}\right)\alpha_1\gamma_1 - \beta_2\alpha_1}{\beta_2\alpha_1}\\
= & \gamma_1-1+ \gamma_1\frac{\alpha_1^2\beta_1\sigma_A^2}{\beta_2\alpha_1(\alpha_1^2\sigma_A^2+\sigma_M^2)} \\
= & \gamma_1-1+ \gamma_1\frac{\alpha_1^2\sigma_A^2}{\alpha_1^2\sigma_A^2+\sigma_M^2}\left(1/\text{MP}_{a,a^*|\bm w}-1\right) \quad \text{(By Noting $\frac{1}{\text{MP}_{a,a^*|\bm w}}-1=\frac{\beta_1}{\beta_2\alpha_1}$)} \\
= & \gamma_1-1+ \gamma_1\rho\left(1/\text{MP}_{a,a^*|\bm w}-1\right).
\end{align*}
The asymptotic relative bias of $\widehat{\text{NDE}}_{a,a^*|\bm w}^*$ is
\begin{align*}
RelBias(\widehat{\text{NDE}}_{a,a^*|\bm w}^*) =& \underset{n\rightarrow \infty }{\text{lim}} \frac{\widehat{\text{NDE}}_{a,a^*|\bm w}^* - \text{NDE}_{a,a^*|\bm w}}{\text{NDE}_{a,a^*|\bm w}} 
\\
\approx & \frac{\frac{\beta_1\gamma_1\sigma_M^2}{\alpha_1^2\sigma_A^2+\sigma_M^2}-\beta_1}{\beta_1}
=  \frac{\gamma_1\sigma_M^2}{\alpha_1^2\sigma_A^2+\sigma_M^2} - 1 \\
=&  (\gamma_1-1)\frac{\sigma_M^2}{\alpha_1^2\sigma_A^2+\sigma_M^2}+\frac{\sigma_M^2}{\alpha_1^2\sigma_A^2+\sigma_M^2} - 1 \\
= & (\gamma_1-1)(1-\rho) - \rho.
\end{align*}
Also, we can show that the asymptotic relative bias of $\widehat{\text{TE}}_{a,a^*|\bm w}^*$ is
\begin{align*}
RelBias(\widehat{\text{TE}}_{a,a^*|\bm w}^*) =& \underset{n\rightarrow \infty }{\text{lim}} \frac{\widehat{\text{TE}}_{a,a^*|\bm w}^* - \text{TE}_{a,a^*|\bm w}}{\text{TE}_{a,a^*|\bm w}} 
\\
\approx & \frac{\frac{\beta_1\gamma_1\sigma_M^2}{\alpha_1^2\sigma_A^2+\sigma_M^2}+\left(\beta_2+\frac{\beta_1\alpha_1\sigma_A^2}{\alpha_1^2\sigma_A^2+\sigma_M^2}\right)\alpha_1\gamma_1 -  (\beta_1+\beta_2\alpha_1)}{\beta_1+\beta_2\alpha_1}\\
= & \frac{\gamma_1(\beta_1+\beta_2\alpha_1)-(\beta_1+\beta_2\alpha_1)}{\beta_1+\beta_2\alpha_1} \\
= & \gamma_1-1.
\end{align*}
Finally, the  asymptotic relative bias of $\widehat{\text{MP}}_{a,a^*|\bm w}^*$ is
\begin{align*}
RelBias(\widehat{\text{MP}}_{a,a^*|\bm w}^*) =& \underset{n\rightarrow \infty }{\text{lim}} \frac{\widehat{\text{MP}}_{a,a^*|\bm w}^* - \text{MP}_{a,a^*|\bm w}}{\text{MP}_{a,a^*|\bm w}} 
\\
\approx & \frac{\frac{\left(\beta_2+\frac{\beta_1\alpha_1\sigma_A^2}{\alpha_1^2\sigma_A^2+\sigma_M^2}\right)\alpha_1\gamma_1}{\frac{\beta_1\gamma_1\sigma_M^2}{\alpha_1^2\sigma_A^2+\sigma_M^2}+\left(\beta_2+\frac{\beta_1\alpha_1\sigma_A^2}{\alpha_1^2\sigma_A^2+\sigma_M^2}\right)\alpha_1\gamma_1} -  \text{MP}_{a,a^*|\bm w}}{\text{MP}_{a,a^*|\bm w}} \\
= & \frac{\frac{\left(\beta_2+\frac{\beta_1\alpha_1\sigma_A^2}{\alpha_1^2\sigma_A^2+\sigma_M^2}\right)\alpha_1\gamma_1}{(\beta_1+\beta_2\alpha_1)\gamma_1} -  \text{MP}_{a,a^*|\bm w}}{\text{MP}_{a,a^*|\bm w}} = \frac{\frac{(\beta_2\alpha_1+\beta_1\rho)\gamma_1}{(\beta_1+\beta_2\alpha_1)\gamma_1} -  \text{MP}_{a,a^*|\bm w}}{\text{MP}_{a,a^*|\bm w}} \\
= &  \frac{\frac{\beta_2\alpha_1+\beta_1\rho}{\beta_1+\beta_2\alpha_1} -  \text{MP}_{a,a^*|\bm w}}{\text{MP}_{a,a^*|\bm w}} =  \frac{ \text{MP}_{a,a^*|\bm w} + \rho\left(1-\text{MP}_{a,a^*|\bm w}\right)-  \text{MP}_{a,a^*|\bm w}}{\text{MP}_{a,a^*|\bm w}} \\
= & \rho\left(\frac{1}{\text{MP}_{a,a^*|\bm w}}-1\right).
\end{align*}
This completes the proof of Theorem 1.

\subsection*{Appendix B.2: Derivation of the expression of the MP reliability index ($\rho$) in the absence of covariates}

In this section, we assume the absence of any covariates (i.e., $\bm W$ is empty) and derive the formula $\rho = \frac{1-\rho_{AA^*}^2}{1/\rho_{AM}^2-\rho_{AA^*}^2}$, where $\rho_{AA^*}=\text{Corr}(A,A^*)$ and $\rho_{AM}=\text{Corr}(A,M)$. We first define a few notation such that $v_A^2=\text{Var}(A)$, $v_{A^*}^2=\text{Var}(A^*)$, and $v_M^2=\text{Var}(M)$ are the variances of $A$, $A^*$, and $M$, respectively. 

In the absence of covariates, the measurement error model (4) simplifies to $A=\gamma_0+\gamma_1A^*+\epsilon_A$, based on which we obtain
\begin{align*}
\left\{\begin{matrix}
\text{Var}(A)  =\gamma_1^2 \text{Var}(A^*) + \text{Var}(\epsilon_A) \\
\text{Cov}(A,A^*)  = \text{Cov}(\gamma_0+\gamma_1A^*,A^*)
\end{matrix}\right. \Rightarrow \left\{\begin{matrix}
v_A^2  =\gamma_1^2 v_{A^*}^2 + \sigma_A^2 \\
\rho_{AA^*}\sqrt{v_{A}^2v_{A^*}^2}  = \gamma_1v_{A^*}^2. 
\end{matrix}\right.
\end{align*}
Solving the above two equations for $v_A^2$ leads to
$$
\sigma_A^2 = (1-\rho_{AA^*}^2)v_A^2.
$$
Similarly, when $\bm W$ is empty, the mediator model (1) simplifies to $M=\alpha_0+\alpha_1A+\epsilon_M$, and therefore 
\begin{align*}
\left\{\begin{matrix}
\text{Var}(M)  =\alpha_1^2 \text{Var}(A) + \text{Var}(\epsilon_M) \\
\text{Cov}(M,A)  = \text{Cov}(\alpha_0+\alpha_1A,A)
\end{matrix}\right. \Rightarrow \left\{\begin{matrix}
v_M^2  =\alpha_1^2 v_{A}^2 + \sigma_M^2 \\
\rho_{AM}\sqrt{v_{A}^2v_{M}^2}  = \alpha_1v_{A}^2
\end{matrix}\right.,
\end{align*}
which suggests 
$$
\rho_{AM}^2 = \frac{\alpha_1^2v_A^2}{\alpha_1^2v_A^2+\sigma_M^2} \Rightarrow \frac{\sigma_M^2}{\alpha_1^2} = (1/\rho_{AM}^2-1)v_A^2.
$$
Now, we have $\frac{\sigma_M^2}{\alpha_1^2} = (1/\rho_{AM}^2-1)v_A^2$ and $\sigma_A^2 = (1-\rho_{AA^*}^2)v_A^2$. Finally, by the definition of $\rho$, we have
$$
\rho = \frac{\sigma_A^2}{\sigma_A^2 + \frac{\sigma_M^2}{\alpha_1^2}} = \frac{(1-\rho_{AA^*}^2)v_A^2}{(1-\rho_{AA^*}^2)v_A^2+(1/\rho_{AM}^2-1)v_A^2}=\frac{1-\rho_{AA^*}^2}{1/\rho_{AM}^2-\rho_{AA^*}^2}.
$$
This completes the derivation.\medskip

\subsection*{Appendix B.3: Connections to bias formulas for mediation analysis with a binary outcome and a mismeasured exposure}

In this section, we compare the bias formula in the present work to the bias formula given in \cite{cheng2022mediation} for mediation analysis with a binary outcome in the absence of an exposure-mediator interaction. 
In \cite{cheng2022mediation}, the mediation effect measures are defined through two outcome regression models (employing the difference-in-coefficients method), and the exposure measurement error process is characterized by two measurement error models. The first measurement error model is same to model (4) in this paper and the second measurement error model is
$$
E[A|A^*,M,\bm W] = \eta_0 +\eta_1 A^* + \eta_2 M + \bm\eta_3^T \bm W. 
$$
Based on the previous two measurement error models, \cite{cheng2022mediation} showed that the asymptotic relative bias formulas of the uncorrected mediation effect measure estimators are
\begin{align*}
& RelBias\left(\widehat{\text{NIE}}^*\right)= \gamma_1-1 + \left(\frac{1}{\text{MP}}-1\right)\left(\gamma_1-\eta_1\right), \\
& RelBias\left(\widehat{\text{NDE}}^*\right)=  \eta_1-1,\\
& RelBias\left(\widehat{\text{TE}}^*\right)= \gamma_1-1, \\
& RelBias\left(\widehat{\text{MP}}^*\right)= \left(\frac{1}{\text{MP}}-1\right)\left(1-\frac{\eta_1}{\gamma_1}\right), 
\end{align*}
where $\widehat{\text{NIE}}^*$, $\widehat{\text{NDE}}^*$, $\widehat{\text{TE}}^*$, and $\widehat{\text{MP}}^*$ are the uncorrected estimators of NIE, NDE, TE, and MP, respectively. 

Comparing the bias formulas in \cite{cheng2022mediation} to Theorem 1, one can observe that the two versions of bias formulas will coincide if and only if $\rho = 1 - \eta_1/\gamma_1$. Below we show that $\rho = 1 - \eta_1/\gamma_1$ holds under  mediator model (1). Specifically, based on (1) and (4), Lemma 1 implies that
\begin{equation*}
   E[A|A^*,M,\bm W] = \frac{\gamma_0\sigma_M^2-\alpha_0\alpha_1\sigma_A^2}{\alpha_1^2\sigma_A^2+\sigma_M^2} + \frac{\gamma_1\sigma_M^2}{\alpha_1^2\sigma_A^2+\sigma_M^2}A^* + \frac{\alpha_1\sigma_A^2}{\alpha_1^2\sigma_A^2+\sigma_M^2}M + \frac{\sigma_M^2\bm\gamma_2^T-\alpha_1\sigma_A^2\bm\alpha_2^T}{\alpha_1^2\sigma_A^2+\sigma_M^2}\bm W,
\end{equation*}
which suggests that $\eta_1 = \displaystyle\frac{\gamma_1\sigma_M^2}{\alpha_1^2\sigma_A^2+\sigma_M^2}$. Substituting this into the formula of $\rho$, we obtain
$$
\rho = \frac{\sigma_A^2}{\sigma_A^2+\sigma_M^2/\alpha_1^2} = \frac{\alpha_1^2\sigma_A^2}{\alpha_1^2\sigma_A^2+\sigma_M^2} = 1-\frac{\displaystyle\frac{\gamma_1\sigma_M^2}{\alpha_1^2\sigma_A^2+\sigma_M^2}}{\gamma_1} = 1-\frac{\eta_1}{\gamma_1}.
$$
This confirms that $\rho=1-\displaystyle\frac{\eta_1}{\gamma_1}$ and reveals that the two versions of bias formulas are identical, despite the apparent differences in the estimation methods and expressions of the mediation effect measures.

\section*{Appendix C: Conditions for valid approximations using the ORC and RRC approaches}

\subsection*{Appendix C.1: Valid approximations in the ORC approach}

In this subsection, we shall elaborate on the four conditions in Section 4.2 for establishing the validity of the approximation $E[e^{ (\beta_1+\beta_3M)A }|A^*,M,\bm W] \approx e^{ (\beta_1+\beta_3M)E[A|A^*,M,\bm W]}$. The four conditions are adapted from \cite{liao2018survival} who considered correcting for bias in the Cox model due to covariates measurement error. Here, we generalize their conditions to the survival mediation setting with a mismeasured exposure. Specifically, the four conditions include
\begin{enumerate}
\item[(i)] There is no exposure-mediator interaction effect ($\beta_3=0$) and $A|\{A^*,M,\bm W\}$ is normally distributed with a constant variance such that $\text{Var}(A|A^*,M,\bm W)=\sigma^2$.
\item[(ii)] The distribution $A|\{A^*,M,\bm W\}$ is normal and $(\beta_1+\beta_3M)^2\text{Var}(A|A^*,M,\bm W)$ is small.
\item[(iii)] $|\beta_1|+|\beta_3|$ is small and $(\beta_1+\beta_3M)^2\text{Var}(A|A^*,M,\bm W)$ is small.
\item[(iv)] $|\beta_1|$ is small with no exposure-mediator interaction ($\beta_3=0$) and $\text{Var}(A|A^*,M,\bm W)=\sigma^2$ is a constant.
\end{enumerate}
In particular, the approximation based on condition (i) is exact. Compared to condition (i), condition (ii) allows for an exposure-mediator interaction and non-constant variance $\text{Var}(A|A^*,M,\bm W)$, but it requires $(\beta_1+\beta_3M)^2\text{Var}(A|A^*,M,\bm W)$ to be small. Conditions (iii) and (iv) relax the normality assumption for the distribution $A|\{A^*,M,\bm W\}$. Compared to condition (iii), condition (iv) does not require the entire term $(\beta_1+\beta_3M)^2\text{Var}(A|A^*,M,\bm W)$ to be small, but additionally needs $\beta_3=0$ and a constant variance assumption. Below, we show that each of these four conditions leads to valid approximation.

Under condition (i), one can show
\begin{align*}
    &E[e^{ (\beta_1+\beta_3M)A }|A^*,M,\bm W] \\
    = & E[e^{ \beta_1 A }|A^*,M,\bm W] \\
    = & \exp\left\{\beta_1 E[A|A^*,M,\bm W] + \beta_1^2\text{Var}(A|A^*,M,\bm W)\right\} \\
    = & e^{0.5\beta_1^2\sigma^2}\times e^{\beta_1 E[A|A^*,M,\bm W]},
\end{align*}
where the first equality follows from $\beta_3=0$, the second equality follows from the moment-generating function of the normal distribution, and the third equality follows from $\text{Var}(A|A^*,\\ M,\bm W)=\sigma^2$. Observing that $e^{0.5\beta_1^2\sigma^2}$ is a constant that does not depend on any random variables, the term $e^{0.5\beta_1^2\sigma^2}$ is absorbed into the baseline hazard and does not affect the estimation of the Cox outcome model coefficients, $\bm\beta$. Therefore the approximation under condition (1) is exact.

Under condition (ii), we have that
\begin{align*}
    &E[e^{ (\beta_1+\beta_3M)A }|A^*,M,\bm W] \\
    = & \exp\left\{(\beta_1+\beta_3M) E[A|A^*,M,\bm W] + 0.5(\beta_1+\beta_3M)^2\text{Var}(A|A^*,M,\bm W)\right\},
\end{align*}
where the equality holds by the moment-generating function of the normal distribution. Moreover, because $(\beta_1+\beta_3M)^2\text{Var}(A|A^*,M,\bm W)$ is assumed to be small in condition (ii), we further obtain
\begin{align*}
    E[e^{ (\beta_1+\beta_3M)A }|A^*,M,\bm W]
    \approx  \exp\left\{(\beta_1+\beta_3M) E[A|A^*,M,\bm W]\right\}.
\end{align*}
This concludes that the approximation holds under the second condition.

Conditions (iii) and (iv) do not require the normality assumption. Approximation under these two conditions use the following second-order Tayler expansion to the moment generating function of the distribution of $A|\{A^*,M,\bm W\}$, which is valid if both $\beta_1$ and $\beta_3$ are small:
\begin{align*}
    & E[e^{ (\beta_1+\beta_3M)A }|A^*,M,\bm W]\\
    \approx & \exp\left\{(\beta_1+\beta_3M) E[A|A^*,M,\bm W] + 0.5(\beta_1+\beta_3M)^2 \text{Var}(A|A^*,M,\bm W)\right\}.
\end{align*}
If we further assume condition (iii) such that $(\beta_1+\beta_3M)^2 \text{Var}(A|A^*,M,\bm W)$ is small, then
\begin{align*}
    E[e^{ (\beta_1+\beta_3M)A }|A^*,M,\bm W]
    \approx  \exp\left\{(\beta_1+\beta_3M) E[A|A^*,M,\bm W]\right\}.
\end{align*} 
Otherwise, if we assume condition (iv) such that $\beta_3=0$ and $\text{Var}(A|A^*,M,\bm W)=\sigma^2$ is a constant, then
\begin{align*}
    E[e^{ (\beta_1+\beta_3M)A }|A^*,M,\bm W]
    & \approx  \exp\left\{ \beta_1 E[A|A^*,M,\bm W] + \beta_1^2\text{Var}(A|A^*,M,\bm W)\right\} \\
    & \approx  e^{0.5\beta_1^2\sigma^2}\times e^{\beta_1 E[A|A^*,M,\bm W]},
\end{align*}
where $e^{0.5\beta_1^2\sigma^2}$ is a constant value and is therefore absorbed into the baseline hazard function in the Cox model. 

\subsection*{Appendix C.2: Valid approximations in the RRC approach}

In this subsection, we elaborate on the four conditions shown in Section 4.3 for establishing the validity of the approximation $E[e^{ (\beta_1+\beta_3M)A }|A^*,M,\bm W,T\geq t] \approx e^{ (\beta_1+\beta_3M)E[A|A^*,M,\bm W,T\geq t]}$. The four conditions resemble those presented in Appendix C.1, and include
\begin{enumerate}
\item[(i)] There is no exposure-mediator interaction effect ($\beta_3=0$), $A|\{A^*,M,\bm W,T\geq t\}$ is normally distributed, and $\text{Var}(A|A^*,M,\bm W,T\geq t)=\sigma^2(t)$ at most depends on $t$.
\item[(ii)] The distribution $A|\{A^*,M,\bm W,T\geq t\}$ is normal and $(\beta_1+\beta_3M)^2\text{Var}(A|A^*,M,\bm W,T\geq t)$ is small.
\item[(iii)] $|\beta_1|+|\beta_3|$ is small and $(\beta_1+\beta_3M)^2\text{Var}(A|A^*,M,\bm W,T\geq t)$ is small.
\item[(iv)] $|\beta_1|$ is small, there is no exposure-mediator interaction ($\beta_3=0$), and $\text{Var}(A|A^*,M,\bm W,T\geq t)=\sigma^2(t)$ is at most only a function of $t$.
\end{enumerate}

These four conditionas are same as those in Appendix C.1 except that conditions (i) and (iv) in Appendix C.1 require that $\text{Var}(A|A^*,M,\bm W)=\sigma^2$ to be a constant whereas now we require $\text{Var}(A|A^*,M,\bm W,T\geq t)$ to be at most a function of $t$ (but free of $\{A^*,M,\bm W\}$). The steps to show these are sufficient conditions for valid approximations are also similar to those in Appendix C.1 and is therefore omitted for brevity.

\section*{Appendix D: Asymptotic properties of the ORC estimators}

This section presents the asymptotic properties of the ORC approach. We shall primarily focus on the ORC1 approach, and an extension to the ORC2 approach is discussed in Appendix D.3. We denote the ORC1 estimators for the mediator model (1) and outcome model parameters (2) as $\widehat{\bm\theta}^{(O1)}=\left[\begin{matrix}\widehat{\bm\alpha} \\ \widehat{\bm\beta}^{(O1)}\end{matrix}\right]$.  

\subsection*{Appendix D.1: Consistency of $\widehat{\bm\alpha}$ and approximate consistency of $\widehat{\bm\beta}^{(O1)}$}

Besides the transportability assumption (given just before equation (5) of the main manuscript) and surrogacy assumption (given in Section 3 of the main manuscript), we additionally assume the following regularity conditions: 
\begin{itemize}
    \item[D1a.] {Assume both the main study and validation study sizes go to infinity such that $n_1\to \infty$ and $n-n_1 \to \infty$.}
    \item[D1b.] {To prove asymptotic normality in Appendix D.2, we additionally assume that the ratio of main study size and total sample size, $n_1/n$, converges to a constant $\omega \in (0,1)$.}
     \item[D2.] Assume that the supports of the unknown parameters $\bm\gamma$, $\bm\eta$, $\bm \alpha$, and $\bm\beta$, denoted by $\bm{\Gamma}$, $\bm{\mathcal H}$, $\bm{\mathcal A}$, and $\bm{\mathcal B}$, are compact subsets of the Euclidean space. Assume that estimators the parameters in the measurement error models (4) and (10) are consistent such that  $\|\widehat{\bm\gamma}-\bm\gamma_0\| \overset{p}{\to} \bm 0$ and $\|\widehat{\bm\eta}-\bm\eta_0\| \overset{p}{\to} \bm 0$, where $\bm\gamma_0$ and $\bm\eta_0$ are true values of $\bm\gamma$ and $\bm\eta$, respectively, and $\|\cdot\|$ denotes the Euclidean norm. By the transportability assumption, the true values $\bm\gamma_0$ and $\bm\eta_0$ are identical for individuals in the main study and validation study.
     \item[D3.] Stack equations (6) and (7) and define the joint estimating equations as $\eU_{\bm\alpha}(\bm\alpha,\widehat{\bm\gamma})=\bm 0$. We further define $\eU_{\bm\alpha}(\bm\alpha,\bm\gamma)$ as $\eU_{\bm\alpha}(\bm\alpha,\widehat{\bm\gamma})$ with $\widehat{\bm\gamma}$ in replacement by $\bm\gamma$, and define $\eU_{\bm\alpha,i}(\bm\alpha,\bm\gamma)$ as its corresponding estimating function (i.e., the $i$-th component of $\eU_{\bm\alpha}(\bm\alpha,{\bm\gamma})$). Let $\bm\alpha_0$ be the true value of $\bm\alpha$. Suppose that $\eU_{\bm\alpha,i}(\bm\alpha,\bm\gamma)$ is continuous, differentiable, and bounded for $(\bm\alpha,\bm\gamma) \in \bm{\mathcal A}\times\bm{\Gamma}$, which is sufficient to show $\frac{1}{n}\frac{\partial}{\partial \bm\alpha}\eU_{\bm\alpha}(\bm\alpha_0,\bm\gamma_0)$ converges weakly to $\bm{\mathcal{I}}_{\bm\alpha}=E[\frac{\partial}{\partial \bm\alpha}\eU_{\bm\alpha,i}(\bm\alpha_0,\bm\gamma_0)]$ by law of large numbers. We further assume $\bm{\mathcal{I}}_{\bm\alpha}$ is negative definite, which ensures that ${\bm\alpha}_0$ is unique solution of $\eU_{\bm\alpha}(\bm\alpha,{\bm\gamma}_0)=\bm 0$.  
    \item[D4.] Define 
    $$
    s^{(k)}(\bm\beta,\bm\eta,t) = E\left[Y(t) \Big(\bm Z(\bm\eta)\Big)^{\otimes k} \exp\Big(\bm\beta^T\bm Z(\bm\eta)\Big)\right],\text{ for } k=0,1,2,
    $$
    where $\bm Z(\bm\eta) = \left [m_A^{(O1)}(\bm\eta),M,m_A^{(O1)}(\bm\eta)M,\bm W^T\right]^T$. Here, for a vector $\bm v$, we define $\bm v^{\otimes k}=1$, $\bm v$, and $\bm v\bm v^T$, if $k=0$, 1, and 2, respectively. Assume that $s^{(k)}(\bm\beta,\bm\eta,t)$, for $k\in\{0,1,2\}$, is a continuous function of $(\bm\beta,\bm\eta) \in \bm{\mathcal B}\times \bm{\mathcal H}$ and uniformly continuous in the time interval $t\in [0,t^*]$. Also, suppose that $s^{(k)}(\bm\beta,\bm\eta,t)$, for $k\in\{0,1,2\}$, is bounded on $(\bm\beta,\bm\eta,t) \in \bm{\mathcal B}\times\bm{\mathcal H} \times [0,t^*]$. Moreover, assume that $s^{(0)}(\bm\beta,\bm\eta,t)$ is bounded away from 0. 
    \item[D5.] Define
    $$
    S^{(k)}(\bm\beta,\bm\eta,t) = \frac{1}{n_1}\sum_{i=1}^{n_1} Y_i(t) \Big(\bm Z_i(\bm\eta)\Big)^{\otimes k} \exp\Big(\bm\beta^T\bm Z_i(\bm\eta)\Big), \text{ for } k=0,1,2.
    $$
    Assume that $\sup_{(\bm\beta,\bm\eta,t)\in \mathcal{\bm B}\times\bm{\mathcal H} \times [0,t^*]} \left\|S^{(k)}(\bm\beta,\bm\eta,t)- s^{(k)}(\bm\beta,\bm\eta,t)\right\| \overset{p}{\to}\bm 0$, for $k=0,1,2$.
\end{itemize}



Before starting the formal proof, we introduce the following lemma, which suggests the estimating equation for $\widehat{\bm\alpha}$, i.e., $\mathbb{U}_{\bm\alpha}(\bm\alpha,\bm\gamma)$ defined in regularity condition (D3), is an unbiased estimating equation.

\begin{lemma}\label{lemma:unbiased}
The estimating equation $\mathbb{U}_{\bm\alpha}(\bm\alpha,\bm\gamma)=\bm 0$ is an unbiased estimating equation for $\bm\alpha$ in the sense that the expectation of its estimating function evaluated under the true parameter values $\{\bm\alpha_0,\bm\gamma_0\}$ is $\bm 0$, that is, $E[\mathbb{U}_{\bm\alpha,i}(\bm\alpha_0,\bm\gamma_0)]=\bm 0$.
\end{lemma}
\begin{proof}
Let $\bm\gamma_0$, $\widetilde{\bm\alpha}_0=[\alpha_{0,0},\alpha_{1,0},\bm\alpha_{2,0}^T]^T$, $\sigma_{M0}^2$ be the true values of $\bm\gamma$, $\widetilde{\bm\alpha}$, $\sigma_M^2$, respectively. From measurement error model (4), we can show the first two moments of $P(M|A^*,\bm W)$ are
\begin{align}
E[M|A^*,\bm W] & = E_A\left[E[M|A,A^*,\bm W]|A^*, \bm W\right]\\
& = E_A\left[\alpha_{0,0} + \alpha_{1,0} A + \bm \alpha_{2,0}^T \bm W |A^*, \bm W\right]\nonumber \\
& = \alpha_{0,0} +  \alpha_{1,0} E[A|A^*,\bm W] + \bm \alpha_{2,0}^T \bm W \nonumber \\
& = \alpha_{0,0} +  \alpha_{1,0} \mu_A(\bm\gamma_0) + \bm \alpha_{2,0}^T \bm W,\label{eq:mean}
\end{align}
and 
\begin{align}
\text{Var}(M|A^*,\bm W) & = \text{Var}\Big(\alpha_{0,0}+\alpha_{1,0}A+\bm\alpha_{2,0}^T\bm W+\epsilon_{M}|A^*,\bm W\Big) \nonumber \\
& = \alpha_{1,0}^2\text{Var}(A|A^*,\bm W) + \text{Var}(A|A^*,\bm W) \nonumber \\
& = \alpha_{1,0}^2\sigma_{A0}^2 +\sigma_{M0}^2.\label{eq:var}
\end{align}
We can rewrite $\mathbb{U}_{\bm\alpha}(\bm\alpha,\bm\gamma)=\displaystyle\sum_{i=1}^n \eU_{\bm\alpha,i}(\bm\alpha,{\bm\gamma})=\displaystyle\sum_{i=1}^n\left[\begin{matrix}\eU_{\widetilde{\bm\alpha},i}(\widetilde{\bm\alpha},{\bm\gamma}) \\ \eU_{\sigma_M^2,i}(\sigma_M^2,\widetilde{\bm\alpha},{\bm\gamma})\end{matrix}\right] = \bm 0$, where $\eU_{\widetilde{\bm\alpha},i}(\widetilde{\bm\alpha},{\bm\gamma})$ and $\eU_{\sigma_M^2,i}(\sigma_M^2,\widetilde{\bm\alpha},{\bm\gamma})$ are defined in equation (6) and (7) of the main manuscript.  In what follows, we show $E[\eU_{\widetilde{\bm\alpha},i}(\widetilde{\bm\alpha}_0,\bm\gamma_0)]=\bm 0$ and $E[\eU_{\sigma_M^2,i}(\sigma_{M0}^2,\widetilde{\bm\alpha}_0,\bm\gamma_0)]=0$, separately, which  concludes our proof. 

For individuals in the validation study ($R_i=0$), one can see that $\eU_{\bm\alpha,i}(\bm\alpha_0,\bm\gamma_0)$ is actually the likelihood score of $\bm\alpha$ in the mediator model (1), and therefore its expectation is zero, if mediator model (1) is correctly specified. Otherwise, for individuals in the main study ($R_i=1$), we can show that
\begin{align*}
    & E[\eU_{\widetilde{\bm\alpha},i}(\widetilde{\bm\alpha}_0,\bm\gamma_0)|R_i=1]  \\
   = &  E\left\{E\left[\eU_{\widetilde{\bm\alpha}_0,i}(\widetilde{\bm\alpha}_0,\bm\gamma_0)\Big|A_i^*,\bm W_i,R_i=1\right]\Big| R_i=1\right\} \\
   = & E\left\{E\left[\left(\left.\begin{matrix}1\\
    \mu_{A_i}(\bm\gamma_0) \\ 
    \bm W_i\end{matrix}\right)(M_i\!-\!\alpha_{0,0}\!-\!\alpha_{1,0} \mu_{A_i}(\bm\gamma_0) \!-\! \bm\alpha_{2,0}^T \bm W_i)\right|A_i^*,\bm W_i,R_i=1\right]\Big| R_i=1\right\} \\
    & \quad \text{(by transportability, we have that)}\\
    = & E\left\{\left(\begin{matrix}1\\
    \mu_{A_i}(\bm\gamma_0) \\ 
    \bm W_i\end{matrix}\right)E\left[\left.(M_i\!-\!\alpha_{0,0}\!-\!\alpha_{1,0} \mu_{A_i}(\bm\gamma_0) \!-\! \bm\alpha_{2,0}^T \bm W_i)\right|A_i^*,\bm W_i\right]\Big| R_i=1 \right\} \\
    & \quad \text{(by equation \eqref{eq:mean}, we have that)}\\
    = & E\left\{\left(\begin{matrix}1\\
    \mu_{A_i}(\bm\gamma_0) \\ 
    \bm W_i\end{matrix}\right)E\left[\left.M_i - E[M_i|A_i^*,\bm W_i]\right|A_i^*,\bm W_i\right]\Big| R_i=1  \right\} \\
    = & E\left\{\left(\begin{matrix}1\\
    \mu_{A_i}(\bm\gamma_0) \\ 
    \bm W_i\end{matrix}\right)\times 0\Big| R_i=1 \right\} = \bm 0,
\end{align*}
and
\begin{align*}
    & E[\eU_{\sigma_M^2,i}(\sigma_{M0}^2,\widetilde{\bm\alpha}_0,\bm\gamma_0)|R_i=1] \\
    = & E\left\{E_{M_i}\left[\sigma_{M0}^2+\alpha_{1,0}^2\sigma_{A0}^2-(M_i\!-\!\alpha_{0,0}\!-\!\alpha_{1,0} \mu_{A}({\bm\gamma_0}) \!-\! {\bm\alpha}_{2,0}^T \bm W_i)^2\Big|A_i^*,\bm W_i,R_i=1\right]\Big|R_i=1\right\} \\
    & \quad \text{(by transportability and equation \eqref{eq:var}, we have that)}\\
    = & E\left\{\sigma_{M0}^2+\alpha_{1,0}^2\sigma_{A0}^2-E\left[(M_i\!-E[M_i|A_i^*,\bm W_i])^2\Big|A_i^*,\bm W_i\right]\Big|R_i=1\right\} \\
    = & E\left\{\sigma_{M0}^2+\alpha_{1,0}^2\sigma_{A0}^2-\text{Var}\left(M_i\big|A_i^*,\bm W_i\right)\Big|R_i=1\right\} \\
    = & E[0|R_i=1] = 0.
\end{align*}
This completes the proof.
\end{proof}

Now, we show consistency of $\widehat{\bm\alpha}$. Notice that $\widehat{\bm\alpha}$ is the solution of $\eU_{\bm\alpha}(\bm\alpha,\widehat{\bm\gamma})=\bm 0$. From  (D2) and (D3) we see that
$
\sup_{\bm\alpha\in \mathcal{\bm A}} \left\|\frac{1}{n}\eU_{\bm\alpha}(\bm\alpha,\widehat{\bm\gamma})-\frac{1}{n}\eU_{\bm\alpha}(\bm\alpha,\bm\gamma_0)\right\| \overset{p}{\to} \bm 0.
$
Observing that $E[\eU_{\bm\alpha,i}(\bm\alpha_0,\bm\gamma_0)]=\bm 0$ (Lemma \ref{lemma:unbiased}) and $E[\frac{\partial}{\partial \bm\alpha}\eU_{\bm\alpha,i}(\bm\alpha_0,\bm\gamma_0)]$ is negative definite (condition (D3)), one can conclude that $\widehat{\bm\alpha}$ is consistent such that $\left\|\widehat{\bm\alpha}-\bm\alpha_0\right\|\overset{p}{\to}\bm 0$ based on standard proofs for M-estimators (for example, Theorem 5.9 in \citealp{van2000asymptotic}).

Next, we sketch the proof for the approximate consistency of $\widehat{\bm\beta}^{(O1)}$, using  similar arguments as in \cite{xie2001risk} and \cite{struthers1986misspecified}. Specifically, we let the left side of (16) with $m_{A_i}=m_{A_i}^{(O1)}(\widehat{\bm\eta})$ be $\eU_{\bm\beta}^{(O1)}(\bm\beta,\widehat{\bm\eta})$ and notice that $\eU_{\bm\beta}(\bm\beta,\widehat{\bm\eta}) = \frac{\partial}{\partial \bm\beta} l^{(O1)}(\bm\beta)$, where $l^{(O1)}(\bm\beta)$ is the logarithm of the partial likelihood for the outcome model given as:
\begin{align*}
l^{(O1)}(\bm\beta) & = \sum_{i=1}^n R_i\int_{0}^{t^*} \Big[\bm\beta^T\bm Z(\widehat{\bm\eta}) - \log\left\{S^{(0)}(\bm\beta,\widehat{\bm\eta},t)\right\}\Big] dN_i(t).
\end{align*}
Observing that $n_1^{-1}l^{(O1)}(\bm\beta) = \frac{1}{n_1}\sum_{i=1}^{n_1} \int_{0}^{t^*} \Big[\bm\beta^T\bm Z(\widehat{\bm\eta}) - \log\left\{S^{(0)}(\bm\beta,\widehat{\bm\eta},t)\right\}\Big] dN_i(t)$ and $n_1\to\infty$ by regularity condition (D1a), one can show
\begin{equation}\label{eq:l_O1}
n_1^{-1}l^{(O1)}(\bm\beta) \overset{p}{\to}  H^{(O1)}(\bm\beta) \quad \text{for any }\bm\beta\in\bm{\mathcal B},
\end{equation}
with
$$
H^{(O1)}(\bm\beta) = \int_{0}^{t^*} \Big[\bm\beta^Ts^{(1)}(t)-\log\left\{s^{(0)}(\bm\beta,\bm\eta_0,t)\right\}s^{(0)}(t)\Big]dt.
$$
Here, 
$$
s^{(k)}(t)=\lambda_0(t)E\left[Y(t) \Big(\bm Z(\bm\eta_0)\Big)^{\otimes k} E\left\{\exp\left(\bm\beta_0[A,M,AM,\bm W^T]^T\right)|A^*,M,\bm W, T\geq t\right\}\right],
$$ 
and $\bm\beta_0=[\beta_{01},\beta_{02},\beta_{03},\bm\beta_{04}^T]^T$ is the true value of $\bm\beta$. The first derivative of $H^{(O1)}(\bm\beta)$ with respective to $\bm\beta$ is
$$
h_1^{(O1)}(\bm\beta) = \int_{0}^{t^*} \left\{s^{(1)}(t) - \frac{s^{(1)}(\bm\beta,\bm\eta_0,t)}{s^{(0)}(\bm\beta,\bm\eta_0,t)}\right\}s^{(0)}(t) dt,
$$
and the second derivative regarding $\bm\beta$ is
\begin{equation}\label{eq:h_2}
h_2^{(O1)}(\bm\beta) = -\int_{0}^{t^*} \left\{\frac{s^{(2)}(\bm\beta,\bm\eta_0,t)}{s^{(0)}(\bm\beta,\bm\eta_0,t)} - \left[\frac{s^{(1)}(\bm\beta,\bm\eta_0,t)}{s^{(0)}(\bm\beta,\bm\eta_0,t)}\right]^{\otimes 2}\right\}s^{(0)}(t) dt.
\end{equation}
Define $\bm\beta_0^{(O1)}$ as the solution to $h_1^{(O1)}(\bm\beta)=0$ such that $h_1(\bm\beta_0^{(O1)})=0$. Also, we assume that $h_2^{(O1)}(\bm\beta_0^{(O1)})$ is negative definite, which ensures that the log partial likelihood $l^{(O1)}(\bm\beta)$ is a concave function of $\bm\beta$ in a small neighborhood around $\bm\beta_0^{(O1)}$. Because $\widehat{\bm\beta}^{(O1)}$ maximizes the concave function $l^{(O1)}(\bm\beta)$, one can conclude that $\widehat{\bm\beta}^{(O1)}$ converges in probability to $\bm\beta_0^{(O1)}$ by using the convex analysis arguments in Section 3 of \cite{andersen1982cox}.

\begin{remark}
(Semi-parametric transportability assumption) In the literature, there exist two forms of transportability assumption, the `full-distributional'  transportability assumption (e.g., \cite{yi2015functional,yi2018parametric}) and the `semi-parametric' transportability assumption (e.g., \cite{liao2011survival,liao2018survival}),  where the former assumes that the entire distribution of the measurement error process must be transportable between the main and validation studies (when parametric measurement error models are used, as in this manuscript, this assumption requires that both the form of the distribution function as well as its paramseter values are identical between the main and validation studies) and the latter assumes the regression parameters in a pre-specified parametric (or semi-parametric) measurement error model are transportable. It should be noted that the full-distributional transportability assumption is stronger than the semi-parametric transportability assumption, because if the full-distributional transportability assumption is assumed, the parameters in the measurement error model are automatically transportable. In this paper, we follow \cite{yi2015functional,yi2018parametric} and make the full-distributional transportablity assumption such that $P(T_i,C_i,M_i,A_i|A_i^*,\bm W_i)$ is assumed to be transportable, which directly implies that measurement error models (4), (10), (14) are transportable between the main and validation studies. Without compromising the validity of our proposed approaches, we  could replace the full-distributional transportablity assumption used in this paper by the semi-parametric transportablity assumption, as clarified below:
\begin{itemize}
\item[(i)] For ORC1 approach, assume regression parameters in measurement error models (4) and (10) are transportable.
\item[(ii)] For ORC2 approach, assume regression parameters in measurement error models (4) and (10) are transportable. Also, additionally assume that $\epsilon_A$ in measurement error model (4) is normally distributed and this normal distribution is transportable.
\item[(iii)] For RRC approach, assume regression parameters in measurement error models (4) and (14) are transportable.
\end{itemize}

\end{remark}

\begin{remark}
(Connection between the full-distributional transportability assumption and ignorability assumption on the validation study selection mechanism) In the proof of consistency of $\widehat{\bm\alpha}$ and  $\widehat{\bm\beta}$, we leverage the transportability assumption to ensure that  measurement error model parameters in the validation study can be transported to the main study. As an alternative to the transportability assumption, some literature invokes the ignorability assumption (see, for example, \cite{spiegelman2001efficient}) to develop the proof, by assuming the selection mechanism into the validation study is conditionally ignorable given the baseline covariates and mismeasured exposure. Interestingly, the full-distributional transportability assumption used in this paper (see Remark 1) is mathematically equivalent to the ignorability assumption. Specifically, the full-distributional transportability assumption requires $P(T_i,C_i,M_i,A_i|A_i^*,\bm W_i,R_i)=P(T_i,C_i,M_i,A_i|A_i^*,\bm W_i)$, where $R_i$ is the indicator for being in the main study. The ignorability assumption, otherwise, requires that $P(R_i|T_i,C_i,M_i,A_i,A_i^*,\bm W_i)=P(R_i|A_i^*,\bm W_i)$ such that $\{T_i,C_i,M_i,A_i\}$ does not affect the validation study selection mechanism once $\{A_i^*,\bm W_i\}$ is given. We show that these two assumptions are equivalent by the Bayes' rule:
\begin{align*}
& \text{full-distributional transportability assumption} \\ 
\Longleftrightarrow & P(T_i,C_i,M_i,A_i|A_i^*,\bm W_i,R_i)=P(T_i,C_i,M_i,A_i|A_i^*,\bm W_i) \\
\Longleftrightarrow & \frac{P(T_i,C_i,M_i,A_i|A_i^*,\bm W_i)P(R_i|T_i,C_i,M_i,A_i,A_i^*,\bm W_i)}{P(R_i|A_i^*,\bm W_i)}=P(T_i,C_i,M_i,A_i|A_i^*,\bm W_i) \\
\Longleftrightarrow & P(R_i|T_i,C_i,M_i,A_i,A_i^*,\bm W_i)=P(R_i|A_i^*,\bm W_i)\\
\Longleftrightarrow & \text{ignorability assumption of the validation study selection mechanism}.
\end{align*}
\end{remark}

\subsection*{Appendix D.2: Asymptotic normality of $\widehat{\bm\theta}^{(O1)}$}

Next, we establish the asymptotic normality of $\sqrt{n}(\widehat{\bm\theta}^{(O1)}-\bm\theta_0^{(O1)})$, where $\bm\theta_0^{(O1)}=\left[\begin{matrix}\bm\alpha_0\\ \bm\beta_0^{(O1)}\end{matrix}\right]$ is the probability limit of $\widehat{\bm\theta}^{(O1)}$. Define $\eU_{\bm\theta}^{(O1)}(\bm\theta,\bm\gamma,\bm\eta)=\left[\begin{matrix}\eU_{\bm\alpha}(\bm\alpha,\bm\gamma)\\
\eU_{\bm\beta}^{(O1)}(\bm\beta,\bm\eta)\end{matrix}\right]$, which is the joint estimating equation for $\bm\theta$ by the ORC1 approach. A first-order Taylor expansion of $\eU_{\bm\theta}^{(O1)}(\widehat{\bm\theta}^{(O1)},\widehat{\bm\gamma},\widehat{\bm\eta})$ around $(\bm\theta_0^{(O1)},\bm\gamma_0,\bm\eta_0)$ results in\begingroup\makeatletter\def\f@size{10}\check@mathfonts
\begin{align*}
    & \sqrt{n}(\widehat{\bm\theta}^{(O1)}\!-\!\bm\theta_0^{(O1)}) \\
   = & \left[-\frac{1}{n}\frac{\partial \eU_{\bm\theta}^{(O1)}(\bm\theta_0^{(O1)},\bm\gamma_0,\bm\eta_0)}{\partial \bm\theta}\right]^{-1} \times \\
    & \quad  \frac{1}{\sqrt n} \Bigg[\eU_{\bm\theta}^{(O1)}(\bm\theta_0^{(O1)},\bm\gamma_0,\bm\eta_0)+\left.\frac{1}{n}\frac{\partial \eU_{\bm\theta}^{(O1)}(\bm\theta_0^{(O1)},\bm\gamma_0,\bm\eta_0)}{\partial \bm\gamma}\sqrt{n}(\widehat{\bm\gamma}-\bm\gamma_0)+\frac{1}{n}\frac{\partial \eU_{\bm\theta}^{(O1)}(\bm\theta_0^{(O1)},\bm\gamma_0,\bm\eta_0)}{\partial \bm\eta}\sqrt{n}(\widehat{\bm\eta}-\bm\eta_0)\right]+o_p(1) 
\end{align*}\endgroup
Define $\bm I_{\bm\theta}^{(O1)}$ and $\bm S_{\bm\theta}^{(O1)}$ as the component in the first and second brackets on the right-side hand of previous formula such that $\sqrt{n}(\widehat{\bm\theta}^{(O1)}\!-\!\bm\theta_0^{(O1)})=\left[\bm I_{\bm\theta}^{(O1)}\right]^{-1} \cdot \frac{1}{\sqrt n}\bm S_{\bm\theta}^{(O1)} + o_p(1)$. We can further write $\bm I_{\bm\theta}^{(O1)}$ as \begingroup\makeatletter\def\f@size{10}\check@mathfonts
$$
\bm I_{\bm\theta}^{(O1)} = -\frac{1}{n}\frac{\partial \eU_{\bm\theta}^{(O1)}(\bm\theta^{(O1)},\bm\gamma_0,\bm\eta_0)}{\partial \bm\theta} = -\frac{1}{n}\frac{\partial}{\partial \bm\theta}\left[\begin{matrix}\eU_{\bm\alpha}(\bm\alpha_0,\bm\gamma_0)\\
\eU_{\bm\beta}^{(O1)}(\bm\beta_0,\bm\eta_0)\end{matrix}\right] = \left[\begin{matrix}-\frac{1}{n}\frac{\partial}{\partial \bm\alpha}\eU_{\bm\alpha}(\bm\alpha_0,\bm\gamma_0) & \bm 0\\
\bm 0 & -\frac{1}{n}\frac{\partial}{\partial \bm\beta}\eU_{\bm\beta}^{(O1)}(\bm\beta_0^{(O1)},\bm\eta_0) \end{matrix}\right], 
$$\endgroup
where $\frac{1}{n}\frac{\partial}{\partial \bm\alpha}\eU_{\bm\alpha}(\bm\alpha_0,\bm\gamma_0) \overset{p}{\to} \bm{\mathcal I}_{\bm\alpha}$ by condition (D2) and
\begin{align*}
\frac{1}{n}\frac{\partial}{\partial \bm\beta}\eU_{\bm\beta}^{(O1)}(\bm\beta_0^{(O1)},\bm\eta_0) & =  \frac{1}{n}\sum_{i=1}^{n} R_i
 \int_{0}^{t^*}\left\{\frac{S^{(2)}(\bm\beta_0,\bm\eta_0,t)}{S^{(0)}(\bm\beta_0,\bm\eta_0,t)} - \left[\frac{S^{(1)}(\bm\beta_0,\bm\eta_0,t)}{S^{(0)}(\bm\beta_0,\bm\eta_0,t)}\right]^{\otimes 2}\right\}dN_i(t) \\
&=  \frac{n_1}{n}\times \frac{1}{n_1}\sum_{i=1}^{n_1}
 \int_{0}^{t^*}\left\{\frac{S^{(2)}(\bm\beta_0,\bm\eta_0,t)}{S^{(0)}(\bm\beta_0,\bm\eta_0,t)} - \left[\frac{S^{(1)}(\bm\beta_0,\bm\eta_0,t)}{S^{(0)}(\bm\beta_0,\bm\eta_0,t)}\right]^{\otimes 2}\right\}dN_i(t) \\
 & \overset{p}{\to} \omega h_2^{(O1)}(\bm\beta_0^{(O1)}),
\end{align*}
where the second to the third row holds by regularity conditions (D1a)--(D1b) and $h_2^{(O1)}(\bm\beta_0^{(O1)})$ is defined in \eqref{eq:h_2}. This suggests 
\begin{equation}\label{eq:I_O1}
\bm I_{\bm\theta}^{(O1)} \overset{p}{\to} \bm{\mathcal I}_{\bm\theta}^{(O1)} = - \left[\begin{matrix}\bm{\mathcal I}_{\bm\alpha} & \bm 0 \\ \bm 0 & \omega h_2^{(O1)}(\bm\beta_0^{(O1)}) \end{matrix}\right]
\end{equation}

In what follows, we explore the asymptotic behavior $\frac{1}{\sqrt n}\bm S_{\bm\theta}^{(O1)}$. By using a first-order Taylor expansion to $\eU_{\bm\gamma}(\widehat{\bm\gamma})$ around $\bm\gamma_0$, one can show
$$
\sqrt{n}(\widehat{\bm\gamma}-\bm\gamma_0) = \left[-\frac{1}{n}\frac{\partial \eU_{\bm\gamma}(\bm\gamma_0)}{\partial \bm\gamma}\right]^{-1} \frac{1}{\sqrt{n}}  \sum_{i=1}^n \eU_{\bm\gamma,i}(\bm\gamma_0) + o_p(1), 
$$
where $\frac{1}{n}\frac{\partial \eU_{\bm\gamma}(\bm\gamma_0)}{\partial \bm\gamma}$ converges in probability to $\bm {\mathcal I}_{\bm\gamma}=E\left[\frac{\partial \eU_{\bm\gamma,i}(\bm\gamma_0)}{\partial \bm\gamma}\right]$. Similarly, using a Taylor expansion to $\eU_{\bm\eta}(\widehat{\bm\eta})$ around $\bm\eta_0$, we have 
$$
\sqrt{n}(\widehat{\bm\eta}-\bm\eta_0) = \left[-\frac{1}{n}\frac{\partial \eU_{\bm\eta}(\bm\eta_0)}{\partial \bm\eta}\right]^{-1} \frac{1}{\sqrt{n}}\sum_{i=1}^n  \eU_{\bm\eta,i}(\bm\eta_0) + o_p(1),
$$
where $\frac{1}{n}\frac{\partial \eU_{\bm\eta}(\bm\eta_0)}{\partial \bm\eta}$ converges in probability to $\bm{\mathcal I}_{\bm\eta}=E\left[\frac{\partial \eU_{\bm\eta,i}(\bm\eta_0)}{\partial \bm\eta}\right]$.
Using a similar proof to the one for Theorem 2.1 in \cite{lin1989robust}, one can show that
$$
\frac{1}{\sqrt n} \eU_{\bm\beta}^{(O1)}(\bm\beta_0^{(O1)},\bm\eta_0) = \frac{1}{\sqrt n} \sum_{i=1}^n \bm G_i(\bm\beta_0^{(O1)},\bm\eta_0) +o_p(1),
$$
where
\begin{align*}
   \bm G_i^{(O1)}(\bm\beta_0^{(O1)},\bm\eta_0) = & R_i\int_0^{t^*} \left\{\bm Z_i(\bm\eta_0) - \frac{s^{(1)}(\bm\beta_0^{(O1)},\bm\eta_0,t)}{s^{(0)}(\bm\beta_0^{(O1)},\bm\eta_0,t)}\right\}dN_i(t)  \\
   & \quad \quad - R_i \int_0^{t^*} \frac{Y_i(t)\exp\left(\bm\beta_0^{(O1)^T}\bm Z_i(\bm\eta_0)\right)}{s^{(0)}(\bm\beta_0^{(O1)},\bm\eta_0,t)}\left\{\bm Z_i(\bm\eta_0)- \frac{s^{(1)}(\bm\beta_0^{(O1)},\bm\eta_0,t)}{s^{(0)}(\bm\beta_0^{(O1)},\bm\eta_0,t)}\right\} d\widetilde{F}(t) ,
\end{align*}
and $\widetilde{F}(t) = E[\sum_{i=1}^{n_1} N_i(t)/n_1]$. The above discussions implies that
\begin{align}
    \frac{1}{\sqrt n}\bm S_{\bm\theta}^{(O1)} =\frac{1}{\sqrt n} \sum_{i=1}^n \left\{\left(\begin{matrix} \mathbb{U}_{\bm\alpha,i}(\bm\alpha_0,\bm\gamma_0) \\ \bm G_i^{(O1)}(\bm\beta_0^{(O1)},\bm\eta_0)\end{matrix}\right) -  \bm{\mathcal I}_{\bm\theta,\bm\gamma}^{(O1)} \bm{\mathcal I}_{\bm\gamma}^{-1}\eU_{\bm\gamma,i}(\bm\gamma_0) - \bm{\mathcal I}_{\bm\theta,\bm\eta}^{(O1)} \bm{\mathcal I}_{\bm\eta}^{-1}\eU_{\bm\eta,i}(\bm\eta_0)\right\} + o_p(1),\label{eq:S_O1}
\end{align}
where $\bm{\mathcal I}_{\bm\theta,\bm\gamma}^{(O1)} = E\left[\frac{\partial}{\partial \bm\gamma}\eU_{\bm\gamma,i}^{(O1)}\left(\bm\theta^{(O1)},\bm\gamma,\bm\eta\right)\right]$ and $\bm{\mathcal I}_{\bm\theta,\bm\eta}^{(O1)} = E\left[\frac{\partial}{\partial \bm\eta}\eU_{\bm\gamma,i}^{(O1)}\left(\bm\theta^{(O1)},\bm\gamma,\bm\eta\right)\right]$. 

Therefore, it follows from the multivariate central limit theorem that
$$
\sqrt{n}\left(\widehat{\bm\theta}^{(O1)}-\bm\theta^{(O1)}\right) \overset{D}{\rightarrow} N\left(\bm 0, \bm V^{(O1)}\right),
$$
where $\bm V^{(O1)} = \left[\bm{\mathcal I}_{\bm\theta}^{(O1)}\right]^{-1} \bm{\mathcal M}_{\bm\theta}^{(O1)} \left[\bm{\mathcal I}_{\bm\theta}^{(O1)}\right]^{-T}$ and
$$
\bm{\mathcal M}_{\bm\theta}^{(O1)} = E\left\{\left(\begin{matrix} \mathbb{U}_{\bm\alpha,i}(\bm\alpha_0,\bm\gamma_0) \\ \bm G_i^{(O1)}(\bm\beta_0^{(O1)},\bm\eta_0)\end{matrix}\right) -   \bm{\mathcal I}_{\bm\theta,\bm\gamma}^{(O1)} \bm{\mathcal I}_{\bm\gamma}^{-1}\eU_{\bm\gamma,i}(\bm\gamma_0) - \bm{\mathcal I}_{\bm\theta,\bm\eta}^{(O1)} \bm{\mathcal I}_{\bm\eta}^{-1}\eU_{\bm\eta,i}(\bm\eta_0)\right\}^{\otimes 2}.
$$
A consistent, plug-in estimator for $\bm V^{(O1)}$ is
$$
\widehat{\bm V}^{(O1)} = \left[\widehat{\bm I}_{\bm\theta}^{(O1)}\right]^{-1} \widehat{\bm M}_{\bm\theta}^{(O1)} \left[\widehat{\bm I}_{\bm\theta}^{(O1)}\right]^{-T},
$$
where $\widehat{\bm I}_{\bm\theta}^{(O1)}=-\frac{1}{n}\frac{\partial}{\partial \bm\theta}\eU_{\bm\theta}^{(O1)}(\widehat{\bm\theta}^{(O1)},\widehat{\bm\gamma},\widehat{\bm\eta})$,
$$
\widehat{\bm M}_{\bm\theta}^{(O1)} = \frac{1}{n}\sum_{i=1}^n \left\{\eU_{\bm\theta,i}^{(O1)}(\widehat{\bm\theta}^{(O1)},\widehat{\bm\gamma},\widehat{\bm\eta}) -\widehat{\bm I}_{\bm\theta,\bm\gamma}^{(O1)} \widehat{\bm I}_{\bm\gamma}^{-1}\eU_{\bm\gamma,i}(\widehat{\bm\gamma}) - \widehat{\bm I}_{\bm\theta,\bm\eta}^{(O1)} \widehat{\bm I}_{\bm\eta}^{-1}\eU_{\bm\eta,i}(\widehat{\bm\eta}) \right\}^{\otimes 2},
$$
$\widehat{\bm I}_{\bm\theta,\bm\gamma}^{(O1)}=\frac{1}{n}\frac{\partial}{\partial \bm\gamma}\eU_{\bm\theta}^{(O1)}(\widehat{\bm\theta}^{(O1)},\widehat{\bm\gamma},\widehat{\bm\eta})$, $\widehat{\bm I}_{\bm\theta,\bm\eta}^{(O1)}=\frac{1}{n}\frac{\partial}{\partial \bm\eta}\eU_{\bm\theta}^{(O1)}(\widehat{\bm\theta}^{(O1)},\widehat{\bm\gamma},\widehat{\bm\eta})$, $\widehat{\bm I}_{\bm\gamma}=-\frac{1}{n}\frac{\partial}{\partial \bm\gamma}\eU_{\bm\gamma}(\widehat{\bm\gamma})$, and $\widehat{\bm I}_{\bm\eta}=-\frac{1}{n}\frac{\partial}{\partial \bm\eta}\eU_{\bm\eta}(\widehat{\bm\eta})$.

\subsection*{Appendix D.3: Asymptotic properties for the ORC2 estimator}

We use similar arguments as in Appendices D.1 and D.2 to prove the approximate consistency and asymptotic normality of the ORC2 estimator, $\widehat{\bm\theta}^{(O2)} = \left[\begin{matrix}\widehat{\bm\alpha}\\ \widehat{\bm\beta}^{(O2)}\end{matrix}\right]$. Here, we only briefly outline a sketch of the proof. We require the following regularity conditions:
\begin{itemize}
    \item[E1a.] {Assume both the main study and validation study sizes go to infinity such that $n_1\to \infty$ and $n-n_1 \to \infty$.}
    \item[E1b.] {To prove asymptotic normality, additionally assume that the ratio of main study size and total sample size, $n_1/n$, converges to a constant $\omega \in (0,1)$.}
    \item[E2.] Assume that the supports of the unknown parameters $\bm\gamma$, $\bm \alpha$, and $\bm\beta$, denoted by $\bm{\Gamma}$, $\bm{\mathcal A}$, and $\bm{\mathcal B}$, are compact subsets of the Euclidean space. Suppose that $\|\widehat{\bm\gamma}-\bm\gamma_0\| \overset{p}{\to} \bm 0$ such that the point estimator of the parameters in the measurement error model (4) is consistent, where $\bm\gamma_0$ is the true value of $\bm\gamma$.
    \item[E3.] Stacking equations (6) and (7) and defining the joint estimating equations as $\eU_{\bm\alpha}(\bm\alpha,\widehat{\bm\gamma})=\bm 0$. We further define $\eU_{\bm\alpha}(\bm\alpha,\bm\gamma)$ as $\eU_{\bm\alpha}(\bm\alpha,\widehat{\bm\gamma})$ with $\widehat{\bm\gamma}$ in replacement by $\bm\gamma$, and define $\eU_{\bm\alpha,i}(\bm\alpha,\bm\gamma)$ as its corresponding estimating function (i.e., the $i$-th component of $\eU_{\bm\alpha}(\bm\alpha,{\bm\gamma})$). Let $\bm\alpha_0$ be the true value of $\bm\alpha$. Suppose that $\eU_{\bm\alpha,i}(\bm\alpha,\bm\gamma)$ is continuous, differentiable, and bounded for $(\bm\alpha,\bm\gamma) \in \bm{\mathcal A}\times\bm{\Gamma}$, which is sufficient to show $\frac{1}{n}\frac{\partial}{\partial \bm\alpha}\eU_{\bm\alpha}(\bm\alpha_0,\bm\gamma_0)$ converges weakly to $\bm{\mathcal{I}}_{\bm\alpha}=E[\frac{\partial}{\partial \bm\alpha}\eU_{\bm\alpha,i}(\bm\alpha_0,\bm\gamma_0)]$ by law of large numbers. We further assume $\bm{\mathcal{I}}_{\bm\alpha}$ is negative definite, which ensures that ${\bm\alpha}_0$ is unique solution of $\eU_{\bm\alpha}(\bm\alpha,{\bm\gamma}_0)=\bm 0$.
    \item[E4.] Define 
    $
    s_{*}^{(k)}(\bm\beta,\bm\alpha,\bm\gamma,t) =  E\left[Y(t) \Big(\bm Z(\bm\alpha,\bm\gamma)\Big)^{\otimes k} \exp\Big(\bm\beta^T\bm Z(\bm\alpha,\bm\gamma)\Big)\right],\text{ for } k=0,1,2,
    $
    where $\bm Z(\bm\alpha,\bm\gamma) = \left [m_A^{(O2)}(\bm\alpha,\bm\gamma),M,m_A^{(O2)}(\bm\alpha,\bm\gamma)M,\bm W^T\right]^T$. Assume that $s_{*}^{(k)}(\bm\beta,\bm\alpha,\bm\gamma,t)$, $k\in\{0,1,2\}$, is a continuous functions of $(\bm\beta,\bm\alpha,\bm\gamma) \in \mathcal{\bm B}\times\mathcal{\bm A} \times \bm \Gamma$ and uniformly continuous in $t\in [0,t^*]$.  Also, assume that $s_{*}^{(k)}(\bm\beta,\bm\alpha,\bm\gamma,t)$, $k\in\{0,1,2\}$, is bounded on $(\bm\beta,\bm\alpha,\bm\gamma,t) \in \mathcal{\bm B}\times\mathcal{\bm A}\times \bm \Gamma \times [0,t^*]$. Moreover, suppose that $s_{*}^{(0)}(\bm\beta,\bm\alpha,\bm\gamma,t)$ is bounded away from 0. 
    \item[E5.] Define $
    S_{*}^{(k)}(\bm\beta,\bm\alpha,\bm\gamma,t) = \frac{1}{n_1}\sum_{i=1}^{n_1} Y_i(t) \Big(\bm Z_i(\bm\alpha,\bm\gamma)\Big)^{\otimes k} \exp\Big(\bm\beta^T\bm Z_i(\bm\alpha,\bm\gamma)\Big), \text{ for } k=0,1,2.
    $
    Assume $\sup_{(\bm\beta,\bm\alpha,\bm\gamma,t)\in \mathcal{\bm B}\times\mathcal{\bm A}\times\bm\Gamma \times [0,t^*]} \left\|S_{*}^{(k)}(\bm\beta,\bm\alpha,\bm\gamma,t)- s_{*}^{(k)}(\bm\beta,\bm\gamma,t)\right\| \overset{p}{\to}\bm 0$, for $k=0,1,2$.
\end{itemize}
Under these conditions, the proof of consistency of $\widehat{\bm\alpha}$ follows Appendix D.1. Below we show approximate consistency of $\widehat{\bm\beta}^{(O2)}$. Let the left-hand side of (16) with $m_{A_i}=m_{A_i}^{(O2)}(\widehat{\bm\alpha},\widehat{\bm\gamma})$ be $\eU_{\bm\beta}^{(O2)}(\bm\beta,\widehat{\bm\alpha},\widehat{\bm\gamma})$, which can be seen as score equation corresponding to the following log partial likelihood:
$$
l^{(O2)}(\bm\beta) = \sum_{i=1}^n R_i\int_{0}^{t^*} \Big[\bm\beta^T\bm Z(\widehat{\bm\alpha},\widehat{\bm\gamma}) - \log\left\{S^{(0)}(\bm\beta,\widehat{\bm\alpha},\widehat{\bm\gamma},t)\right\}\Big] dN_i(t).
$$
Similar to the proof of \eqref{eq:l_O1} in Appendix D.1, one can show $n_1^{-1}l^{(O2)}(\bm\beta) \overset{p}{\to}   H^{(O2)}(\bm\beta)$ for any $\bm\beta \in \bm{\mathcal B}$, where $H^{(O2)}(\bm\beta)=\int_{0}^{t^*} \Big[\bm\beta^Ts_*^{(1)}(t)-\log\left\{s^{(0)}(\bm\beta,\bm\alpha_0,\bm\gamma_0,t)\right\}s_*^{(0)}(t)\Big]dt$ and 
\begin{equation*}
s_{*}^{(k)}(t)=\lambda_0(t)E\left[Y(t) \Big(\bm Z(\bm\alpha_0,\bm\gamma_0)\Big)^{\otimes k} E\left\{\exp\left(\bm\beta_0[A,M,AM,\bm W^T]^T\right)|A^*,M,\bm W, T\geq t\right\}\right]
\end{equation*}
for $k\in\{0,1,2\}$. The first and second derivative of $H^{(O2)}(\bm\beta)$ with respective to $\bm\beta$ are
$
h_1^{(O2)}(\bm\beta) = \int_{0}^{t^*} \left\{s_*^{(1)}(t) - \frac{s^{(1)}(\bm\beta,\bm\alpha_0,\bm\gamma_0,t)}{s^{(0)}(\bm\beta,\bm\alpha_0,\bm\gamma_0,t)}\right\}s_*^{(0)}(t) dt,
$
and 
\begin{equation}\label{eq:h2_O2}
h_2^{(O2)}(\bm\beta) = -\int_{0}^{t^*} \left\{\frac{s^{(2)}(\bm\beta,\bm\alpha_0,\bm\gamma_0,t)}{s^{(0)}(\bm\beta,\bm\alpha_0,\bm\gamma_0,t)} - \left[\frac{s^{(1)}(\bm\beta,\bm\alpha_0,\bm\gamma_0,t)}{s^{(0)}(\bm\beta,\bm\alpha_0,\bm\gamma_0,t)}\right]^{\otimes 2}\right\}s_*^{(0)}(t) dt,
\end{equation}
respectively. 
We assume that $h_2^{(O2)}(\bm\beta)$ is negative definite. Then one can use the convex analysis in \cite{andersen1982cox} to show $\widehat{\bm\beta}^{(O2)} \overset{p}{\to} \bm\beta_0^{(O2)}$, where $\bm\beta_0^{(O2)}$ is the solution of $h_1^{(O2)}(\bm\beta)=0$.


Next, we show that $\sqrt{n}(\widehat{\bm\theta}^{(O2)}-\bm\theta_0^{(O2)})$ converges to a zero-mean multivariate normal distribution. Let $\eU_{\bm\theta}^{(O2)}(\bm\theta,\bm\gamma)=\left[\begin{matrix}\eU_{\bm\alpha}(\bm\alpha,\bm\gamma)\\
\eU_{\bm\beta}^{(O2)}(\bm\beta,\bm\alpha,\bm\gamma)\end{matrix}\right]$ be the joint estimating equation for $\bm\theta$ by the ORC2 approach. A first-order Taylor expansion of $\eU_{\bm\theta}^{(O2)}(\widehat{\bm\theta}^{(O2)},\widehat{\bm\gamma})$ around $(\bm\theta_0^{(O2)},\bm\gamma_0)$ results in\begingroup\makeatletter\def\f@size{10}\check@mathfonts
\begin{align*}
    &\sqrt{n}(\widehat{\bm\theta}^{(O2)}\!-\!\bm\theta_0^{(O2)}) \!
    \\ 
    = & \Big[\underbrace{-\frac{1}{n}\frac{\partial \eU_{\bm\theta}^{(O2)}(\bm\theta^{(O2)},\bm\gamma_0)}{\partial \bm\theta}}_{\text{defined as }\bm I_{\bm\theta}^{(O2)}}\Big]^{-1} \cdot \frac{1}{\sqrt n} \Big[\underbrace{\eU_{\bm\theta}^{(O2)}(\bm\theta^{(O2)},\bm\gamma_0)+\frac{1}{n}\frac{\partial \eU_{\bm\theta}^{(O2)}(\bm\theta^{(O2)},\bm\gamma_0)}{\partial \bm\gamma}\sqrt{n}(\widehat{\bm\gamma}-\bm\gamma_0)}_{\text{defined as }\bm S_{\bm\theta}^{(O2)}}\Big]+o_p(1).
\end{align*}\endgroup
Next we explore the asymptotic behavior of $\bm I_{\bm\theta}^{(O2)}$ and $\bm S_{\bm\theta}^{(O2)}$. Applying similar strategy to the derivation of \eqref{eq:I_O1} in Appendix D.2, one can easily deduce that $\bm I_{\bm\theta}^{(O2)} \overset{p}{\to} -\bm{\mathcal I}_{\bm\theta}^{(O2)} = \left[\begin{matrix}\bm{\mathcal I}_{\bm\alpha} & \bm 0 \\ \bm 0 & \omega h_2^{(O2)}(\bm\beta_0^{(O2)}) \end{matrix}\right]$, where $h_2^{(O2)}(\bm\beta_0^{(O2)})$ is defined in \eqref{eq:h2_O2}. Also, similar to the derivation of \eqref{eq:S_O1}, one can show 
\begin{align*}
    \frac{1}{\sqrt n}\bm S_{\bm\theta}^{(O2)} =\frac{1}{\sqrt n}\sum_{i=1}^n \left\{\left(\begin{matrix}\mathbb{U}_{\bm\alpha,i}(\bm\alpha_0,\bm\gamma_0) \\ \bm G_i^{(O2)}\left(\bm\beta_0^{(O2)},\bm\alpha_0,\bm\gamma_0\right)\end{matrix}\right) -  \bm{\mathcal I}_{\bm\theta,\bm\gamma}^{(O2)} \bm{\mathcal I}_{\bm\gamma}^{-1}\eU_{\bm\gamma,i}(\bm\gamma_0)\right\} + o_p(1),
\end{align*}
where $\bm{\mathcal I}_{\bm\theta,\bm\gamma}^{(O2)} = E\left[\frac{\partial}{\partial \bm\gamma}\eU_{\bm\theta,i}^{(O2)}\left(\bm\theta^{(O2)},\bm\gamma\right)\right]$, \begingroup\makeatletter\def\f@size{10}\check@mathfonts
\begin{align*}
   \bm G_i^{(O2)}(\bm\beta_0^{(O2)},\bm\alpha_0,\bm\gamma_0) = & R_i\int_0^{t^*} \left\{\bm Z_i(\bm\alpha_0,\bm\gamma_0) - \frac{s_{*}^{(1)}(\bm\beta_0^{(O2)},\bm\alpha_0,\bm\gamma_0,t)}{s_{*}^{(0)}(\bm\beta_0^{(O2)},\bm\alpha_0,\bm\gamma_0,t)}\right\}dN_i(t)  \\
   & - R_i \int_0^{t^*} \frac{Y_i(t)\exp\left(\bm\beta^{(O2)^T}\bm Z_i(\bm\alpha_0,\bm\gamma_0)\right)}{s_{*}^{(0)}(\bm\beta_0^{(O2)},\bm\alpha_0,\bm\gamma_0,t)}\left\{\bm Z_i(\bm\alpha_0,\bm\gamma_0)- \frac{s_{*}^{(1)}(\bm\beta_0^{(O2)},\bm\alpha_0,\bm\gamma_0,t)}{s_{*}^{(0)}(\bm\beta_0^{(O2)},\bm\alpha_0,\bm\gamma_0,t)}\right\} d\widetilde{F}(t).
\end{align*}\endgroup
Then, using the central limit theorem, one can show
$
\sqrt{n}\left(\widehat{\bm\theta}^{(O2)}-\bm\theta_0^{(O2)}\right) \overset{D}{\rightarrow} N\left(\bm 0, \bm V^{(O2)}\right),
$
where $\bm V^{(O2)} = \left[\bm{\mathcal I}_{\bm\theta}^{(O2)}\right]^{-1} \bm{\mathcal M}_{\bm\theta}^{(O2)} \left[\bm{\mathcal I}_{\bm\theta}^{(O2)}\right]^{-T}$ and
$$
\bm{\mathcal M}_{\bm\theta}^{(O2)} = E\left\{\left(\begin{matrix}\mathbb{U}_{\bm\alpha,i}(\bm\alpha_0,\bm\gamma_0) \\ \bm G_i^{(O2)}\left(\bm\beta_0^{(O2)},\bm\alpha_0,\bm\gamma_0\right)\end{matrix}\right) -  \bm{\mathcal I}_{\bm\theta,\bm\gamma}^{(O2)} \bm{\mathcal I}_{\bm\gamma}^{-1}\eU_{\bm\gamma,i}(\bm\gamma_0)\right\}^{\otimes 2}.
$$

The variance matrix, $\bm V^{(O2)}$, can be consistently estimated by
$$
\widehat{\bm V}^{(O2)} = \left[\widehat{\bm I}_{\bm\theta}^{(O2)}\right]^{-1} \widehat{\bm M}_{\bm\theta}^{(O2)} \left[\widehat{\bm I}_{\bm\theta}^{(O2)}\right]^{-T},
$$
where $\widehat{\bm I}_{\bm\theta}^{(O2)}=-\frac{1}{n}\frac{\partial}{\partial \bm\theta}\eU_{\bm\theta}^{(O2)}(\widehat{\bm\theta}^{(O2)},\widehat{\bm\gamma})$,
$
\widehat{\bm M}_{\bm\theta}^{(O2)} = \frac{1}{n}\sum_{i=1}^n \left\{\eU_{\bm\theta,i}^{(O2)}(\widehat{\bm\theta}^{(O2)},\widehat{\bm\gamma}) - \widehat{\bm I}_{\bm\theta,\bm\gamma}^{(O2)} \widehat{\bm I}_{\bm\gamma}^{-1}\eU_{\bm\gamma,i}(\widehat{\bm\gamma})  \right\}^{\otimes 2},
$
$\widehat{\bm I}_{\bm\theta,\bm\gamma}^{(O2)}=\frac{1}{n}\frac{\partial}{\partial \bm\gamma}\eU_{\bm\theta}^{(O2)}(\widehat{\bm\theta}^{(O2)},\widehat{\bm\gamma})$, and $\widehat{\bm I}_{\bm\gamma}=\frac{1}{n}\frac{\partial}{\partial \bm\gamma}\eU_{\bm\gamma}(\widehat{\bm\gamma})$.

\section*{Appendix E: Asymptotic distribution theory for the RRC estimator}

This section proves that $\widehat{\bm\theta}^{(R)}$ is an approximately consistent and asymptotically normal estimator of $\bm\theta$. The proof is similar to those presented for the ORC1 and ORC2 estimators in Appendix D. We require the following regularity conditions:
\begin{itemize}
    \item[F1a.] {Assume both the main study and validation study sizes go to infinity such that $n_1\to \infty$ and $n-n_1 \to \infty$.}
    \item[F1b.] {To prove asymptotic normality, additionally assume that the ratio of main study size and total sample size, $n_1/n$, converges to a constant $\omega \in (0,1)$.}
    \item[F2.] Define $\bm\xi=[\bm\xi^T(t_1),\bm\xi^T(t_2),\dots,\bm\xi^T(t_K)]^T$ as all regression coefficients in model (14) across all risk sets. Then, $\widehat{\bm\xi}$ can be viewed as the solusion of the joint estimating equation $\eU_{\bm\xi}(\bm\xi) = \Big[\eU_{\bm\xi(t_1)}^T\left(\bm\xi(t_1)\right),\dots,\eU_{\bm\xi(t_K)}^T\left(\bm\xi(t_K)\right)\Big]^T=\bm 0$, where $\eU_{\bm\xi(t_k)}\left(\bm\xi(t_k)\right)$ is defined as the left-hand side of (15).
    \item[F3.] Assume estimators of the parameters in the measurement error models (4) and (14) are consistent so that $\|\widehat{\bm\gamma}-\bm\gamma_0\| \overset{p}{\to} \bm 0$ and $\|\widehat{\bm\xi}-\bm\xi_0\| \overset{p}{\to} \bm 0$, where $\bm\gamma_0$ and $\bm\xi_0=[\bm\xi_0^T(t_1),\bm\xi_0^T(t_2),\dots,\bm\xi_0^T(t_K)]^T$ are the true values of $\bm\gamma$ and $\bm\xi$, respectively. Assume that the supports of the unknown parameters $\bm\gamma$, $\bm \xi$, $\bm\alpha$, and $\bm\beta$, denoted by $\bm{\Gamma}$, $\bm{\Xi}$, $\bm{\mathcal A}$, and $\bm{\mathcal B}$, are compact subsets of the Euclidean space.
    \item[F4.] Stacking equations (6) and (7) and defining the joint estimating equations as $\eU_{\bm\alpha}(\bm\alpha,\widehat{\bm\gamma})=\bm 0$. We further define $\eU_{\bm\alpha}(\bm\alpha,\bm\gamma)$ as $\eU_{\bm\alpha}(\bm\alpha,\widehat{\bm\gamma})$ with $\widehat{\bm\gamma}$ in replacement by $\bm\gamma$, and define $\eU_{\bm\alpha,i}(\bm\alpha,\bm\gamma)$ as its corresponding estimating function (i.e., the $i$-th component of $\eU_{\bm\alpha}(\bm\alpha,{\bm\gamma})$). Let $\bm\alpha_0$ be the true value of $\bm\alpha$. Suppose that $\eU_{\bm\alpha,i}(\bm\alpha,\bm\gamma)$ is continuous, differentiable, and bounded for $(\bm\alpha,\bm\gamma) \in \bm{\mathcal A}\times\bm{\Gamma}$, which is sufficient to show $\frac{1}{n}\frac{\partial}{\partial \bm\alpha}\eU_{\bm\alpha}(\bm\alpha_0,\bm\gamma_0)$ converges weakly to $\bm{\mathcal{I}}_{\bm\alpha}=E[\frac{\partial}{\partial \bm\alpha}\eU_{\bm\alpha,i}(\bm\alpha_0,\bm\gamma_0)]$ by law of large numbers. We further assume $\bm{\mathcal{I}}_{\bm\alpha}$ is negative definite, which ensures that ${\bm\alpha}_0$ is unique solution of $\eU_{\bm\alpha}(\bm\alpha,{\bm\gamma}_0)=\bm 0$.
    \item[F5.] Define 
    $$
    s_{**}^{(k)}(\bm\beta,\bm\xi,t) =  E\left[Y(t) \Big(\bm Z(\bm\xi,t)\Big)^{\otimes k} \exp\Big(\bm\beta^T\bm Z(\bm\xi,t)\Big)\right],\text{ for } k=0,1,2,
    $$
    where $\bm Z(\bm\xi,t) = \left [m_A^{(R)}(\bm\xi,t),M,m_A^{(R)}(\bm\xi,t)\times M,\bm W^T\right]^T$. Assume that $s_{**}^{(k)}(\bm\beta,\bm\eta,t)$ is a continuous function of $(\bm\beta,\bm\xi) \in \mathcal{\bm B}\times\bm \Xi$ and uniformly continuous in $t\in [0,t^*]$, for all $m=0,1,2$. Also, assume that $s_{**}^{(k)}(\bm\beta,\bm\xi,t)$ is bounded on $(\bm\beta,\bm\xi,t) \in \mathcal{\bm B}\times\bm \Xi \times [0,t^*]$, for $k=0,1,2$. In addition, suppose that $s_{**}^{(0)}(\bm\beta,\bm\xi,t)$ is bounded away from 0.
    \item[F6.] Define
    $$
    S_{**}^{(k)}(\bm\beta,\bm\xi,t) = \frac{1}{n_1}\sum_{i=1}^{n_1} Y_i(t) \Big(\bm Z_i(\bm\xi,t)\Big)^{\otimes k} \exp\Big(\bm\beta^T\bm Z_i(\bm\xi,t)\Big), \text{ for } k=0,1,2.
    $$
    Assume that $\sup_{(\bm\beta,\bm\xi,t)\in \mathcal{\bm B}\times\bm \Xi \times [0,t^*]} \left\|S_{**}^{(k)}(\bm\beta,\bm\xi,t)- s_{**}^{(k)}(\bm\beta,\bm\xi,t)\right\| \overset{p}{\to}\bm 0$, for $k=0,1,2$.
\end{itemize}

Proof of consistency of $\widehat{\bm\alpha}$ follows Appendix D.1. Therefore, here we only provide the proof for the approximate consistency of $\widehat{\bm\beta}^{(R)}$. Rewrite the left-hand side of (16) with $m_{A_i}=m_{A_i}^{(R)}(\widehat{\bm\xi})$ as $\eU_{\bm\beta}^{(R)}(\bm\beta,\widehat{\bm\xi})$, which is the score equation corresponding to the following partial likelihood function
$$
l^{(R)}(\bm\beta) = \sum_{i=1}^n R_i\int_{0}^{t^*} \Big[\bm\beta^T\bm Z(\widehat{\bm\xi}) - \log\left\{S_{**}^{(0)}(\bm\beta,\widehat{\bm\xi},t)\right\}\Big] dN_i(t).
$$
Similar to our proof in \eqref{eq:l_O1}, one can show that $n_1^{-1}l^{(R)}(\bm\beta)\overset{p}{\to} H^{(R)}(\bm\beta)$ for $\bm\beta \in \mathcal{\bm B}$, where
$$
H^{(R)}(\bm\beta) = \int_{0}^{t^*} \Big[\bm\beta^Ts_{**}^{(1)}(t)-\log\left\{s_{**}^{(0)}(\bm\beta,\bm\eta_0,t)\right\}s_{**}^{(0)}(t)\Big]dt,
$$
and $
s_{**}^{(k)}(t)=\lambda_0(t)E\left[Y(t) \Big(\bm Z(\bm\xi_0,t)\Big)^{\otimes k} E\left\{\exp\left(\bm\beta_0[A,M,AM,\bm W^T]^T\right)|A^*,M,\bm W, T\geq t\right\}\right].
$
The first and second derivatives of $H^{(O1)}(\bm\beta)$ with respective to $\bm\beta$ are
\begin{align}
h_1^{(R)}(\bm\beta) &= \int_{0}^{t^*} \left\{s_{**}^{(1)}(t) - \frac{s_{**}^{(1)}(\bm\beta,\bm\xi_0,t)}{s_{**}^{(0)}(\bm\beta,\bm\xi_0,t)}\right\}s_{**}^{(0)}(t) dt,\nonumber \\
h_2^{(R)}(\bm\beta) & = -\int_{0}^{t^*} \left\{\frac{s_{**}^{(2)}(\bm\beta,\bm\xi_0,t)}{s_{**}^{(0)}(\bm\beta,\bm\xi_0,t)} - \left[\frac{s_{**}^{(1)}(\bm\beta,\bm\xi_0,t)}{s_{**}^{(0)}(\bm\beta,\bm\xi_0,t)}\right]^{\otimes 2}\right\}s_{**}^{(0)}(t) dt,\label{eq:h2_R}
\end{align}
respectively. Let $\bm\beta_0^{(R)}$ be the solution of $h_1^{(R)}(\bm\beta)=0$. Suppose that $h_2^{(R)}(\bm\beta)$ is negative definite, then we can use the convex analysis in \cite{andersen1982cox} to show $\widehat{\bm\beta}^{(R)}$ converges in probability to $\bm\beta_0^{(R)}$.

In what follows, we show that $\widehat{\bm\theta}^{(R)}$ is asymptotically normal so that $\sqrt{n}(\widehat{\bm\theta}^{(R)}-\bm\theta_0^{(R)})$ converges to a zero-mean multivariate normal distribution. Note that the joint estimating equation for $\bm\theta$ by RRC  is $\eU_{\bm\theta}^{(R)}(\bm\theta,\bm\gamma,\bm\xi)=\left[\begin{matrix}\eU_{\bm\alpha}(\bm\alpha,\bm\gamma)\\
\eU_{\bm\beta}^{(R)}(\bm\beta,\bm\xi)\end{matrix}\right]$. Then, applying a first-order Taylor expansion to $\eU_{\bm\theta}^{(R)}(\widehat{\bm\theta}^{(R)},\widehat{\bm\gamma},\widehat{\bm\xi})$ around $(\bm\theta_0^{(R)},\bm\gamma_0,\bm\xi_0)$, we have that
\begin{align*}
    \sqrt{n}(\widehat{\bm\theta}^{(R)}\!-\!\bm\theta_0^{(R)}) \!=& \left[-\frac{1}{n}\frac{\partial \eU_{\bm\theta}^{(R)}(\bm\theta^{(R)},\bm\gamma_0,\bm\xi_0)}{\partial \bm\theta}\right]^{-1} \cdot \frac{1}{\sqrt n} \Bigg[\eU_{\bm\theta}^{(R)}(\bm\theta^{(R)},\bm\gamma_0,\bm\xi_0)+\\
    & \quad \left.\frac{1}{n}\frac{\partial \eU_{\bm\theta}^{(R)}(\bm\theta^{(R)},\bm\gamma_0,\bm\xi_0)}{\partial \bm\gamma}\sqrt{n}(\widehat{\bm\gamma}-\bm\gamma_0)+\frac{1}{n}\frac{\partial \eU_{\bm\theta}^{(R)}(\bm\theta^{(R)},\bm\gamma_0,\bm\xi_0)}{\partial \bm\xi}\sqrt{n}(\widehat{\bm\xi}-\bm\xi_0)\right]+o_p(1)
\end{align*}
Define $\bm I_{\bm\theta}^{(R)}$ and $\bm S_{\bm\theta}^{(R)}$ as the component in the first and second brackets on the right-side hand of previous formula such that $\sqrt{n}(\widehat{\bm\theta}^{(R)}\!-\!\bm\theta_0^{(R)})=\left[\bm I_{\bm\theta}^{(R)}\right]^{-1} \cdot \frac{1}{\sqrt n}\bm S_{\bm\theta}^{(R)} + o_p(1)$. Applying similar strategy to derivation of \eqref{eq:I_O1} in Appendix D.2, one can infer that $\bm I_{\bm\theta}^{(R)} \overset{p}{\to} \bm{\mathcal I}_{\bm\theta}^{(R)} = -\left[\begin{matrix}\bm{\mathcal I}_{\bm\alpha} & \bm 0 \\ \bm 0 & \omega h_2^{(R)}(\bm\beta_0^{(R)}) \end{matrix}\right]$, where $h_2^{(R)}(\bm\beta_0^{(R)})$ is defined in \eqref{eq:h2_R}. Also, similar to the derivation of \eqref{eq:S_O1}, one can show 
\begin{align*}
    \frac{1}{\sqrt n}\bm S_{\bm\theta}^{(R)} =\frac{1}{\sqrt n}\sum_{i=1}^n \left\{\left(\begin{matrix}\mathbb U_{\bm\alpha, i}(\bm\alpha_0,\bm\gamma_0) \\ \bm G_i^{(R)}\left(\bm\beta_0^{(R)},\bm\alpha_0,\bm\xi_0\right)\end{matrix}\right) -  \bm{\mathcal I}_{\bm\theta,\bm\gamma}^{(R)} \bm{\mathcal I}_{\bm\gamma}^{-1}\eU_{\bm\gamma,i}(\bm\gamma_0) - \bm{\mathcal I}_{\bm\theta,\bm\xi}^{(R)} \bm{\mathcal I}_{\bm\xi}^{-1}\eU_{\bm\xi,i}(\bm\xi_0)\right\} + o_p(1),
\end{align*}
with $\bm{\mathcal I}_{\bm\theta,\bm\gamma}^{(R)} = E\left[\frac{\partial}{\partial \bm\gamma}\eU_{\bm\theta,i}^{(R)}\left(\bm\theta_0^{(R)},\bm\gamma_0,\bm\xi_0\right)\right]$, $\bm{\mathcal I}_{\bm\theta,\bm\xi}^{(R)} = E\left[\frac{\partial}{\partial \bm\xi}\eU_{\bm\theta,i}^{(R)}\left(\bm\theta_0^{(R)},\bm\gamma_0,\bm\xi_0\right)\right]$, and
\begin{align*}
   \bm G_i^{(R)}\left(\bm\beta_0^{(R)},\bm\alpha_0,\bm\xi_0\right) = & R_i\int_0^{t^*} \left\{\bm Z_i(\bm\xi_0) - \frac{s_{**}^{(1)}(\bm\beta_0^{(R)},\bm\xi_0,t)}{s_{**}^{(0)}(\bm\beta_0^{(R)},\bm\xi_0,t)}\right\}dN_i(t)  \\
   &  - R_i \int_0^{t^*} \frac{Y_i(t)\exp\left(\bm\beta^{(R)^T}\bm Z_i(\bm\xi_0)\right)}{s_{**}^{(0)}(\bm\beta_0^{(R)},\bm\xi_0,t)}\left\{\bm Z_i(\bm\xi_0)- \frac{s_{**}^{(1)}(\bm\beta_0^{(R)},\bm\xi_0,t)}{s_{**}^{(0)}(\bm\beta_0^{(R)},\bm\xi_0,t)}\right\} d\widetilde{F}(t).
\end{align*}
Therefore, by using a multivariate central limit theorem, we have that
$$
\sqrt{n}\left(\widehat{\bm\theta}^{(R)}-\bm\theta_0^{(R)}\right) \overset{D}{\rightarrow} N\left(\bm 0, \bm V^{(R)}\right),
$$
where $\bm V^{(R)} = \left[\bm{\mathcal I}_{\bm\theta}^{(R)}\right]^{-1} \bm{\mathcal M}_{\bm\theta}^{(R)} \left[\bm{\mathcal I}_{\bm\theta}^{(R)}\right]^{-T}$ and
$$
\bm{\mathcal M}_{\bm\theta}^{(R)} = E\left\{\left(\begin{matrix}\bm C_i(\bm\alpha_0) \\ \bm G_i\left(\bm\beta_0^{(R)}\right)\end{matrix}\right) -   \bm{\mathcal I}_{\bm\theta,\bm\gamma}^{(R)} \bm{\mathcal I}_{\bm\gamma}^{-1}\eU_{\bm\gamma,i}(\bm\gamma_0) - \bm{\mathcal I}_{\bm\theta,\bm\xi}^{(R)} \bm{\mathcal I}_{\bm\xi}^{-1}\eU_{\bm\xi,i}(\bm\xi_0)\right\}^{\otimes 2}.
$$

A consistent estimator for $\bm V^{(R)}$ can be
$
\widehat{\bm V}^{(R)} = \left[\widehat{\bm I}_{\bm\theta}^{(R)}\right]^{-1} \widehat{\bm M}_{\bm\theta}^{(R)} \left[\widehat{\bm I}_{\bm\theta}^{(R)}\right]^{-T},
$
where $\widehat{\bm I}_{\bm\theta}^{(R)}=-\frac{1}{n}\frac{\partial}{\partial \bm\theta}\eU_{\bm\theta}^{(R)}(\widehat{\bm\theta}^{(R)},\widehat{\bm\gamma},\widehat{\bm\xi})$,
$$
\widehat{\bm M}_{\bm\theta}^{(R)} = \frac{1}{n}\sum_{i=1}^n \left\{\eU_{\bm\theta,i}^{(R)}(\widehat{\bm\theta}^{(R)},\widehat{\bm\gamma},\widehat{\bm\xi}) - \widehat{\bm I}_{\bm\theta,\bm\gamma}^{(R)} \widehat{\bm I}_{\bm\gamma}^{-1}\eU_{\bm\gamma,i}(\widehat{\bm\gamma}) - \widehat{\bm I}_{\bm\theta,\bm\xi}^{(R)} \widehat{\bm I}_{\bm\xi}^{-1}\eU_{\bm\xi,i}(\widehat{\bm\xi}) \right\}^{\otimes 2},
$$
$\widehat{\bm I}_{\bm\theta,\bm\gamma}^{(R)}=\frac{1}{n}\frac{\partial}{\partial \bm\gamma}\eU_{\bm\theta}^{(R)}(\widehat{\bm\theta}^{(R)},\widehat{\bm\gamma},\widehat{\bm\xi})$, $\widehat{\bm I}_{\bm\theta,\bm\xi}^{(R)}=\frac{1}{n}\frac{\partial}{\partial \bm\xi}\eU_{\bm\theta}^{(R)}(\widehat{\bm\theta}^{(R)},\widehat{\bm\gamma},\widehat{\bm\xi})$, $\widehat{\bm I}_{\bm\gamma}=\frac{1}{n}\frac{\partial}{\partial \bm\gamma}\eU_{\bm\gamma}(\widehat{\bm\gamma})$, and $\widehat{\bm I}_{\bm\xi}=\frac{1}{n}\frac{\partial}{\partial \bm\xi}\eU_{\bm\xi}(\widehat{\bm\xi})$. 


\section*{Appendix F: Derivation of $\bm I_{\tau,\bm\theta}$.}

This Appendix derives the expressions of $\bm I_{\tau,\bm\theta}=\frac{\partial \tau_{a,a^*|\bm w}}{\partial \bm\theta}$, where $\widehat{\bm I}_{\tau,\bm\theta}^{(m)}$ is $\bm I_{\tau,\bm\theta}$ evaluated at $\widehat{\bm\theta}^{(m)}$, for $m\in\{O1,O2,R\}$.

Recall that $\text{NIE}_{a,a^*|\bm w} = (\beta_2\alpha_1+\beta_3\alpha_1a)(a-a^*)$ and $\bm\theta=[\alpha_0,\alpha_1,\bm\alpha_2^T,\sigma_M^2,\beta_1,\beta_2,\beta_3,\bm\beta_4^T]^T$, we have that, for $\tau=\text{NIE}$,
\begin{align*}
\bm I_{\text{NIE},\bm\theta} = (a-a^*)\times [0,\beta_2+\beta_3a,\bm 0^T,0,0,\alpha_1,\alpha_1a,\bm 0^T]^T.
\end{align*}
Similarly, we have that
\begin{align*}
\bm I_{\text{NDE},\bm\theta} = \left[\begin{matrix}
\beta_3(a-a^*)\\ 
\beta_3 (a-a^*)a^* \\ 
\beta_3 (a-a^*)\bm w \\ 
\beta_2\beta_3 (a-a^*)+0.5\beta_3^2(a^2-{a^*}^2)\\ 
(a-a^*)\\ 
\beta_3\sigma_M^2(a-a^*)\\ 
(\alpha_0+\alpha_1a^*+\bm\alpha_2^T\bm w+\beta_2\sigma_M^2)(a-a^*)+\beta_3\sigma_M^2(a^2-{a^*}^2)\\ 
\bm 0\\ 
\end{matrix}\right].
\end{align*}
Because $\text{TE}_{a,a^*|\bm w}=\text{NIE}_{a,a^*|\bm w}+\text{NDE}_{a,a^*|\bm w}$, we obtain $\bm I_{\text{TE},\bm\theta}=\bm I_{\text{NDE},\bm\theta}+\bm I_{\text{NIE},\bm\theta}$ such that
\begin{align*}
\bm I_{\text{TE},\bm\theta} = \left[\begin{matrix}
\beta_3(a-a^*)\\ 
\beta_2+\beta_3a+\beta_3 (a-a^*)a^* \\ 
\beta_3 (a-a^*)\bm w \\ 
\beta_2\beta_3 (a-a^*)+0.5\beta_3^2(a^2-{a^*}^2)\\ 
(a-a^*)\\ 
\alpha_1+\beta_3\sigma_M^2(a-a^*)\\ 
\alpha_1a+(\alpha_0+\alpha_1a^*+\bm\alpha_2^T\bm w+\beta_2\sigma_M^2)(a-a^*)+\beta_3\sigma_M^2(a^2-{a^*}^2)\\ 
\bm 0\\ 
\end{matrix}\right].
\end{align*}
Finally, because $\text{MP}_{a,a^*|\bm w}=\text{NIE}_{a,a^*|\bm w}/\text{TE}_{a,a^*|\bm w}$, we have that
\begin{align*}
\bm I_{\text{MP},\bm\theta} & = \frac{\partial \text{MP}_{a,a^*|\bm w}}{\partial \bm\theta} \\
& = \frac{1}{\text{TE}_{a,a^*|\bm w}}\times \frac{\partial \text{NIE}_{a,a^*|\bm w}}{\partial \bm\theta} - \frac{\text{NIE}_{a,a^*|\bm w}}{\text{TE}_{a,a^*|\bm w}^2}\times \frac{\partial \text{TE}_{a,a^*|\bm w}}{\partial \bm\theta} \\
& = \frac{1}{\text{TE}_{a,a^*|\bm w}}\times \bm I_{\text{NIE},\bm\theta} - \frac{\text{NIE}_{a,a^*|\bm w}}{\text{TE}_{a,a^*|\bm w}^2}\times\bm I_{\text{TE},\bm\theta}
\end{align*}
where $\bm I_{\text{NIE},\bm\theta}$ and $\bm I_{\text{TE},\bm\theta}$ are defined above.

\section*{Appendix G: Supporting information for the simulation studies}

\subsection*{Appendix G.1: Simulations when $\epsilon_A$ in the true measurement error model (4) is not normally distributed.}

The ORC2 approach assumes $\epsilon_A$ in measurement error model (4) is normally distributed. In Section 6.3 of the manuscript, we conducted additional simulations to examine the sensitivity of ORC2 under departure from the normality assumption for $\epsilon_A$. We introduce the data generation process for this additional simulation here. Specifically, the original data generation process used in Section 6.1 follows the following three steps:
\begin{enumerate}
    \item[i.] Generate $\left(\begin{matrix}A \\ M \\ W\end{matrix}\right) \sim N\left(\left[\begin{matrix}0 \\ 0 \\ 0\end{matrix}\right], 0.5 \times \left[\begin{matrix} 1 & 0.2 & 0.2 \\ 0.2 & 1 & 0.2 \\  0.2 & 0.2 & 1\end{matrix}\right] \right)$.
    \item[ii.] Generate $A^*$ by $N(\rho_{AA^*} A, 0.5(1-\rho_{AA^*}^2))$ to ensure that the correlation  between $A$ and $A^*$ is $\rho_{AA^*}$ and $A^*$ has the same marginal distribution to it of $A$.
    \item[iii.] Given $(A,M,W)$, generate $\widetilde T$ based on the Cox model (2), where choice of $\bm\beta$ and $\lambda_0(t)$ are specified in Section 6 of the main manuscript. Then, the observed failure time $T$ and the censoring time $C$ are also generated  accordingly.
\end{enumerate}
In this additional simulation, we aim to introduce non-normality in the conditional distribution of $A|\{A^*,W\}$, while maintaining all other aspects of the data generation process. To do this, we rewrite the previous 3 steps to the following 5 steps:
\begin{enumerate}
    \item  Generate $W\sim N(0,0.5)$.
    \item  Given $W$, generate $A^* = \sqrt{0.02}W+\delta_1$, where $\delta_1 \sim N(0,0.5/\rho_{AA^*}^2 -0.02)$.
    \item  Given $(W,A^*)$, generate $A=\frac{2\sqrt{2}\rho_{AA^*}^2}{5-\rho_{AA^*}^2}A^*+\frac{5-5\rho_{AA^*}^2}{5-\rho_{AA^*}^2}\sqrt{0.5}W+\epsilon_A$, where $\epsilon_A \sim N\left(0,\frac{12(1-\rho_{AA^*}^2)\rho_{AA^*}^2}{(5-\rho_{AA^*}^2)^2}\right)$.
    \item  Given $(A,W)$, generate $M = \frac{\rho_{MA}-0.04}{0.96}\sqrt{0.5} A + \frac{1 - \rho_{MA}}{4.8}\sqrt{0.5} W+\epsilon_M$, where $\epsilon_M\sim N(0,0.5-\frac{\rho_{MA}^2 - 0.08 \rho_{MA}+0.04}{1.92})$.
    \item Same to Step iii in the previous procedure. That is, given $(A,M,W)$, we generate $\widetilde T$ based on the Cox model (2), with the same choice of $\bm\beta$ and $\lambda_0(t)$. Then, the observed failure time $T$ and the censoring time $C$ are also generated accordingly.
\end{enumerate}
Step 1--5 are equivalent to Steps i--iii, but the new data generation process generates $W$, $A^*|\{A,W\}$, $A|\{A^*,W\}$, and $M|\{A,W\}$ sequentially, which allows us to modify the distribution of $A|\{A^*,W\}$, but still maintain the distributions of all other variables. In the main simulation study presented in Section 6.1, the error term $\epsilon_A$ in the measurement error model (4), $A|\{A^*,W\}$, followed a normal distribution. Now, we changed $\epsilon_A$ in Step 3 from a normal distribution to a skew normal distribution (\citealp{fernandez1998bayesian}) or a uniform distribution. When a skew normal distribution is considered, we generated $\epsilon_A \sim SN\left(0,\displaystyle\frac{12(1-\rho_{AA^*}^2)\rho_{AA^*}^2}{(5-\rho_{AA^*}^2)^2},5\right)$, where $SN(\mu,\sigma^2,\xi)$ denotes a skew normal distribution with mean $\mu$, variance $\sigma^2$, and $\xi$ is the skewness parameter controlling the degree of skewness. We set $\xi=5$ such that the coefficient of the skewness of $\epsilon_A$ approximate to 1. When the uniform distribution is considered, we generated $\epsilon_A \sim \text{Unif}\left(-0.5\times \sqrt{\frac{144(1-\rho_{AA^*}^2)\rho_{AA^*}^2}{(5-\rho_{AA^*}^2)^2}},0.5\times \sqrt{\frac{144(1-\rho_{AA^*}^2)\rho_{AA^*}^2}{(5-\rho_{AA^*}^2)^2}}\right)$ such that it has mean 0 and the same variance under the original simulations with normality. Web Figure \ref{fig:skew} in the supplementary material illustrates the density plots of $\epsilon_A$ used in this additional simulation study.


\subsection*{Appendix G.2: Simulations under misspecified mediator or outcome models.}

{We extend the expressions for the mediation effect measures to allow for a more general specification of the mediator and outcome models. Specifically, let $f_{M|A,\bm W}$ be the conditional density for the mediator given $\{A,\bm W\}$ and $f_{\widetilde T|A,M,\bm W}$ and $S_{\widetilde T|A,M,\bm W}$ be the conditional density and survival function of $\widetilde T$ given $\{A,M,\bm W\}$. Then, \cite{vanderweele2011causal} has shown that
\begin{equation}\label{e:general_mediation_formula}
\lambda_{\widetilde T_{aM_{a^*}}}(t|\bm w) = \frac{\int_m f_{\widetilde T|A,M,\bm W}(t|a,m,\bm w) f_{M|A,\bm W}(m|a^*,\bm w) dm }{\int_m S_{\widetilde T|A,M,\bm W}(t|a,m,\bm w) f_{M|A,\bm W}(m|a^*,\bm w) dm}
\end{equation}
under the identifiability assumptions listed in Web Appendix A. It follows that $$
\text{NIE}_{a,a^*|\bm w}(t) = \log(\lambda_{\widetilde T_{aM_a}}(t|\bm w))-\log(\lambda_{\widetilde T_{aM_{a^*}}}(t|\bm w)),
$$ 
$$
\text{NDE}_{a,a^*|\bm w}(t) = \log(\lambda_{\widetilde T_{aM_{a^*}}}(t|\bm w))-\log(\lambda_{\widetilde T_{a^*M_{a^*}}}(t|\bm w)),
$$ and $\text{MP}_{a,a^*|\bm w}(t) = \text{NIE}_{a,a^*|\bm w}(t) \Big/\left\{\text{NIE}_{a,a^*|\bm w}(t)+\text{NDE}_{a,a^*|\bm w}(t)\right\}$.

We first consider the scenario when $\epsilon_M$ does not follow a normal distribution. Specifically, we shall follow our main simulation setup to generate $\{W,A^*,A,M,\widetilde T,C\}$ based on Steps 1--5 in Web Appendix G.1. The only difference is that we now generate $\epsilon_M$ in Step 4 based on a skew normal distribution or a uniform distribution in order to violate the normality assumption. When normality assumption holds, we considered $\epsilon_M\sim N(0,\sigma_M^2)$, where $\sigma_M^2 = 0.5-\frac{\rho_{MA}^2 - 0.08 \rho_{MA}+0.04}{1.92}$.   When the skew normal distribution was used, we simulated $\epsilon_M\sim SN\left(0,\sigma_M^2,5\right)$ such that its mean and variance matched the one used in the original normal distribution, but we now set the skew parameter to 5 so that the coefficient of the skewness of $\epsilon_M$ approximate to 1.  Similarly, when uniform distribution was considered, we simulated $\epsilon_M\sim \text{Unif}\left(-0.5\sqrt{12\sigma_M^2},0.5\sqrt{12\sigma_M^2}\right)$ so that its mean and variance aligned with the original normal distribution. The true values of the mediation effect measures can be calculated based on \eqref{e:general_mediation_formula} with
\begin{align*}
f_{M|A,W}(m|a,w) & = f_{\epsilon_M}(m -\alpha_0 - \alpha_1 a - \alpha_2^T w)\\
S_{\widetilde T|A,M,W}(t|a,m,w) & = \exp\Big(-\Lambda_0(t)e^{\beta_1a+\beta_2m+\beta_3am+\beta_4w}\Big) \\
f_{\widetilde T|A,M,W}(t|a,m,w) & = \lambda_0(t)e^{\beta_1a+\beta_2m+\beta_3am+\beta_4w} S_{\widetilde T|A,M,W}(t|a,m,w), 
\end{align*}
where $f_{\epsilon_M}$ is the density of $\epsilon_M$, $\lambda_0(t) = \theta v (vt)^{\theta-1}$, and $\Lambda_0(t) = (vt)^{\theta}$.

We next considered the scenario where the proportional hazards assumption in the Cox outcome model does not hold. In the original data generation process (see Steps 1--5 in Web Appendix G.1), we simulated $\widetilde T|\{A,M,W\}$ based on a Weibull-Cox model with $\lambda_{\widetilde T|A,M,W} = \theta v (vt)^{\theta-1} \exp(\beta_1A+\beta_2M+\beta_3AM+\beta_4W)$. This Weibull-Cox model is equivalent to the following accelerated failure time model:
\begin{equation}\label{m:atf}
\log(\widetilde T) = \psi_0 + \psi_1 A + \psi_2 M + \psi_3 AM + \psi_4 W + \sigma\epsilon_{\widetilde T},
\end{equation}
where $f_{\epsilon_{\widetilde T}}(\epsilon)=\exp\{\epsilon-e^{-\epsilon}\}$, $\sigma=\frac{1}{\theta}$, $\psi_0=-\log(v)$, and $\psi_j = -\frac{\beta_j}{\theta}$ for $j\in\{1,2,3,4\}$. To violate the proportional hazards assumption, we instead simulated $\widetilde T$ based on \eqref{m:atf} with $f_{\epsilon_{\widetilde T}}(\epsilon)=\frac{e^\epsilon}{(1+e^\epsilon)^2}$ following a logistic distribution so that $\widetilde T$ follows a proportional odds model. Specifically, we maintained the values of $\{\psi_0,\psi_1,\psi_2,\psi_3,\psi_4\}$ in the Weibull-Cox model and set $\sigma=0.706$ in order to ensure that the cumulative event rate approximately equal 5\% at the maximum follow-up time $t^*=50$. The true values of the mediation effect measures were then calculated based on \eqref{e:general_mediation_formula} as
\begin{align*}
f_{M|A,W}(m|a,w) & = f_{\epsilon_M}(m-\alpha_0 - \alpha_1 a - \alpha_2^T w)\\
f_{\widetilde T|A,M, W}(t|a,m, w) & = f_{\epsilon_{\widetilde T}}\left(\frac{\log(t)-\psi_1 a - \psi_2 m - \psi_3 am - \psi_4 w}{\sigma}\right)\times \frac{1}{\sigma t} \\
S_{\widetilde T|A,M,W}(t|a,m,w) & = 1 - F_{\epsilon_{\widetilde T}}\left(\frac{\log(t)-\psi_1 a - \psi_2 m - \psi_3 am - \psi_4 w}{\sigma}\right)
\end{align*}
where $f_{\epsilon_M}$ is the density of $N(0,\sigma_M^2)$, $f_{\epsilon_{\widetilde T}}(\epsilon)=\frac{e^\epsilon}{(1+e^\epsilon)^2}$, and $F_{\epsilon_{\widetilde T}}(\epsilon) = \frac{e^{\epsilon}}{1+e^{\epsilon}}$.}

\section*{Appendix H: Mediation analysis with a survival outcome when the Cox outcome model includes piece-wise constant coefficients}

{To describe the Cox model with piece-wise constant coefficients, we split the follow-up duration $[0,t^*]$ into $J$ intervals, $\mathcal T_1=[t_1,t_2)$, $\mathcal T_2 = [t_2,t_3)$, $\dots$, $\mathcal T_J=[t_J,t_{J+1}]$, where $t_1=0$ and $t_{J+1}=t^*$. We consider the following Cox outcome model with piece-wise constant coefficients
\begin{equation}\label{m:cox_piecewise_constant2}
\lambda_{\widetilde T|A,M,\bm W}(t|A,M,\bm W) = \lambda_0(t)\exp\left(\sum_{j=1}^J \mathbb{I}(t\in \mathcal T_j)\left\{\beta_{1,j}A+\beta_{2,j}M+\beta_{3,j}AM+\bm\beta_{4,j}^T \bm W\right\}\right),
\end{equation}
where $\bm\beta_{\mathcal T_j} = [\beta_{1,j},\beta_{2,j},\beta_{3,j},\bm\beta_{4,j}^T]^T$ contains the coefficients in $j$-th interval. Note that this is a generalization of Cox model (2) to allow for time-dependent coefficients. One can also enforce a coefficient to be a constant value over the different time intervals if needed, and the mediation analysis result will become a special case of our general development below. The piece-wise constant coefficients in the Cox model \eqref{m:cox_piecewise_constant2}, $\bm\beta=[\bm\beta_{\mathcal T_1}^T,\dots,\bm\beta_{\mathcal T_J}^T]^T$, can be estimated based on several off-the-shelf software; for example, the \texttt{survival} package in R software \citep{therneau2017using}. Similarly, the cumulative baseline hazard, $\Lambda_0(t)$, can be obtained by the Breslow estimator, which can also be performed through the \texttt{survival} package.

Next, we derive expressions of the mediation effect measures based on the mediator model (1) and the Cox outcome model \eqref{m:cox_piecewise_constant2}. According to \eqref{m:cox_piecewise_constant2}, the conditional survival function of $\widetilde T|\{A,M,\bm W\}$ is
\begin{align*}
 & S_{\widetilde T|A,M,\bm W}(t|a,m,\bm w) \\
= &\left\{\begin{aligned}
& \exp\left(-\Lambda_0(t)e^{g_1} \right) \text{ if $t\in\mathcal T_1$,}  \\
& \exp\left(-\left\{\Lambda_0(t)-\Lambda_0(t_j)\right\}e^{g_j}-\sum_{l=1}^{j-1} \left\{\Lambda_0(t_{l+1})-\Lambda_0(t_l)\right\}e^{g_l} \right) \text{ if $t\in\mathcal T_j$ with $j>1$,} 
\end{aligned}\right.
\end{align*}
where $g_l \equiv g_l(a,m,\bm w) =  \beta_{1,l}a+\beta_{2,l}m+\beta_{3,l}am+\bm\beta_{4,l}^T \bm w$. The density of $\widetilde T|\{A,M,\bm W\}$ is therefore
\begin{align*}
f_{\widetilde T|A,M,\bm W}(t|A,M,\bm W) & = \lambda_{\widetilde T|A,M,\bm W}(t|A,M,\bm W) S_{\widetilde T|A,M,\bm W}(t|A,M,\bm W) \\
& = \lambda_0(t)\exp\left(\sum_{j=1}^J \mathbb{I}(t\in \mathcal T_j) g_j(A,M,\bm W)\right) S_{\widetilde T|A,M,\bm W}(t|A,M,\bm W)
\end{align*}
Observing $f_{M|A,M,\bm W}(m|a,m,\bm w) = \frac{1}{\sqrt{2\pi}\sigma_M^2}\exp\left(-\frac{(m-\alpha_0-\alpha_1a-\bm\alpha_2^T\bm w)^2}{2\sigma_M^2}\right)$, and substituting  $S_{\widetilde T|A,M,\bm W}$, $f_{\widetilde T|A,M,\bm W}$, and $f_{M|A,M,\bm W}$ into \eqref{e:general_mediation_formula}, one obtains 
\begin{align*}
\text{NIE}_{a,a^*|\bm w}(t) & = \widetilde \delta_{a,a|\bm w}\left(\Lambda_0(t),\bm\theta\right) - \widetilde \delta_{a,a^*|\bm w}\left(\Lambda_0(t),\bm\theta\right), \\
\text{NDE}_{a,a^*|\bm w}(t) & = \widetilde \delta_{a,a^*|\bm w}\left(\Lambda_0(t),\bm\theta\right) - \widetilde \delta_{a^*,a^*|\bm w}\left(\Lambda_0(t),\bm\theta\right),
\end{align*}
where $\bm\theta = [\bm\alpha^T,\bm\beta^T]^T$ and
$$
\widetilde \delta_{a,a^*|\bm w}\left(\Lambda_0(t),\bm\theta\right) = \log\left(\frac{\int_{m} e^{\sum_{j=1}^J \mathbb{I}(t\in\mathcal T_j) g_j(a,m,\bm w)} S_{\widetilde T|A,M,\bm W}(t|a,m,\bm w)\exp\left(-\frac{(m-\alpha_0-\alpha_1a^*-\bm\alpha_2^T\bm w)^2}{2\sigma_M^2}\right) dm}{\int_{m}  S_{\widetilde T|A,M,\bm W}(t|a,m,\bm w)\exp\left(-\frac{(m-\alpha_0-\alpha_1a^*-\bm\alpha_2^T\bm w)^2}{2\sigma_M^2}\right) dm}\right). 
$$
Notice that the expressions of $\text{NIE}_{a,a^*|\bm w}(t)$ and $\text{NDE}_{a,a^*|\bm w}(t)$ are functions of coefficients in the mediator and Cox models (i.e., $\bm\theta$) and the cumulative baseline hazard (i.e., $\Lambda_0(t)$). Therefore, we can simply substitute $\widehat{\bm\theta}$ and $\widehat\Lambda_0(t)$ into their expressions to obtain estimates of the corresponding mediation effect measures.

The ORC1, ORC2, and RRC approaches can be easily extended to fit into the scenario with piece-wise constant coefficients. Unknown parameters in the mediator model can be obtained through our procedures in Section 4.1. For the Cox outcome model coefficients, we can run \eqref{m:cox_piecewise_constant2} in the main study samples with $A_i$ replaced by $m_{A_i}^{(O1)}(\widehat{\bm\eta})$, $m_{A_i}^{(O2)}(\widehat{\bm\alpha},\widehat{\bm\gamma})$, and $m_{A_i}^{(R)}(\widehat{\bm\xi},t)$ for the ORC1, ORC2, and RRC approaches, respectively. The cumulative baseline hazard $\widehat \Lambda_0(t)$ is then obtained by the Breslow estimator. Finally, we plug in the coefficient and cumulative baseline hazard estimates into the expressions of the mediation effect measures to obtain the final mediation effect estimate.}

\section*{Appendix I: Web Figures and Tables}

\renewcommand{\figurename}{Web Figure}
\renewcommand{\tablename}{Web Table}

\begin{figure}[htbp] 
\centering 
\includegraphics[width=0.9\textwidth]{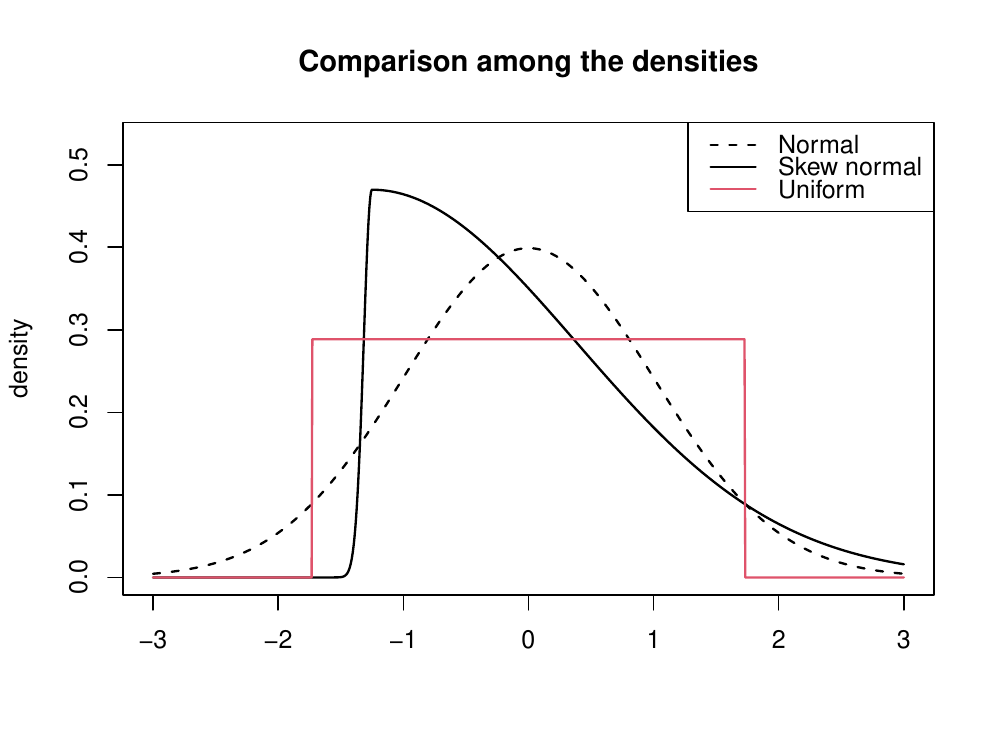} 
\caption{Density plot of the skew normal and uniform distributions for the additional simulation study in Section 6.3. A normal density curve (dotted line) was added as reference.} 
\label{fig:skew}
\end{figure}

\begin{figure}[htbp]
\centering\includegraphics[width=0.9\textwidth]{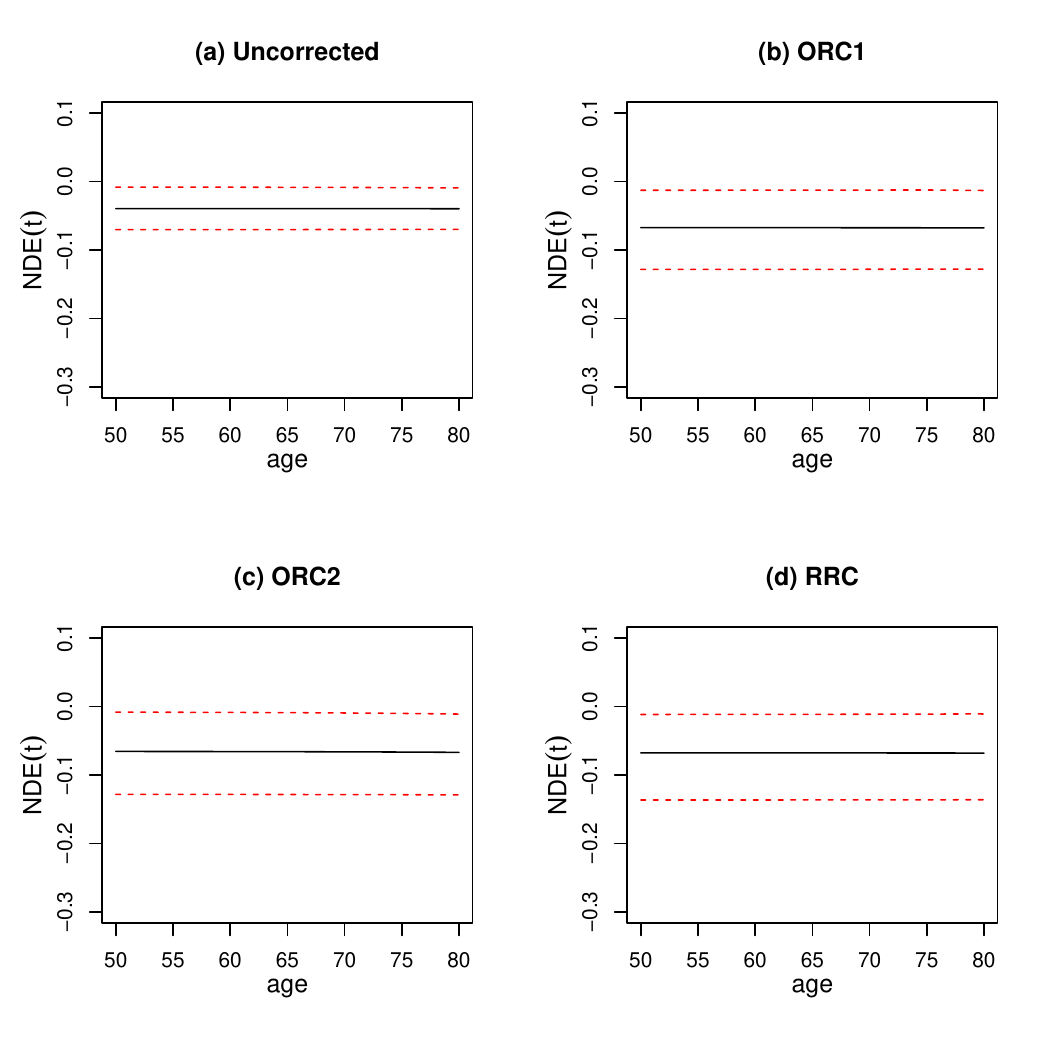}
\caption{NDE estimates based on the exact mediation effect measure expressions for the HPFS, 1986--2016. The black solid line is the point estimates and the red dotted lines are the 95\% confidence interval obtained via  bootstrapping.}
\label{fig:nde}
\end{figure}

\clearpage

\begin{table}[htbp]
\caption{Simulation results for a common outcome with a 50\% cumulative event rate at the maximum follow-up time $t^*=50$. Here, $\widehat{\tau}^{(U)}(t)$, $\widehat{\tau}^{(G)}(t)$, $\widehat{\tau}^{(O1)}(t)$, $\widehat{\tau}^{(O2)}(t)$, and $\widehat{\tau}^{(R)}(t)$ denote the estimators given by the uncorrected, gold-standard, ORC1, ORC2, and RRC approaches, respectively, where the RRC uses 4 split points to determine the validation study risk sets. Upper panel: performance of the NIE estimator; middle panel: performance of the NDE estimator; lower panel: performance of the MP estimator.}
\centering
\scalebox{0.66}[0.66]{
\begin{threeparttable}
\label{tab:sim_exact}
\setlength\tabcolsep{4pt}
\begin{tabular}{cc ccccc ccccc }
  \hline
& & \multicolumn{5}{c}{Percent Bias (SE$\times$100)} &   \multicolumn{5}{c}{Bootstrap Coverage Rate} \\
\cmidrule(lr){3-7} \cmidrule(lr){8-12}
$t$ &   NIE$(t)$  & 
 $\widehat{\text{NIE}}^{(U)}(t)$ &  $\widehat{\text{NIE}}^{(G)}(t)$ & $\widehat{\text{NIE}}^{(O1)}(t)$ & $\widehat{\text{NIE}}^{(O2)}(t)$ & $\widehat{\text{NIE}}^{(R)}(t)$ &
 $\widehat{\text{NIE}}^{(U)}(t)$ &  $\widehat{\text{NIE}}^{(G)}(t)$ & $\widehat{\text{NIE}}^{(O1)}(t)$ & $\widehat{\text{NIE}}^{(O2)}(t)$ & $\widehat{\text{NIE}}^{(R)}(t)$ \\ 
  \hline
10 & log(1.129) & -42.6 (0.8) & -0.4 (1.0) & -2.0 (1.8) & -1.9 (1.8) & 0.1 (2.1) & 0.1 & 95.0 & 94.9 & 94.7 & 95.3 \\ 
  15 & log(1.129) & -42.6 (0.8) & -0.4 (1.0) & -2.0 (1.8) & -1.9 (1.8) & 0.1 (2.1) & 0.1 & 95.0 & 94.9 & 94.7 & 95.3 \\ 
  20 & log(1.129) & -42.6 (0.8) & -0.4 (1.0) & -2.0 (1.8) & -1.9 (1.8) & 0.1 (2.1) & 0.1 & 95.0 & 94.9 & 94.7 & 95.3 \\ 
  25 & log(1.129) & -42.6 (0.8) & -0.4 (1.0) & -2.0 (1.8) & -1.9 (1.8) & 0.0 (2.1) & 0.1 & 95.0 & 94.8 & 94.7 & 95.3 \\ 
  30 & log(1.128) & -42.5 (0.8) & -0.4 (0.9) & -2.0 (1.8) & -1.9 (1.7) & -0.0 (2.0) & 0.1 & 94.8 & 94.7 & 94.8 & 95.2 \\ 
  35 & log(1.125) & -42.4 (0.8) & -0.4 (0.9) & -2.0 (1.7) & -1.9 (1.7) & -0.2 (1.9) & 0.1 & 94.8 & 94.7 & 94.7 & 95.2 \\ 
  40 & log(1.121) & -42.2 (0.8) & -0.4 (0.9) & -1.9 (1.6) & -1.7 (1.6) & -0.4 (1.8) & 0.1 & 94.9 & 94.5 & 94.5 & 95.3 \\ 
  45 & log(1.114) & -41.9 (0.7) & -0.5 (0.8) & -1.7 (1.5) & -1.6 (1.5) & -0.8 (1.6) & 0.1 & 94.5 & 94.5 & 94.2 & 95.3 \\
   \hline
   \hline
 & & \multicolumn{5}{c}{Percent Bias (SE$\times$100)} &   \multicolumn{5}{c}{Bootstrap Coverage Rate} \\
\cmidrule(lr){3-7} \cmidrule(lr){8-12}
$t$ &   NDE$(t)$ & 
 $\widehat{\text{NDE}}^{(U)}(t)$ &  $\widehat{\text{NDE}}^{(G)}(t)$ & $\widehat{\text{NDE}}^{(O1)}(t)$ & $\widehat{\text{NDE}}^{(O2)}(t)$ & $\widehat{\text{NDE}}^{(R)}(t)$ &
 $\widehat{\text{NDE}}^{(U)}(t)$ &  $\widehat{\text{NDE}}^{(G)}(t)$ & $\widehat{\text{NDE}}^{(O1)}(t)$ & $\widehat{\text{NDE}}^{(O2)}(t)$ & $\widehat{\text{NDE}}^{(R)}(t)$  \\ 
  \hline
 10 & log(1.328) & -43.0 (2.5) & 0.2 (2.5) & -1.3 (5.0) & -1.3 (5.0) & 7.0 (7.8) & 0.5 & 95.5 & 95.0 & 95.2 & 94.8 \\ 
  15 & log(1.328) & -43.0 (2.5) & 0.2 (2.5) & -1.3 (5.0) & -1.3 (5.0) & 7.0 (7.8) & 0.5 & 95.5 & 95.0 & 95.2 & 94.8 \\ 
  20 & log(1.327) & -43.0 (2.5) & 0.2 (2.5) & -1.3 (5.0) & -1.3 (5.0) & 7.0 (7.8) & 0.5 & 95.6 & 95.0 & 95.2 & 94.8 \\ 
  25 & log(1.323) & -43.0 (2.5) & 0.2 (2.5) & -1.3 (5.0) & -1.2 (5.0) & 7.0 (7.7) & 0.5 & 95.6 & 95.0 & 95.2 & 94.7 \\ 
  30 & log(1.314) & -43.0 (2.4) & 0.1 (2.4) & -1.2 (4.9) & -1.2 (4.9) & 7.0 (7.5) & 0.5 & 95.8 & 95.0 & 95.1 & 94.8 \\ 
  35 & log(1.294) & -43.1 (2.4) & 0.1 (2.4) & -1.1 (4.8) & -1.3 (4.8) & 6.8 (7.2) & 0.6 & 95.8 & 94.9 & 95.1 & 95.0 \\ 
  40 & log(1.257) & -43.3 (2.3) & 0.0 (2.3) & -0.9 (4.7) & -1.1 (4.7) & 5.9 (6.7) & 1.1 & 96.1 & 95.2 & 95.2 & 95.0 \\ 
  45 & log(1.197) & -43.4 (2.3) & -0.0 (2.4) & -0.7 (4.6) & -0.8 (4.6) & 4.8 (6.2) & 2.3 & 95.7 & 95.3 & 95.2 & 94.7 \\ 
   \hline
   \hline
    & & \multicolumn{5}{c}{Percent Bias (SE$\times$100)} &   \multicolumn{5}{c}{Bootstrap Coverage Rate} \\
\cmidrule(lr){3-7} \cmidrule(lr){8-12}
$t$ &   MP$(t)$  & 
 $\widehat{\text{MP}}^{(U)}(t)$ &  $\widehat{\text{MP}}^{(G)}(t)$ & $\widehat{\text{MP}}^{(O1)}(t)$ & $\widehat{\text{MP}}^{(O2)}(t)$ & $\widehat{\text{MP}}^{(R)}(t)$ &
 $\widehat{\text{MP}}^{(U)}(t)$ &  $\widehat{\text{MP}}^{(G)}(t)$ & $\widehat{\text{MP}}^{(O1)}(t)$ & $\widehat{\text{MP}}^{(O2)}(t)$ & $\widehat{\text{MP}}^{(R)}(t)$  \\ 
  \hline
  10 & 0.300 & 0.8 (4.2) & -0.3 (2.5) & -0.2 (4.3) & -0.3 (4.2) & -4.4 (5.1) & 95.1 & 95.5 & 95.1 & 95.2 & 94.3 \\ 
  15 & 0.300 & 0.8 (4.2) & -0.3 (2.5) & -0.2 (4.3) & -0.3 (4.2) & -4.4 (5.1) & 95.1 & 95.5 & 95.1 & 95.2 & 94.3 \\ 
  20 & 0.300 & 0.8 (4.2) & -0.3 (2.5) & -0.2 (4.3) & -0.3 (4.2) & -4.4 (5.1) & 95.1 & 95.5 & 95.1 & 95.2 & 94.3 \\ 
  25 & 0.302 & 0.8 (4.2) & -0.3 (2.5) & -0.2 (4.3) & -0.3 (4.3) & -4.3 (5.1) & 95.2 & 95.5 & 95.1 & 95.2 & 94.3 \\ 
  30 & 0.305 & 0.9 (4.2) & -0.4 (2.5) & -0.2 (4.4) & -0.3 (4.3) & -4.3 (5.1) & 95.2 & 95.5 & 95.0 & 95.2 & 94.4 \\ 
  35 & 0.314 & 1.1 (4.3) & -0.3 (2.6) & -0.3 (4.5) & -0.4 (4.4) & -4.2 (5.2) & 95.2 & 95.6 & 95.1 & 95.0 & 94.2 \\ 
  40 & 0.333 & 1.3 (4.6) & -0.3 (2.7) & -0.5 (4.7) & -0.5 (4.7) & -4.0 (5.5) & 95.2 & 95.5 & 94.9 & 95.1 & 94.3 \\ 
  45 & 0.375 & 1.8 (5.2) & -0.1 (3.0) & -0.7 (5.3) & -0.5 (5.2) & -3.9 (6.0) & 95.1 & 95.5 & 95.2 & 94.9 & 94.2 \\
 \hline
   \hline
\end{tabular}
\begin{tablenotes}
\item[1] The data generation process is given in Section 6, where the mediator follows a linear regression (1) and outcome follows the Cox model (2), with a linear measurement error process characterized in (4). In this simulation, we considered a moderate sample size of the validation study ($n_2=250$) and a moderate exposure measurement error ($\rho_{AA^*}=0.6$). The percent bias is defined as the of the ratio of bias to the true value over 2,000 replications, i.e., $mean(\frac{\widehat{\tau}-\tau}{\tau}) \times 100\%$. The standard error (SE) is defined as the square root of empirical variance of mediation effect measure estimates from the 2,000 replications. The coverage rate is obtained based on a non-parametric bootstrap confidence interval.
\end{tablenotes}
\end{threeparttable}}
\end{table}

\begin{table}[ht]
\caption{Simulation results for a common outcome with a large exposure effect in the Cox outcome model ($\beta_1=1$). The simulation setting is identical to Web Table 1, except that a large exposure effect ($\beta_1=1$) is introduced in the Cox model. Upper panel: performance of the NIE estimator; middle panel: performance of the NDE estimator; lower panel: performance of the MP estimator.}
\centering
\scalebox{0.66}[0.66]{
\begin{threeparttable}
\label{tab:sim_exact}
\setlength\tabcolsep{4pt}
\begin{tabular}{cc ccccc ccccc }
  \hline
& & \multicolumn{5}{c}{Percent Bias (SE$\times$100)} &   \multicolumn{5}{c}{Bootstrap Coverage Rate} \\
\cmidrule(lr){3-7} \cmidrule(lr){8-12}
$t$ &   NIE$(t)$  & 
 $\widehat{\text{NIE}}^{(U)}(t)$ &  $\widehat{\text{NIE}}^{(G)}(t)$ & $\widehat{\text{NIE}}^{(O1)}(t)$ & $\widehat{\text{NIE}}^{(O2)}(t)$ & $\widehat{\text{NIE}}^{(R)}(t)$ &
 $\widehat{\text{NIE}}^{(U)}(t)$ &  $\widehat{\text{NIE}}^{(G)}(t)$ & $\widehat{\text{NIE}}^{(O1)}(t)$ & $\widehat{\text{NIE}}^{(O2)}(t)$ & $\widehat{\text{NIE}}^{(R)}(t)$ \\ 
  \hline
10 & log(1.129) & -43.3 (0.8) & -0.6 (0.9) & -16.0 (1.7) & -15.8 (1.4) & -7.0 (2.5) & 0.0 & 95.5 & 83.7 & 78.4 & 95.3 \\ 
  15 & log(1.129) & -43.3 (0.8) & -0.6 (0.9) & -16.0 (1.7) & -15.8 (1.4) & -7.0 (2.5) & 0.0 & 95.5 & 83.8 & 78.3 & 95.3 \\ 
  20 & log(1.129) & -43.3 (0.8) & -0.6 (0.9) & -16.0 (1.7) & -15.8 (1.4) & -7.0 (2.5) & 0.0 & 95.6 & 83.7 & 78.5 & 95.3 \\ 
  25 & log(1.128) & -43.2 (0.8) & -0.5 (0.9) & -15.8 (1.7) & -15.6 (1.4) & -7.1 (2.4) & 0.0 & 95.6 & 83.6 & 78.5 & 95.1 \\ 
  30 & log(1.126) & -42.9 (0.8) & -0.5 (0.9) & -15.5 (1.6) & -15.2 (1.3) & -7.3 (2.2) & 0.0 & 95.8 & 83.6 & 78.8 & 94.9 \\ 
  35 & log(1.121) & -42.3 (0.7) & -0.6 (0.8) & -14.9 (1.5) & -14.4 (1.3) & -7.6 (2.0) & 0.0 & 95.8 & 84.0 & 79.9 & 94.2 \\ 
  40 & log(1.113) & -41.1 (0.7) & -0.6 (0.7) & -13.4 (1.3) & -12.8 (1.2) & -7.8 (1.7) & 0.0 & 96.0 & 85.5 & 82.0 & 93.7 \\ 
  45 & log(1.102) & -39.5 (0.6) & -0.6 (0.6) & -11.3 (1.2) & -10.8 (1.0) & -7.6 (1.4) & 0.0 & 96.3 & 87.6 & 85.8 & 93.4 \\ 
   \hline
   \hline
 & & \multicolumn{5}{c}{Percent Bias (SE$\times$100)} &   \multicolumn{5}{c}{Bootstrap Coverage Rate} \\
\cmidrule(lr){3-7} \cmidrule(lr){8-12}
$t$ &   NDE$(t)$ & 
 $\widehat{\text{NDE}}^{(U)}(t)$ &  $\widehat{\text{NDE}}^{(G)}(t)$ & $\widehat{\text{NDE}}^{(O1)}(t)$ & $\widehat{\text{NDE}}^{(O2)}(t)$ & $\widehat{\text{NDE}}^{(R)}(t)$ &
 $\widehat{\text{NDE}}^{(U)}(t)$ &  $\widehat{\text{NDE}}^{(G)}(t)$ & $\widehat{\text{NDE}}^{(O1)}(t)$ & $\widehat{\text{NDE}}^{(O2)}(t)$ & $\widehat{\text{NDE}}^{(R)}(t)$  \\ 
  \hline
 10 & log(2.802) & -48.8 (2.5) & -0.2 (2.8) & -11.1 (9.5) & -11.0 (9.6) & 5.5 (18.0) & 0.0 & 95.2 & 82.5 & 82.7 & 94.0 \\ 
  15 & log(2.801) & -48.8 (2.5) & -0.2 (2.8) & -11.1 (9.5) & -11.0 (9.6) & 5.4 (18.0) & 0.0 & 95.2 & 82.5 & 82.7 & 94.1 \\ 
  20 & log(2.797) & -48.8 (2.5) & -0.2 (2.8) & -11.0 (9.5) & -10.9 (9.6) & 5.4 (18.0) & 0.0 & 95.3 & 82.4 & 82.8 & 94.2 \\ 
  25 & log(2.783) & -48.8 (2.5) & -0.2 (2.7) & -10.9 (9.4) & -10.8 (9.5) & 5.3 (17.7) & 0.0 & 95.2 & 82.8 & 83.0 & 94.2 \\ 
  30 & log(2.747) & -48.7 (2.5) & -0.2 (2.7) & -10.6 (9.3) & -10.6 (9.5) & 5.0 (17.2) & 0.0 & 95.3 & 83.5 & 83.6 & 94.2 \\ 
  35 & log(2.672) & -48.5 (2.4) & -0.2 (2.6) & -9.9 (9.2) & -9.9 (9.3) & 4.5 (16.3) & 0.0 & 95.1 & 85.0 & 85.2 & 94.3 \\ 
  40 & log(2.546) & -48.0 (2.4) & -0.2 (2.6) & -8.5 (9.0) & -8.5 (9.2) & 4.0 (15.3) & 0.0 & 95.2 & 88.0 & 88.3 & 94.7 \\ 
  45 & log(2.372) & -47.1 (2.5) & -0.2 (2.8) & -6.3 (9.0) & -6.1 (9.1) & 3.4 (14.6) & 0.0 & 95.6 & 92.1 & 92.3 & 94.7 \\
   \hline
   \hline
    & & \multicolumn{5}{c}{Percent Bias (SE$\times$100)} &   \multicolumn{5}{c}{Bootstrap Coverage Rate} \\
\cmidrule(lr){3-7} \cmidrule(lr){8-12}
$t$ &   MP$(t)$  & 
 $\widehat{\text{MP}}^{(U)}(t)$ &  $\widehat{\text{MP}}^{(G)}(t)$ & $\widehat{\text{MP}}^{(O1)}(t)$ & $\widehat{\text{MP}}^{(O2)}(t)$ & $\widehat{\text{MP}}^{(R)}(t)$ &
 $\widehat{\text{MP}}^{(U)}(t)$ &  $\widehat{\text{MP}}^{(G)}(t)$ & $\widehat{\text{MP}}^{(O1)}(t)$ & $\widehat{\text{MP}}^{(O2)}(t)$ & $\widehat{\text{MP}}^{(R)}(t)$  \\ 
  \hline
  10 & 0.106 & 9.1 (1.2) & -0.1 (0.7) & -5.1 (1.4) & -5.0 (1.2) & -9.1 (1.9) & 89.1 & 96.5 & 94.8 & 93.6 & 92.8 \\ 
  15 & 0.106 & 9.1 (1.2) & -0.1 (0.7) & -5.1 (1.4) & -5.0 (1.2) & -9.1 (1.9) & 89.1 & 96.5 & 94.8 & 93.6 & 92.8 \\ 
  20 & 0.106 & 9.2 (1.2) & -0.1 (0.7) & -5.1 (1.4) & -5.0 (1.2) & -9.1 (1.9) & 89.1 & 96.5 & 94.8 & 93.6 & 92.7 \\ 
  25 & 0.105 & 9.3 (1.2) & -0.1 (0.7) & -5.0 (1.4) & -4.9 (1.2) & -9.2 (1.8) & 89.0 & 96.5 & 94.8 & 93.6 & 92.7 \\ 
  30 & 0.105 & 9.7 (1.2) & -0.1 (0.7) & -4.9 (1.4) & -4.8 (1.2) & -9.3 (1.8) & 88.7 & 96.5 & 94.8 & 93.6 & 92.5 \\ 
  35 & 0.104 & 10.4 (1.2) & -0.1 (0.7) & -4.8 (1.4) & -4.6 (1.2) & -9.5 (1.8) & 87.7 & 96.5 & 94.8 & 93.8 & 92.5 \\ 
  40 & 0.102 & 11.5 (1.2) & -0.1 (0.7) & -4.6 (1.4) & -4.4 (1.2) & -9.4 (1.8) & 86.4 & 96.3 & 94.8 & 94.0 & 92.2 \\ 
  45 & 0.101 & 12.5 (1.3) & -0.1 (0.7) & -4.6 (1.4) & -4.5 (1.3) & -9.3 (1.9) & 85.7 & 96.2 & 94.6 & 94.1 & 92.8 \\
 \hline
   \hline
\end{tabular}
\begin{tablenotes}
\item[1]  Here, $\widehat{\tau}^{(U)}(t)$, $\widehat{\tau}^{(G)}(t)$, $\widehat{\tau}^{(O1)}(t)$, $\widehat{\tau}^{(O2)}(t)$, and $\widehat{\tau}^{(R)}(t)$ denote the estimators given by the uncorrected, gold-standard, ORC1, ORC2, RRC approach with 4 split points, respectively. The data generation process is given in Section 6, where the mediator follows a linear regression (1) and outcome follows the Cox model (2), with a linear measurement error process characterized in (4). The percent bias is defined as the ratio of bias to the true value over 2,000 replications, i.e., $mean(\frac{\widehat{\tau}-\tau}{\tau}) \times 100\%$.  The standard error (SE) is defined as the square root of empirical variance of mediation effect measure estimates from the 2,000 replications. The coverage rate is obtained based on a non-parametric bootstrap confidence interval.
\end{tablenotes}
\end{threeparttable}}
\end{table}

\begin{table}[ht]
\caption{Simulation results for a common outcome with a large exposure effect in the Cox outcome model ($\beta_1=1.5$). The simulation setting is identical to Web Table 1, except that a large exposure effect ($\beta_1=1.5$) is introduced in the Cox model. Upper panel: performance of the NIE estimator; middle panel: performance of the NDE estimator; lower panel: performance of the MP estimator.}
\centering
\scalebox{0.66}[0.66]{
\begin{threeparttable}
\label{tab:sim_exact}
\setlength\tabcolsep{4pt}
\begin{tabular}{cc ccccc ccccc }
  \hline
& & \multicolumn{5}{c}{Percent Bias (SE$\times$100)} &   \multicolumn{5}{c}{Bootstrap Coverage Rate} \\
\cmidrule(lr){3-7} \cmidrule(lr){8-12}
$t$ &   NIE$(t)$  & 
 $\widehat{\text{NIE}}^{(U)}(t)$ &  $\widehat{\text{NIE}}^{(G)}(t)$ & $\widehat{\text{NIE}}^{(O1)}(t)$ & $\widehat{\text{NIE}}^{(O2)}(t)$ & $\widehat{\text{NIE}}^{(R)}(t)$ &
 $\widehat{\text{NIE}}^{(U)}(t)$ &  $\widehat{\text{NIE}}^{(G)}(t)$ & $\widehat{\text{NIE}}^{(O1)}(t)$ & $\widehat{\text{NIE}}^{(O2)}(t)$ & $\widehat{\text{NIE}}^{(R)}(t)$ \\ 
  \hline
10 & log(1.129) & -47.0 (0.8) & -0.6 (0.9) & -28.9 (1.8) & -28.6 (1.2) & -16.0 (3.0) & 0.0 & 94.4 & 57.9 & 35.2 & 89.4 \\ 
  15 & log(1.129) & -47.0 (0.8) & -0.6 (0.9) & -28.9 (1.8) & -28.6 (1.2) & -16.0 (3.0) & 0.0 & 94.4 & 57.9 & 35.2 & 89.4 \\ 
  20 & log(1.129) & -46.9 (0.8) & -0.7 (0.9) & -28.8 (1.8) & -28.5 (1.2) & -16.1 (2.9) & 0.0 & 94.3 & 57.8 & 35.2 & 89.3 \\ 
  25 & log(1.127) & -46.6 (0.8) & -0.7 (0.9) & -28.4 (1.7) & -28.1 (1.2) & -16.2 (2.8) & 0.0 & 94.3 & 57.8 & 35.5 & 88.7 \\ 
  30 & log(1.123) & -46.0 (0.7) & -0.6 (0.9) & -27.5 (1.6) & -27.2 (1.2) & -16.5 (2.5) & 0.0 & 94.3 & 58.2 & 36.9 & 87.6 \\ 
  35 & log(1.116) & -44.6 (0.7) & -0.6 (0.8) & -25.4 (1.5) & -25.3 (1.1) & -16.5 (2.1) & 0.0 & 94.4 & 60.1 & 41.0 & 85.1 \\ 
  40 & log(1.106) & -42.2 (0.7) & -0.5 (0.7) & -22.1 (1.3) & -22.2 (1.0) & -16.4 (1.8) & 0.0 & 94.1 & 64.2 & 49.4 & 82.6 \\ 
  45 & log(1.093) & -39.0 (0.6) & -0.3 (0.6) & -17.9 (1.2) & -17.9 (0.9) & -15.1 (1.5) & 0.1 & 94.6 & 72.7 & 62.9 & 82.8 \\ 
   \hline
   \hline
 & & \multicolumn{5}{c}{Percent Bias (SE$\times$100)} &   \multicolumn{5}{c}{Bootstrap Coverage Rate} \\
\cmidrule(lr){3-7} \cmidrule(lr){8-12}
$t$ &   NDE$(t)$ & 
 $\widehat{\text{NDE}}^{(U)}(t)$ &  $\widehat{\text{NDE}}^{(G)}(t)$ & $\widehat{\text{NDE}}^{(O1)}(t)$ & $\widehat{\text{NDE}}^{(O2)}(t)$ & $\widehat{\text{NDE}}^{(R)}(t)$ &
 $\widehat{\text{NDE}}^{(U)}(t)$ &  $\widehat{\text{NDE}}^{(G)}(t)$ & $\widehat{\text{NDE}}^{(O1)}(t)$ & $\widehat{\text{NDE}}^{(O2)}(t)$ & $\widehat{\text{NDE}}^{(R)}(t)$  \\ 
  \hline
10 & log(4.620) & -53.9 (2.6) & -0.3 (3.1) & -19.5 (11.9) & -19.6 (12.2) & 2.3 (26.8) & 0.0 & 94.9 & 50.3 & 51.5 & 94.3 \\ 
  15 & log(4.617) & -53.9 (2.6) & -0.3 (3.1) & -19.5 (11.9) & -19.6 (12.2) & 2.3 (26.7) & 0.0 & 95.0 & 50.5 & 51.5 & 94.2 \\ 
  20 & log(4.604) & -53.9 (2.6) & -0.3 (3.1) & -19.4 (11.9) & -19.5 (12.1) & 2.3 (26.6) & 0.0 & 94.8 & 50.6 & 51.8 & 94.5 \\ 
  25 & log(4.561) & -53.9 (2.6) & -0.3 (3.0) & -19.3 (11.8) & -19.3 (12.1) & 2.1 (26.3) & 0.0 & 95.0 & 51.1 & 52.4 & 94.7 \\ 
  30 & log(4.454) & -53.7 (2.5) & -0.3 (3.0) & -18.8 (11.8) & -18.8 (12.0) & 1.7 (25.7) & 0.0 & 95.0 & 52.5 & 54.6 & 94.6 \\ 
  35 & log(4.242) & -53.3 (2.5) & -0.3 (2.9) & -17.6 (11.7) & -17.5 (12.0) & 1.4 (24.9) & 0.0 & 95.3 & 57.1 & 59.6 & 94.7 \\ 
  40 & log(3.911) & -52.3 (2.4) & -0.3 (3.0) & -15.4 (11.7) & -15.2 (11.9) & 1.6 (24.3) & 0.0 & 94.8 & 68.2 & 68.8 & 94.7 \\ 
  45 & log(3.495) & -50.6 (2.6) & -0.2 (3.4) & -11.5 (11.8) & -11.4 (11.9) & 2.8 (24.2) & 0.0 & 94.6 & 82.0 & 81.8 & 94.5 \\ 
   \hline
   \hline
    & & \multicolumn{5}{c}{Percent Bias (SE$\times$100)} &   \multicolumn{5}{c}{Bootstrap Coverage Rate} \\
\cmidrule(lr){3-7} \cmidrule(lr){8-12}
$t$ &   MP$(t)$  & 
 $\widehat{\text{MP}}^{(U)}(t)$ &  $\widehat{\text{MP}}^{(G)}(t)$ & $\widehat{\text{MP}}^{(O1)}(t)$ & $\widehat{\text{MP}}^{(O2)}(t)$ & $\widehat{\text{MP}}^{(R)}(t)$ &
 $\widehat{\text{MP}}^{(U)}(t)$ &  $\widehat{\text{MP}}^{(G)}(t)$ & $\widehat{\text{MP}}^{(O1)}(t)$ & $\widehat{\text{MP}}^{(O2)}(t)$ & $\widehat{\text{MP}}^{(R)}(t)$  \\ 
  \hline
 10 & 0.074 & 13.8 (0.9) & -0.4 (0.5) & -11.0 (1.2) & -10.7 (0.9) & -15.4 (1.7) & 82.3 & 94.7 & 90.5 & 88.1 & 89.2 \\ 
  15 & 0.074 & 13.9 (0.9) & -0.4 (0.5) & -11.0 (1.2) & -10.7 (0.9) & -15.4 (1.7) & 82.3 & 94.7 & 90.5 & 88.2 & 89.2 \\ 
  20 & 0.073 & 14.0 (0.9) & -0.4 (0.5) & -10.9 (1.2) & -10.6 (0.9) & -15.5 (1.7) & 82.2 & 94.7 & 90.6 & 88.3 & 88.8 \\ 
  25 & 0.073 & 14.3 (0.9) & -0.4 (0.5) & -10.8 (1.2) & -10.4 (0.9) & -15.6 (1.7) & 81.8 & 94.6 & 90.7 & 88.5 & 88.6 \\ 
  30 & 0.072 & 15.3 (0.9) & -0.4 (0.5) & -10.3 (1.2) & -9.9 (0.9) & -15.9 (1.7) & 79.8 & 94.5 & 91.3 & 89.3 & 88.1 \\ 
  35 & 0.071 & 17.0 (0.9) & -0.3 (0.5) & -9.4 (1.2) & -9.0 (0.9) & -16.3 (1.6) & 76.3 & 94.5 & 91.7 & 90.1 & 87.6 \\ 
  40 & 0.069 & 19.3 (0.9) & -0.3 (0.5) & -8.2 (1.1) & -7.8 (0.9) & -16.3 (1.6) & 72.0 & 94.7 & 92.2 & 91.8 & 88.0 \\ 
  45 & 0.067 & 21.2 (0.9) & -0.3 (0.5) & -7.5 (1.1) & -7.2 (0.9) & -15.9 (1.6) & 68.3 & 94.7 & 92.5 & 92.3 & 89.2 \\ 
 \hline
   \hline
\end{tabular}
\begin{tablenotes}
\item[1] Here, $\widehat{\tau}^{(U)}(t)$, $\widehat{\tau}^{(G)}(t)$, $\widehat{\tau}^{(O1)}(t)$, $\widehat{\tau}^{(O2)}(t)$, and $\widehat{\tau}^{(R)}(t)$ denote the estimators given by the uncorrected, gold-standard, ORC1, ORC2, RRC approach with 4 split points, respectively. The data generation process is given in Section 6, where the mediator follows a linear regression (1) and outcome follows the Cox model (2), with a linear measurement error process characterized in (4). The percent bias is defined as the ratio of bias to the true value over 2,000 replications, i.e., $mean(\frac{\widehat{\tau}-\tau}{\tau}) \times 100\%$.  The standard error (SE) is defined as the square root of empirical variance of mediation effect measure estimates from the 2,000 replications. The coverage rate is obtained based on a non-parametric bootstrap confidence interval.
\end{tablenotes}
\end{threeparttable}}
\end{table}

\begin{table}[htbp]
\caption{Simulation results with a constant baseline hazard. Here, $\widehat{\tau}^{(U)}(t)$, $\widehat{\tau}^{(G)}(t)$, $\widehat{\tau}^{(O1)}(t)$, $\widehat{\tau}^{(O2)}(t)$, and $\widehat{\tau}^{(R)}(t)$ denote the estimators given by the uncorrected, gold-standard, ORC1, ORC2, and RRC approaches, respectively, where the RRC uses 4 split points to determine the validation study risk sets. Upper panel: performance of the NIE estimator; middle panel: performance of the NDE estimator; lower panel: performance of the MP estimator.}
\centering
\scalebox{0.66}[0.66]{
\begin{threeparttable}
\label{tab:sim_exact_constant}
\setlength\tabcolsep{4pt}
\begin{tabular}{cc ccccc ccccc }
  \hline
& & \multicolumn{5}{c}{Percent Bias (SE$\times$100)} &   \multicolumn{5}{c}{Bootstrap Coverage Rate} \\
\cmidrule(lr){3-7} \cmidrule(lr){8-12}
$t$ &   NIE$(t)$  & 
 $\widehat{\text{NIE}}^{(U)}(t)$ &  $\widehat{\text{NIE}}^{(G)}(t)$ & $\widehat{\text{NIE}}^{(O1)}(t)$ & $\widehat{\text{NIE}}^{(O2)}(t)$ & $\widehat{\text{NIE}}^{(R)}(t)$ &
 $\widehat{\text{NIE}}^{(U)}(t)$ &  $\widehat{\text{NIE}}^{(G)}(t)$ & $\widehat{\text{NIE}}^{(O1)}(t)$ & $\widehat{\text{NIE}}^{(O2)}(t)$ & $\widehat{\text{NIE}}^{(R)}(t)$ \\ 
  \hline
10 & log(1.123) & -42.0 (0.7) & 0.1 (0.8) & -1.1 (1.7) & -1.1 (1.6) & -0.6 (1.7) & 0.0 & 95.7 & 95.1 & 95.2 & 95.3 \\ 
  15 & log(1.120) & -41.8 (0.7) & 0.2 (0.8) & -1.0 (1.6) & -1.1 (1.5) & -0.6 (1.6) & 0.0 & 95.5 & 95.2 & 95.3 & 95.2 \\ 
  20 & log(1.118) & -41.7 (0.7) & 0.2 (0.8) & -1.0 (1.6) & -1.1 (1.5) & -0.6 (1.6) & 0.0 & 95.6 & 95.0 & 95.2 & 95.1 \\ 
  25 & log(1.115) & -41.6 (0.7) & 0.2 (0.8) & -1.0 (1.5) & -1.1 (1.5) & -0.7 (1.5) & 0.0 & 95.8 & 94.9 & 95.0 & 95.0 \\ 
  30 & log(1.113) & -41.4 (0.7) & 0.2 (0.7) & -1.0 (1.5) & -1.0 (1.4) & -0.7 (1.5) & 0.0 & 95.8 & 94.9 & 95.0 & 95.2 \\ 
  35 & log(1.111) & -41.3 (0.7) & 0.1 (0.7) & -0.9 (1.4) & -1.0 (1.4) & -0.6 (1.5) & 0.0 & 95.8 & 95.0 & 95.0 & 95.0 \\ 
  40 & log(1.110) & -41.3 (0.7) & 0.1 (0.7) & -1.0 (1.4) & -1.0 (1.4) & -0.7 (1.4) & 0.0 & 95.6 & 95.0 & 95.0 & 95.1 \\ 
  45 & log(1.108) & -41.2 (0.6) & 0.1 (0.7) & -0.9 (1.4) & -0.9 (1.4) & -0.7 (1.4) & 0.0 & 95.8 & 95.0 & 95.0 & 95.0 \\ 
   \hline
   \hline
 & & \multicolumn{5}{c}{Percent Bias (SE$\times$100)} &   \multicolumn{5}{c}{Bootstrap Coverage Rate} \\
\cmidrule(lr){3-7} \cmidrule(lr){8-12}
$t$ &   NDE$(t)$ & 
 $\widehat{\text{NDE}}^{(U)}(t)$ &  $\widehat{\text{NDE}}^{(G)}(t)$ & $\widehat{\text{NDE}}^{(O1)}(t)$ & $\widehat{\text{NDE}}^{(O2)}(t)$ & $\widehat{\text{NDE}}^{(R)}(t)$ &
 $\widehat{\text{NDE}}^{(U)}(t)$ &  $\widehat{\text{NDE}}^{(G)}(t)$ & $\widehat{\text{NDE}}^{(O1)}(t)$ & $\widehat{\text{NDE}}^{(O2)}(t)$ & $\widehat{\text{NDE}}^{(R)}(t)$  \\ 
  \hline
 10 & log(1.301) & -42.8 (2.2) & 0.3 (2.2) & -0.6 (4.5) & -0.8 (4.5) & -1.1 (4.6) & 0.5 & 95.0 & 95.1 & 95.3 & 95.7 \\ 
  15 & log(1.290) & -42.7 (2.1) & 0.3 (2.2) & -0.3 (4.4) & -0.6 (4.5) & -1.1 (4.6) & 0.6 & 95.3 & 95.5 & 95.4 & 95.8 \\ 
  20 & log(1.279) & -42.7 (2.1) & 0.4 (2.2) & -0.3 (4.4) & -0.4 (4.4) & -0.9 (4.5) & 0.6 & 95.1 & 95.5 & 95.5 & 95.7 \\ 
  25 & log(1.269) & -42.8 (2.1) & 0.4 (2.2) & -0.0 (4.4) & -0.1 (4.4) & -0.6 (4.5) & 0.9 & 95.0 & 96.0 & 95.7 & 95.8 \\ 
  30 & log(1.260) & -42.8 (2.1) & 0.4 (2.3) & 0.2 (4.4) & 0.2 (4.4) & -0.3 (4.5) & 1.8 & 95.3 & 96.0 & 95.7 & 95.5 \\ 
  35 & log(1.251) & -42.9 (2.2) & 0.5 (2.3) & 0.4 (4.4) & 0.4 (4.4) & -0.4 (4.5) & 2.6 & 95.3 & 95.6 & 95.5 & 95.5 \\ 
  40 & log(1.243) & -43.0 (2.2) & 0.4 (2.3) & 0.7 (4.4) & 0.5 (4.4) & -0.3 (4.5) & 3.8 & 95.5 & 95.5 & 95.5 & 95.5 \\ 
  45 & log(1.235) & -43.1 (2.2) & 0.4 (2.3) & 0.7 (4.4) & 0.8 (4.4) & -0.1 (4.5) & 4.9 & 95.2 & 95.4 & 95.5 & 95.7 \\ 
   \hline
   \hline
    & & \multicolumn{5}{c}{Percent Bias (SE$\times$100)} &   \multicolumn{5}{c}{Bootstrap Coverage Rate} \\
\cmidrule(lr){3-7} \cmidrule(lr){8-12}
$t$ &   MP$(t)$  & 
 $\widehat{\text{MP}}^{(U)}(t)$ &  $\widehat{\text{MP}}^{(G)}(t)$ & $\widehat{\text{MP}}^{(O1)}(t)$ & $\widehat{\text{MP}}^{(O2)}(t)$ & $\widehat{\text{MP}}^{(R)}(t)$ &
 $\widehat{\text{MP}}^{(U)}(t)$ &  $\widehat{\text{MP}}^{(G)}(t)$ & $\widehat{\text{MP}}^{(O1)}(t)$ & $\widehat{\text{MP}}^{(O2)}(t)$ & $\widehat{\text{MP}}^{(R)}(t)$  \\ 
  \hline
  10 & 0.306 & 1.2 (4.1) & 0.0 (2.5) & -0.5 (4.2) & -0.6 (4.2) & -0.1 (4.3) & 95.0 & 94.8 & 95.0 & 95.3 & 96.0 \\ 
  15 & 0.308 & 1.4 (4.2) & -0.0 (2.6) & -0.6 (4.4) & -0.7 (4.3) & -0.0 (4.5) & 95.0 & 94.7 & 94.9 & 95.3 & 96.0 \\ 
  20 & 0.311 & 1.7 (4.4) & -0.1 (2.7) & -0.6 (4.5) & -0.8 (4.5) & -0.2 (4.6) & 95.0 & 94.5 & 95.2 & 95.3 & 95.8 \\ 
  25 & 0.314 & 1.7 (4.5) & -0.1 (2.8) & -0.8 (4.7) & -1.0 (4.6) & -0.2 (4.8) & 95.0 & 94.5 & 95.3 & 95.5 & 95.7 \\ 
  30 & 0.317 & 1.9 (4.7) & -0.0 (2.9) & -0.8 (4.9) & -1.2 (4.8) & -0.2 (5.0) & 95.2 & 94.7 & 95.5 & 95.7 & 95.7 \\ 
  35 & 0.320 & 2.0 (4.9) & -0.1 (3.0) & -0.8 (5.1) & -1.3 (5.0) & -0.4 (5.2) & 95.4 & 94.8 & 95.5 & 95.8 & 95.5 \\ 
  40 & 0.324 & 2.2 (5.1) & -0.2 (3.1) & -1.0 (5.2) & -1.2 (5.2) & -0.4 (5.4) & 95.6 & 94.8 & 95.4 & 95.8 & 95.6 \\ 
  45 & 0.327 & 2.2 (5.4) & -0.2 (3.2) & -1.1 (5.4) & -1.2 (5.3) & -0.5 (5.6) & 95.7 & 94.9 & 95.5 & 95.8 & 95.7 \\ 
 \hline
   \hline
\end{tabular}
\begin{tablenotes}
\item[1] The data generation process is given in Section 6, where the mediator follows a linear regression (1) and outcome follows the Cox model (2), with a linear measurement error process characterized in (4). In this simulation, we considered a moderate sample size of the validation study ($n_2=250$) and a moderate exposure measurement error ($\rho_{AA^*}=0.6$). The percent bias is defined as the ratio of bias to the true value over 2,000 replications, i.e., $mean(\frac{\widehat{\tau}-\tau}{\tau}) \times 100\%$. The standard error (SE) is defined as the square root of empirical variance of mediation effect measure estimates from the 2,000 replications. The coverage rate is obtained based on a non-parametric bootstrap confidence interval.
\end{tablenotes}
\end{threeparttable}}
\end{table}

\begin{table}[ht]
\caption{Simulation results with a constant baseline hazard and a large exposure effect in the Cox outcome model ($\beta_1=1$). The simulation setting is identical to Web Table 4, except that a large exposure effect ($\beta_1=1$) is introduced in the Cox model. Upper panel: performance of the NIE estimator; middle panel: performance of the NDE estimator; lower panel: performance of the MP estimator.}
\centering
\scalebox{0.66}[0.66]{
\begin{threeparttable}
\label{tab:sim_exact_constant_large1}
\setlength\tabcolsep{4pt}
\begin{tabular}{cc ccccc ccccc }
  \hline
& & \multicolumn{5}{c}{Percent Bias (SE$\times$100)} &   \multicolumn{5}{c}{Bootstrap Coverage Rate} \\
\cmidrule(lr){3-7} \cmidrule(lr){8-12}
$t$ &   NIE$(t)$  & 
 $\widehat{\text{NIE}}^{(U)}(t)$ &  $\widehat{\text{NIE}}^{(G)}(t)$ & $\widehat{\text{NIE}}^{(O1)}(t)$ & $\widehat{\text{NIE}}^{(O2)}(t)$ & $\widehat{\text{NIE}}^{(R)}(t)$ &
 $\widehat{\text{NIE}}^{(U)}(t)$ &  $\widehat{\text{NIE}}^{(G)}(t)$ & $\widehat{\text{NIE}}^{(O1)}(t)$ & $\widehat{\text{NIE}}^{(O2)}(t)$ & $\widehat{\text{NIE}}^{(R)}(t)$ \\ 
  \hline
10 & log(1.117) & -41.5 (0.7) & -0.4 (0.8) & -13.9 (1.4) & -13.6 (1.2) & -11.4 (1.5) & 0.1 & 95.3 & 84.7 & 79.5 & 88.8 \\ 
  15 & log(1.113) & -40.9 (0.7) & -0.4 (0.7) & -13.1 (1.4) & -12.7 (1.2) & -10.9 (1.4) & 0.1 & 95.1 & 84.8 & 81.2 & 89.0 \\ 
  20 & log(1.109) & -40.3 (0.7) & -0.4 (0.7) & -12.4 (1.3) & -11.9 (1.1) & -10.6 (1.3) & 0.1 & 95.3 & 85.5 & 82.2 & 89.0 \\ 
  25 & log(1.105) & -39.8 (0.7) & -0.3 (0.7) & -11.7 (1.3) & -11.2 (1.1) & -10.2 (1.3) & 0.1 & 95.2 & 86.7 & 83.3 & 89.0 \\ 
  30 & log(1.103) & -39.3 (0.6) & -0.3 (0.6) & -11.1 (1.2) & -10.6 (1.0) & -9.9 (1.2) & 0.1 & 95.1 & 87.3 & 84.0 & 89.4 \\ 
  35 & log(1.100) & -38.9 (0.6) & -0.3 (0.6) & -10.5 (1.2) & -9.9 (1.0) & -9.6 (1.2) & 0.1 & 95.2 & 88.0 & 85.0 & 89.9 \\ 
  40 & log(1.098) & -38.6 (0.6) & -0.3 (0.6) & -10.0 (1.1) & -9.4 (1.0) & -9.4 (1.1) & 0.1 & 95.2 & 88.6 & 86.3 & 90.1 \\ 
  45 & log(1.095) & -38.2 (0.6) & -0.3 (0.6) & -9.6 (1.1) & -9.0 (1.0) & -9.3 (1.1) & 0.1 & 95.2 & 89.0 & 87.1 & 90.1 \\ 
   \hline
   \hline
 & & \multicolumn{5}{c}{Percent Bias (SE$\times$100)} &   \multicolumn{5}{c}{Bootstrap Coverage Rate} \\
\cmidrule(lr){3-7} \cmidrule(lr){8-12}
$t$ &   NDE$(t)$ & 
 $\widehat{\text{NDE}}^{(U)}(t)$ &  $\widehat{\text{NDE}}^{(G)}(t)$ & $\widehat{\text{NDE}}^{(O1)}(t)$ & $\widehat{\text{NDE}}^{(O2)}(t)$ & $\widehat{\text{NDE}}^{(R)}(t)$ &
 $\widehat{\text{NDE}}^{(U)}(t)$ &  $\widehat{\text{NDE}}^{(G)}(t)$ & $\widehat{\text{NDE}}^{(O1)}(t)$ & $\widehat{\text{NDE}}^{(O2)}(t)$ & $\widehat{\text{NDE}}^{(R)}(t)$  \\ 
  \hline
 10 & log(2.620) & -47.8 (2.2) & -0.2 (2.4) & -8.9 (8.7) & -8.8 (8.9) & -6.2 (9.6) & 0.0 & 96.0 & 86.0 & 87.2 & 90.2 \\ 
  15 & log(2.547) & -47.6 (2.1) & -0.2 (2.4) & -8.1 (8.7) & -8.1 (8.8) & -5.7 (9.5) & 0.0 & 96.2 & 87.5 & 88.6 & 91.5 \\ 
  20 & log(2.484) & -47.3 (2.2) & -0.2 (2.4) & -7.4 (8.6) & -7.4 (8.7) & -5.1 (9.5) & 0.0 & 95.8 & 89.3 & 89.8 & 91.8 \\ 
  25 & log(2.427) & -47.0 (2.2) & -0.2 (2.5) & -6.7 (8.6) & -6.6 (8.7) & -4.6 (9.5) & 0.0 & 95.8 & 90.5 & 91.1 & 92.5 \\ 
  30 & log(2.377) & -46.7 (2.3) & -0.1 (2.6) & -6.0 (8.6) & -5.9 (8.7) & -4.0 (9.5) & 0.0 & 96.0 & 91.6 & 92.0 & 93.2 \\ 
  35 & log(2.332) & -46.4 (2.3) & -0.1 (2.7) & -5.2 (8.6) & -5.1 (8.7) & -3.5 (9.5) & 0.0 & 95.9 & 92.5 & 92.7 & 93.7 \\ 
  40 & log(2.290) & -46.2 (2.4) & -0.1 (2.7) & -4.6 (8.7) & -4.4 (8.6) & -3.1 (9.5) & 0.0 & 96.0 & 93.3 & 93.6 & 94.0 \\ 
  45 & log(2.252) & -45.8 (2.4) & -0.2 (2.8) & -4.0 (8.7) & -3.7 (8.7) & -2.6 (9.5) & 0.0 & 95.8 & 94.0 & 93.9 & 94.5 \\ 
   \hline
   \hline
    & & \multicolumn{5}{c}{Percent Bias (SE$\times$100)} &   \multicolumn{5}{c}{Bootstrap Coverage Rate} \\
\cmidrule(lr){3-7} \cmidrule(lr){8-12}
$t$ &   MP$(t)$  & 
 $\widehat{\text{MP}}^{(U)}(t)$ &  $\widehat{\text{MP}}^{(G)}(t)$ & $\widehat{\text{MP}}^{(O1)}(t)$ & $\widehat{\text{MP}}^{(O2)}(t)$ & $\widehat{\text{MP}}^{(R)}(t)$ &
 $\widehat{\text{MP}}^{(U)}(t)$ &  $\widehat{\text{MP}}^{(G)}(t)$ & $\widehat{\text{MP}}^{(O1)}(t)$ & $\widehat{\text{MP}}^{(O2)}(t)$ & $\widehat{\text{MP}}^{(R)}(t)$  \\ 
  \hline
 10 & 0.103 & 11.1 (1.2) & -0.1 (0.7) & -4.7 (1.4) & -4.9 (1.2) & -4.8 (1.5) & 86.7 & 95.5 & 94.2 & 94.0 & 94.5 \\ 
  15 & 0.102 & 11.4 (1.2) & -0.2 (0.7) & -4.6 (1.4) & -4.9 (1.2) & -5.0 (1.5) & 86.1 & 95.6 & 94.2 & 93.8 & 94.4 \\ 
  20 & 0.102 & 11.9 (1.3) & -0.1 (0.7) & -4.6 (1.4) & -4.9 (1.2) & -5.1 (1.5) & 85.4 & 95.7 & 94.2 & 93.8 & 94.3 \\ 
  25 & 0.102 & 12.2 (1.3) & -0.1 (0.7) & -4.6 (1.4) & -4.9 (1.2) & -5.2 (1.5) & 85.0 & 95.8 & 94.0 & 93.8 & 94.3 \\ 
  30 & 0.101 & 12.4 (1.3) & -0.1 (0.7) & -4.6 (1.4) & -5.0 (1.2) & -5.4 (1.5) & 85.0 & 95.5 & 94.0 & 93.8 & 94.2 \\ 
  35 & 0.101 & 12.6 (1.3) & -0.1 (0.7) & -4.8 (1.4) & -5.1 (1.2) & -5.6 (1.5) & 84.7 & 95.5 & 94.0 & 93.9 & 94.1 \\ 
  40 & 0.101 & 12.7 (1.3) & -0.1 (0.7) & -4.9 (1.4) & -5.3 (1.2) & -5.8 (1.5) & 84.5 & 95.5 & 94.0 & 93.8 & 94.0 \\ 
  45 & 0.101 & 12.7 (1.3) & -0.1 (0.8) & -5.1 (1.4) & -5.5 (1.2) & -6.0 (1.5) & 84.7 & 95.6 & 93.9 & 93.7 & 94.0 \\ 
 \hline
   \hline
\end{tabular}
\begin{tablenotes}
\item[1]  Here, $\widehat{\tau}^{(U)}(t)$, $\widehat{\tau}^{(G)}(t)$, $\widehat{\tau}^{(O1)}(t)$, $\widehat{\tau}^{(O2)}(t)$, and $\widehat{\tau}^{(R)}(t)$ denote the estimators given by the uncorrected, gold-standard, ORC1, ORC2, RRC approach with 4 split points, respectively. The data generation process is given in Section 6, where the mediator follows a linear regression (1) and outcome follows the Cox model (2), with a linear measurement error process characterized in (4). The percent bias is defined as the ratio of bias to the true value over 2,000 replications, i.e., $mean(\frac{\widehat{\tau}-\tau}{\tau}) \times 100\%$.  The standard error (SE) is defined as the square root of empirical variance of mediation effect measure estimates from the 2,000 replications. The coverage rate is obtained based on a non-parametric bootstrap confidence interval.
\end{tablenotes}
\end{threeparttable}}
\end{table}

\begin{table}[ht]
\caption{Simulation results with a constant baseline hazard and a large exposure effect in the Cox outcome model ($\beta_1=1.5$). The simulation setting is identical to Web Table 4, except that a large exposure effect ($\beta_1=1.5$) is introduced in the Cox model. Upper panel: performance of the NIE estimator; middle panel: performance of the NDE estimator; lower panel: performance of the MP estimator.}
\centering
\scalebox{0.66}[0.66]{
\begin{threeparttable}
\label{tab:sim_exact_constant_large2}
\setlength\tabcolsep{4pt}
\begin{tabular}{cc ccccc ccccc }
  \hline
& & \multicolumn{5}{c}{Percent Bias (SE$\times$100)} &   \multicolumn{5}{c}{Bootstrap Coverage Rate} \\
\cmidrule(lr){3-7} \cmidrule(lr){8-12}
$t$ &   NIE$(t)$  & 
 $\widehat{\text{NIE}}^{(U)}(t)$ &  $\widehat{\text{NIE}}^{(G)}(t)$ & $\widehat{\text{NIE}}^{(O1)}(t)$ & $\widehat{\text{NIE}}^{(O2)}(t)$ & $\widehat{\text{NIE}}^{(R)}(t)$ &
 $\widehat{\text{NIE}}^{(U)}(t)$ &  $\widehat{\text{NIE}}^{(G)}(t)$ & $\widehat{\text{NIE}}^{(O1)}(t)$ & $\widehat{\text{NIE}}^{(O2)}(t)$ & $\widehat{\text{NIE}}^{(R)}(t)$ \\ 
  \hline
10 & log(1.111) & -42.7 (0.7) & -0.2 (0.7) & -23.6 (1.4) & -23.6 (1.0) & -21.2 (1.4) & 0.0 & 95.7 & 61.8 & 41.2 & 69.5 \\ 
  15 & log(1.106) & -41.4 (0.6) & -0.1 (0.7) & -21.7 (1.3) & -21.9 (1.0) & -19.8 (1.3) & 0.0 & 95.7 & 64.3 & 46.6 & 71.3 \\ 
  20 & log(1.101) & -40.3 (0.6) & -0.0 (0.6) & -20.3 (1.2) & -20.4 (0.9) & -18.9 (1.2) & 0.1 & 95.3 & 67.2 & 51.7 & 72.5 \\ 
  25 & log(1.097) & -39.2 (0.6) & -0.0 (0.6) & -19.0 (1.2) & -19.0 (0.9) & -18.0 (1.2) & 0.1 & 95.5 & 69.8 & 56.5 & 74.0 \\ 
  30 & log(1.094) & -38.4 (0.6) & -0.1 (0.6) & -17.7 (1.1) & -17.8 (0.9) & -17.3 (1.1) & 0.2 & 95.6 & 72.8 & 60.2 & 74.9 \\ 
  35 & log(1.091) & -37.5 (0.6) & -0.1 (0.5) & -16.7 (1.1) & -16.7 (0.8) & -16.7 (1.1) & 0.2 & 95.6 & 74.9 & 63.8 & 76.1 \\ 
  40 & log(1.088) & -36.8 (0.6) & -0.1 (0.5) & -15.7 (1.1) & -15.7 (0.8) & -16.1 (1.0) & 0.2 & 95.7 & 77.5 & 66.8 & 76.7 \\ 
  45 & log(1.086) & -36.2 (0.6) & -0.1 (0.5) & -14.8 (1.1) & -14.8 (0.8) & -15.7 (1.0) & 0.3 & 95.7 & 79.3 & 70.0 & 77.5 \\ 
   \hline
   \hline
 & & \multicolumn{5}{c}{Percent Bias (SE$\times$100)} &   \multicolumn{5}{c}{Bootstrap Coverage Rate} \\
\cmidrule(lr){3-7} \cmidrule(lr){8-12}
$t$ &   NDE$(t)$ & 
 $\widehat{\text{NDE}}^{(U)}(t)$ &  $\widehat{\text{NDE}}^{(G)}(t)$ & $\widehat{\text{NDE}}^{(O1)}(t)$ & $\widehat{\text{NDE}}^{(O2)}(t)$ & $\widehat{\text{NDE}}^{(R)}(t)$ &
 $\widehat{\text{NDE}}^{(U)}(t)$ &  $\widehat{\text{NDE}}^{(G)}(t)$ & $\widehat{\text{NDE}}^{(O1)}(t)$ & $\widehat{\text{NDE}}^{(O2)}(t)$ & $\widehat{\text{NDE}}^{(R)}(t)$  \\ 
  \hline
10 & log(4.099) & -52.2 (2.2) & -0.1 (2.7) & -15.9 (11.7) & -15.7 (12.0) & -10.7 (13.2) & 0.0 & 94.7 & 65.5 & 66.7 & 80.4 \\ 
  15 & log(3.913) & -51.6 (2.3) & -0.1 (2.8) & -14.5 (11.7) & -14.3 (12.0) & -9.5 (13.2) & 0.0 & 94.9 & 71.3 & 72.8 & 84.2 \\ 
  20 & log(3.757) & -51.0 (2.3) & -0.0 (2.9) & -13.2 (11.7) & -13.0 (12.0) & -8.6 (13.2) & 0.0 & 94.8 & 75.9 & 77.8 & 86.5 \\ 
  25 & log(3.623) & -50.4 (2.3) & -0.0 (3.0) & -11.9 (11.7) & -11.8 (12.0) & -7.5 (13.3) & 0.0 & 94.8 & 80.0 & 81.7 & 89.0 \\ 
  30 & log(3.507) & -49.9 (2.4) & -0.0 (3.1) & -10.7 (11.8) & -10.7 (12.0) & -6.6 (13.3) & 0.0 & 95.2 & 83.2 & 84.2 & 90.7 \\ 
  35 & log(3.405) & -49.4 (2.5) & 0.0 (3.2) & -9.5 (11.8) & -9.6 (12.0) & -5.6 (13.4) & 0.0 & 95.2 & 85.8 & 86.8 & 91.8 \\ 
  40 & log(3.315) & -48.9 (2.5) & 0.0 (3.2) & -8.5 (11.9) & -8.4 (12.0) & -4.8 (13.4) & 0.0 & 95.1 & 88.2 & 88.8 & 92.9 \\ 
  45 & log(3.234) & -48.4 (2.6) & 0.0 (3.3) & -7.4 (11.9) & -7.3 (12.0) & -4.1 (13.5) & 0.0 & 95.3 & 90.5 & 90.7 & 93.8 \\  
   \hline
   \hline
    & & \multicolumn{5}{c}{Percent Bias (SE$\times$100)} &   \multicolumn{5}{c}{Bootstrap Coverage Rate} \\
\cmidrule(lr){3-7} \cmidrule(lr){8-12}
$t$ &   MP$(t)$  & 
 $\widehat{\text{MP}}^{(U)}(t)$ &  $\widehat{\text{MP}}^{(G)}(t)$ & $\widehat{\text{MP}}^{(O1)}(t)$ & $\widehat{\text{MP}}^{(O2)}(t)$ & $\widehat{\text{MP}}^{(R)}(t)$ &
 $\widehat{\text{MP}}^{(U)}(t)$ &  $\widehat{\text{MP}}^{(G)}(t)$ & $\widehat{\text{MP}}^{(O1)}(t)$ & $\widehat{\text{MP}}^{(O2)}(t)$ & $\widehat{\text{MP}}^{(R)}(t)$  \\ 
  \hline
 10 & 0.070 & 18.2 (0.9) & -0.2 (0.5) & -8.7 (1.1) & -8.7 (0.9) & -10.6 (1.1) & 73.5 & 95.7 & 92.4 & 90.5 & 91.7 \\ 
  15 & 0.069 & 19.4 (0.9) & -0.2 (0.5) & -8.2 (1.1) & -8.1 (0.8) & -10.5 (1.1) & 71.2 & 95.8 & 92.7 & 91.0 & 91.7 \\ 
  20 & 0.068 & 20.2 (0.9) & -0.2 (0.5) & -7.9 (1.1) & -7.8 (0.8) & -10.4 (1.1) & 69.8 & 96.0 & 92.7 & 91.5 & 91.7 \\ 
  25 & 0.067 & 20.8 (0.9) & -0.2 (0.5) & -7.8 (1.1) & -7.6 (0.8) & -10.3 (1.1) & 68.8 & 96.0 & 93.0 & 91.7 & 91.8 \\ 
  30 & 0.067 & 21.1 (0.9) & -0.2 (0.5) & -7.7 (1.1) & -7.6 (0.8) & -10.7 (1.1) & 68.2 & 95.9 & 93.0 & 92.0 & 91.5 \\ 
  35 & 0.066 & 21.4 (0.9) & -0.2 (0.5) & -7.7 (1.1) & -7.6 (0.8) & -10.8 (1.1) & 67.8 & 96.0 & 93.0 & 92.0 & 91.5 \\ 
  40 & 0.066 & 21.4 (0.9) & -0.2 (0.5) & -7.8 (1.1) & -7.7 (0.8) & -11.1 (1.1) & 67.8 & 95.8 & 93.1 & 92.0 & 91.5 \\ 
  45 & 0.066 & 21.5 (0.9) & -0.2 (0.5) & -7.8 (1.1) & -7.8 (0.8) & -11.3 (1.1) & 67.7 & 95.9 & 93.1 & 92.0 & 91.4 \\ 
 \hline
   \hline
\end{tabular}
\begin{tablenotes}
\item[1] Here, $\widehat{\tau}^{(U)}(t)$, $\widehat{\tau}^{(G)}(t)$, $\widehat{\tau}^{(O1)}(t)$, $\widehat{\tau}^{(O2)}(t)$, and $\widehat{\tau}^{(R)}(t)$ denote the estimators given by the uncorrected, gold-standard, ORC1, ORC2, RRC approach with 4 split points, respectively. The data generation process is given in Section 6, where the mediator follows a linear regression (1) and outcome follows the Cox model (2), with a linear measurement error process characterized in (4). The percent bias is defined as the ratio of bias to the true value over 2,000 replications, i.e., $mean(\frac{\widehat{\tau}-\tau}{\tau}) \times 100\%$.  The standard error (SE) is defined as the square root of empirical variance of mediation effect measure estimates from the 2,000 replications. The coverage rate is obtained based on a non-parametric bootstrap confidence interval.
\end{tablenotes}
\end{threeparttable}}
\end{table}

\begin{table}[ht]
\caption{Simulation results when the error term in measurement error model (4) follows a skew normal distribution. For each mediation effect measure $\tau\in\{\text{NIE},\text{NDE},\text{MP}\}$, we use $\widehat{\tau}^{(U)}$, $\widehat{\tau}^{(G)}$, $\widehat{\tau}^{(O1)}$, $\widehat{\tau}^{(O2)}$, and $\widehat{\tau}^{(R)}$ to denote the estimator given by the uncorrected, gold-standard, ORC1, ORC2, and RRC approaches, respectively. Upper panel: performance of the NIE estimator; middle panel: performance of the NDE estimator; lower panel: performance of the MP estimator.}
\centering
\scalebox{0.72}[0.72]{
\begin{threeparttable}
\label{tab:sim_skew}
\setlength\tabcolsep{4pt}
\begin{tabular}{l rrrrr rrrrr }
  \hline
&  \multicolumn{5}{c}{Percent Bias (SE$\times$100)} &   \multicolumn{5}{c}{Empirical Coverage Rate v1/v2} \\
\cmidrule(lr){2-6} \cmidrule(lr){7-11}
  $\rho_{AA^*}$  & 
 $\widehat{\text{NIE}}^{(U)}$ &  $\widehat{\text{NIE}}^{(G)}$ & $\widehat{\text{NIE}}^{(O1)}$ & $\widehat{\text{NIE}}^{(O2)}$ & $\widehat{\text{NIE}}^{(R)}$ &
 $\widehat{\text{NIE}}^{(U)}$ &  $\widehat{\text{NIE}}^{(G)}$ & $\widehat{\text{NIE}}^{(O1)}$ & $\widehat{\text{NIE}}^{(O2)}$ & $\widehat{\text{NIE}}^{(R)}$ \\ 
  \hline
  0.4 & -61.3 (1.1) & 0.2 (1.9) & -0.5 (4.4) & -0.4 (4.3) & -0.3 (4.4) & 0.0/0.1 & 95.2/95.2 & 94.5/96.2 & 94.4/96.5 & 94.3/96.3 \\ 
    0.6 & -41.3 (1.4) & -0.8 (2.0) & 0.0 (3.3) & 0.2 (3.2) & -0.1 (3.3) & 9.7/14.6 & 94.2/94.8 & 94.0/95.0 & 94.2/95.5 & 94.0/95.3 \\ 
    0.8 & -20.8 (1.7) & -0.4 (2.0) & -0.2 (2.5) & 0.0 (2.5) & -0.3 (2.5) & 67.5/74.3 & 94.4/94.7 & 95.3/95.8 & 95.5/95.6 & 95.5/95.5 \\ 
   \hline
   \hline
  & \multicolumn{5}{c}{Percent Bias (SE$\times$100)} &   \multicolumn{5}{c}{Empirical Coverage Rate v1/v2} \\
\cmidrule(lr){2-6} \cmidrule(lr){7-11}
   $\rho_{AA^*}$  & 
 $\widehat{\text{NDE}}^{(U)}$ &  $\widehat{\text{NDE}}^{(G)}$ & $\widehat{\text{NDE}}^{(O1)}$ & $\widehat{\text{NDE}}^{(O2)}$ & $\widehat{\text{NDE}}^{(R)}$ &
 $\widehat{\text{NDE}}^{(U)}$ &  $\widehat{\text{NDE}}^{(G)}$ & $\widehat{\text{NDE}}^{(O1)}$ & $\widehat{\text{NDE}}^{(O2)}$ & $\widehat{\text{NDE}}^{(R)}$  \\ 
  \hline
  0.4 & -61.7 (7.4) & 0.2 (6.8) & -0.2 (21.2) & -1.1 (21.4) & -1.5 (21.6) & 35.6/44.0 & 96.0/95.4 & 94.8/95.7 & 94.7/95.7 & 94.5/96.0 \\ 
    0.6 & -40.7 (7.7) & 0.3 (7.1) & 3.4 (13.9) & 3.4 (14.0) & 2.7 (14.1) & 67.8/72.7 & 95.5/95.8 & 94.3/95.0 & 94.2/95.2 & 94.2/95.3 \\ 
    0.8 & -21.9 (7.9) & 0.2 (7.5) & 0.4 (10.2) & 0.5 (10.2) & -0.0 (10.4) & 86.8/88.7 & 94.7/95.0 & 94.5/95.3 & 94.5/95.4 & 94.2/95.2 \\ 
   \hline
   \hline
     & \multicolumn{5}{c}{Percent Bias (SE$\times$100)} &   \multicolumn{5}{c}{Empirical Coverage Rate v1/v2} \\
\cmidrule(lr){2-6} \cmidrule(lr){7-11}
  $\rho_{AA^*}$  & 
 $\widehat{\text{MP}}^{(U)}$ &  $\widehat{\text{MP}}^{(G)}$ & $\widehat{\text{MP}}^{(O1)}$ & $\widehat{\text{MP}}^{(O2)}$ & $\widehat{\text{MP}}^{(R)}$ &
 $\widehat{\text{MP}}^{(U)}$ &  $\widehat{\text{MP}}^{(G)}$ & $\widehat{\text{MP}}^{(O1)}$ & $\widehat{\text{MP}}^{(O2)}$ & $\widehat{\text{MP}}^{(R)}$  \\ 
  \hline
  0.4 & 0.6 (481.7) & 0.3 (7.2) & -2.1 (278.1) & -1.7 (1572.0) & -1.9 (1518.9) & 91.0/98.0 & 96.1/95.5 & 91.5/97.2 & 91.4/97.5 & 91.3/97.9 \\ 
    0.6 & -0.4 (70.5) & -1.0 (7.6) & -2.9 (37.9) & -3.1 (37.8) & -2.4 (48.2) & 92.2/98.2 & 94.5/95.5 & 92.5/97.8 & 92.4/97.7 & 92.7/97.8 \\ 
    0.8 & 2.2 (10.7) & 0.0 (7.7) & -0.1 (11.0) & 0.3 (10.9) & -0.1 (11.0) & 94.4/96.1 & 94.5/95.1 & 94.0/96.0 & 94.3/96.0 & 94.0/96.0 \\ 
 \hline
   \hline
\end{tabular}
\begin{tablenotes}
\item[1]  The data generation process is given in Web Appendix H, where the mediator follows a linear regression (1) and outcome follows the Cox model (2), with measurement error process characterized in (4) with a skew normal distribution. The percent bias is defined as the ratio of bias to the true value over 2,000 replications, i.e., $mean(\frac{\widehat{\tau}-\tau}{\tau}) \times 100\%$.  The empirical coverage rate v1 is obtained based on a Wald-type 95\% confidence interval using the derived asymptotic variance formula. The empirical coverage rate v2 is obtained based on a non-parametric bootstrap confidence interval.  
\end{tablenotes}
\end{threeparttable}}
\end{table}

\begin{table}[ht]
\caption{Simulation results when the error term in measurement error model (4) follows a uniform distribution. For each mediation effect measure $\tau\in\{\text{NIE},\text{NDE},\text{MP}\}$, we use $\widehat{\tau}^{(U)}$, $\widehat{\tau}^{(G)}$, $\widehat{\tau}^{(O1)}$, $\widehat{\tau}^{(O2)}$, and $\widehat{\tau}^{(R)}$ to denote the estimator given by the uncorrected, gold-standard, ORC1, ORC2, and RRC approaches, respectively. Upper panel: performance of the NIE estimator; middle panel: performance of the NDE estimator; lower panel: performance of the MP estimator.}
\centering
\scalebox{0.72}[0.72]{
\begin{threeparttable}
\label{tab:sim_normal}
\setlength\tabcolsep{4pt}
\begin{tabular}{l rrrrr rrrrr }
  \hline
&  \multicolumn{5}{c}{Percent Bias (SE$\times$100)} &   \multicolumn{5}{c}{Empirical Coverage Rate v1/v2} \\
\cmidrule(lr){2-6} \cmidrule(lr){7-11}
  $\rho_{AA^*}$  & 
 $\widehat{\text{NIE}}^{(U)}$ &  $\widehat{\text{NIE}}^{(G)}$ & $\widehat{\text{NIE}}^{(O1)}$ & $\widehat{\text{NIE}}^{(O2)}$ & $\widehat{\text{NIE}}^{(R)}$ &
 $\widehat{\text{NIE}}^{(U)}$ &  $\widehat{\text{NIE}}^{(G)}$ & $\widehat{\text{NIE}}^{(O1)}$ & $\widehat{\text{NIE}}^{(O2)}$ & $\widehat{\text{NIE}}^{(R)}$ \\ 
  \hline
0.4 & -61.6 (1.1) & -0.5 (2.1) & -1.7 (4.4) & -1.2 (4.2) & -2.3 (4.4) & 0.0/0.1 & 95.0/95.0 & 93.8/95.0 & 93.4/95.3 & 93.2/95.2 \\ 
    0.6 & -41.4 (1.4) & -0.4 (2.0) & -1.1 (3.2) & -1.1 (3.1) & -1.1 (3.3) & 7.5/14.1 & 95.0/95.4 & 94.5/94.9 & 94.4/95.5 & 94.2/95.2 \\ 
    0.8 & -20.7 (1.7) & -1.2 (2.0) & -1.1 (2.4) & -1.0 (2.4) & -1.3 (2.5) & 66.6/73.6 & 95.0/95.6 & 95.7/96.0 & 95.8/96.0 & 95.7/96.1 \\ 
   \hline
   \hline
  & \multicolumn{5}{c}{Percent Bias (SE$\times$100)} &   \multicolumn{5}{c}{Empirical Coverage Rate v1/v2} \\
\cmidrule(lr){2-6} \cmidrule(lr){7-11}
   $\rho_{AA^*}$  & 
 $\widehat{\text{NDE}}^{(U)}$ &  $\widehat{\text{NDE}}^{(G)}$ & $\widehat{\text{NDE}}^{(O1)}$ & $\widehat{\text{NDE}}^{(O2)}$ & $\widehat{\text{NDE}}^{(R)}$ &
 $\widehat{\text{NDE}}^{(U)}$ &  $\widehat{\text{NDE}}^{(G)}$ & $\widehat{\text{NDE}}^{(O1)}$ & $\widehat{\text{NDE}}^{(O2)}$ & $\widehat{\text{NDE}}^{(R)}$  \\ 
  \hline
  0.4 & -59.8 (7.6) & -0.3 (7.8) & 7.0 (21.5) & 6.9 (21.6) & 7.6 (22.3) & 38.6/46.7 & 95.5/95.6 & 94.6/95.3 & 94.7/95.4 & 94.5/95.4 \\ 
    0.6 & -40.7 (7.8) & 0.0 (7.7) & 2.8 (14.1) & 3.1 (14.1) & 2.5 (14.2) & 66.9/71.8 & 95.4/95.0 & 93.8/94.5 & 94.2/94.7 & 94.2/95.0 \\ 
    0.8 & -21.7 (7.7) & 0.1 (7.8) & 0.4 (10.1) & 0.6 (10.0) & 0.2 (10.1) & 87.2/88.3 & 94.7/94.5 & 95.0/95.5 & 95.1/95.5 & 95.3/95.8 \\  
   \hline
   \hline
     & \multicolumn{5}{c}{Percent Bias (SE$\times$100)} &   \multicolumn{5}{c}{Empirical Coverage Rate v1/v2} \\
\cmidrule(lr){2-6} \cmidrule(lr){7-11}
  $\rho_{AA^*}$  & 
 $\widehat{\text{MP}}^{(U)}$ &  $\widehat{\text{MP}}^{(G)}$ & $\widehat{\text{MP}}^{(O1)}$ & $\widehat{\text{MP}}^{(O2)}$ & $\widehat{\text{MP}}^{(R)}$ &
 $\widehat{\text{MP}}^{(U)}$ &  $\widehat{\text{MP}}^{(G)}$ & $\widehat{\text{MP}}^{(O1)}$ & $\widehat{\text{MP}}^{(O2)}$ & $\widehat{\text{MP}}^{(R)}$  \\ 
  \hline
  0.4 & -4.0 (741.4) & -0.5 (8.0) & -6.3 (161.2) & -6.7 (11407.7) & -6.4 (4881.4) & 89.2/97.9 & 94.5/96.2 & 88.9/97.0 & 88.8/97.0 & 88.8/97.5 \\ 
    0.6 & 0.6 (304.1) & -0.1 (7.7) & -1.9 (69.6) & -2.4 (92.0) & -1.6 (67.0) & 92.5/97.0 & 95.0/95.5 & 92.2/96.7 & 91.8/96.6 & 91.7/97.0 \\ 
    0.8 & 0.8 (16.7) & -0.0 (8.1) & -0.7 (15.9) & -0.6 (16.0) & -0.5 (15.7) & 94.9/96.3 & 94.0/95.0 & 94.7/96.4 & 94.6/96.3 & 94.5/96.2 \\ 
 \hline
   \hline
\end{tabular}
\begin{tablenotes}
\item[1]  The data generation process is given in Web Appendix H, where the mediator follows a linear regression (1) and outcome follows the Cox model (2), with measurement error process characterized in (4) with a uniform distribution. The percent bias is defined as the ratio of bias to the true value over 2,000 replications, i.e., $mean(\frac{\widehat{\tau}-\tau}{\tau}) \times 100\%$.  The standard error (SE) is defined as the square root of empirical variance of mediation effect measure estimates from the 2,000 replications. The empirical coverage rate v1 is obtained based on a Wald-type 95\% confidence interval using the derived asymptotic variance formula. The empirical coverage rate v2 is obtained based on a non-parametric bootstrap confidence interval.
\end{tablenotes}
\end{threeparttable}}
\end{table}

\begin{table}[htbp]
\caption{Simulation results when the error term in the mediator model follows a skew normal distribution. For each mediation effect measure $\tau\in\{\text{NIE},\text{NDE},\text{MP}\}$, we use $\widehat{\tau}^{(U)}$, $\widehat{\tau}^{(G)}$, $\widehat{\tau}^{(O1)}$, $\widehat{\tau}^{(O2)}$, and $\widehat{\tau}^{(R)}$ to denote the estimator given by the uncorrected, gold-standard, ORC1, ORC2, and RRC approaches, respectively, where the RRC uses 4 split points to determine the validation study risk sets. Upper panel: performance of the NIE estimator; middle panel: performance of the NDE estimator; lower panel: performance of the MP estimator.}
\centering
\scalebox{0.70}[0.70]{
\begin{threeparttable}
\label{tab:sim_M_skew}
\setlength\tabcolsep{4pt}
\begin{tabular}{cc ccccc ccccc }
  \hline
& & \multicolumn{5}{c}{Percent Bias (SE$\times$100)} &   \multicolumn{5}{c}{Empirical Coverage Rate v1/v2} \\
\cmidrule(lr){3-7} \cmidrule(lr){8-12}
$t$ &   NIE$(t)$  & 
 $\widehat{\text{NIE}}^{(U)}$ &  $\widehat{\text{NIE}}^{(G)}$ & $\widehat{\text{NIE}}^{(O1)}$ & $\widehat{\text{NIE}}^{(O2)}$ & $\widehat{\text{NIE}}^{(R)}$ &
 $\widehat{\text{NIE}}^{(U)}$ &  $\widehat{\text{NIE}}^{(G)}$ & $\widehat{\text{NIE}}^{(O1)}$ & $\widehat{\text{NIE}}^{(O2)}$ & $\widehat{\text{NIE}}^{(R)}$ \\ 
  \hline
10 & log(1.129) & -42.2 (1.2) & -0.7 (1.7) & -3.7 (2.7) & -3.3 (2.6) & -3.6 (2.7) & 2.5/5.7 & 95.6/95.8 & 92.8/94.9 & 93.3/94.8 & 92.9/94.8 \\ 
  15 & log(1.129) & -42.2 (1.2) & -0.7 (1.7) & -3.7 (2.7) & -3.3 (2.6) & -3.6 (2.7) & 2.5/5.7 & 95.6/95.8 & 92.8/95.0 & 93.3/94.8 & 92.9/94.8 \\ 
  20 & log(1.129) & -42.2 (1.2) & -0.7 (1.7) & -3.7 (2.7) & -3.3 (2.6) & -3.5 (2.7) & 2.6/5.7 & 95.6/95.8 & 92.8/95.0 & 93.3/94.8 & 92.9/94.8 \\ 
  25 & log(1.129) & -42.2 (1.2) & -0.7 (1.7) & -3.7 (2.7) & -3.2 (2.6) & -3.5 (2.7) & 2.6/5.8 & 95.6/95.8 & 92.8/95.0 & 93.3/94.8 & 92.9/94.8 \\ 
  30 & log(1.129) & -42.1 (1.2) & -0.6 (1.7) & -3.6 (2.7) & -3.1 (2.6) & -3.4 (2.7) & 2.6/5.8 & 95.7/95.8 & 92.9/95.0 & 93.3/94.8 & 92.9/95.0 \\ 
  35 & log(1.129) & -42.0 (1.2) & -0.3 (1.7) & -3.3 (2.7) & -2.9 (2.6) & -3.2 (2.7) & 2.8/6.0 & 95.7/95.8 & 93.2/95.0 & 93.8/94.7 & 93.0/95.0 \\ 
  40 & log(1.128) & -41.7 (1.2) & 0.2 (1.7) & -2.9 (2.7) & -2.4 (2.6) & -2.7 (2.7) & 3.0/6.3 & 95.5/95.7 & 93.4/94.9 & 94.0/94.8 & 93.2/95.2 \\ 
  45 & log(1.127) & -41.1 (1.2) & 1.1 (1.7) & -2.0 (2.7) & -1.5 (2.6) & -1.8 (2.7) & 3.8/7.3 & 95.2/95.5 & 93.8/95.2 & 94.1/95.2 & 93.7/95.4 \\ 
   \hline
   \hline
 & & \multicolumn{5}{c}{Percent Bias (SE$\times$100)} &   \multicolumn{5}{c}{Empirical Coverage Rate v1/v2} \\
\cmidrule(lr){3-7} \cmidrule(lr){8-12}
$t$ &   NDE$(t)$ & 
 $\widehat{\text{NDE}}^{(U)}$ &  $\widehat{\text{NDE}}^{(G)}$ & $\widehat{\text{NDE}}^{(O1)}$ & $\widehat{\text{NDE}}^{(O2)}$ & $\widehat{\text{NDE}}^{(R)}$ &
 $\widehat{\text{NDE}}^{(U)}$ &  $\widehat{\text{NDE}}^{(G)}$ & $\widehat{\text{NDE}}^{(O1)}$ & $\widehat{\text{NDE}}^{(O2)}$ & $\widehat{\text{NDE}}^{(R)}$  \\ 
  \hline
 10 & log(1.338) & -44.9 (7.6) & -3.4 (7.6) & -2.7 (13.8) & -3.1 (13.8) & -2.9 (13.9) & 60.4/67.2 & 95.6/95.5 & 94.2/95.1 & 94.6/95.1 & 94.3/95.3 \\ 
  15 & log(1.338) & -44.9 (7.6) & -3.4 (7.6) & -2.7 (13.8) & -3.2 (13.8) & -3.0 (13.9) & 60.4/67.1 & 95.6/95.5 & 94.2/95.0 & 94.5/95.1 & 94.3/95.3 \\ 
  20 & log(1.339) & -45.0 (7.6) & -3.5 (7.6) & -2.8 (13.8) & -3.3 (13.8) & -3.1 (13.9) & 60.2/66.9 & 95.6/95.5 & 94.2/95.0 & 94.5/95.2 & 94.2/95.3 \\ 
  25 & log(1.339) & -45.0 (7.6) & -3.5 (7.6) & -2.8 (13.8) & -3.3 (13.8) & -3.0 (13.9) & 60.2/67.0 & 95.6/95.5 & 94.2/95.0 & 94.5/95.2 & 94.2/95.3 \\ 
  30 & log(1.337) & -44.7 (7.6) & -3.1 (7.6) & -2.4 (13.8) & -2.8 (13.8) & -2.6 (13.9) & 61.2/67.7 & 95.6/95.3 & 94.4/95.1 & 94.6/95.2 & 94.3/95.3 \\ 
  35 & log(1.336) & -44.6 (7.6) & -2.9 (7.6) & -2.2 (13.8) & -2.7 (13.8) & -2.4 (13.9) & 61.4/68.0 & 95.6/95.2 & 94.5/95.1 & 94.6/95.2 & 94.3/95.3 \\ 
  40 & log(1.332) & -44.0 (7.6) & -1.9 (7.6) & -1.2 (13.8) & -1.7 (13.8) & -1.4 (13.9) & 63.1/69.2 & 95.5/95.5 & 94.5/95.0 & 94.7/95.2 & 94.2/95.2 \\ 
  45 & log(1.328) & -43.3 (7.6) & -0.7 (7.6) & 0.0 (13.8) & -0.4 (13.8) & -0.2 (13.9) & 64.9/70.5 & 95.6/95.8 & 94.5/95.0 & 94.5/94.8 & 94.4/95.2 \\ 
   \hline
   \hline
    & & \multicolumn{5}{c}{Percent Bias (SE$\times$100)} &   \multicolumn{5}{c}{Empirical Coverage Rate v1/v2} \\
\cmidrule(lr){3-7} \cmidrule(lr){8-12}
$t$ &   MP$(t)$  & 
 $\widehat{\text{MP}}^{(U)}$ &  $\widehat{\text{MP}}^{(G)}$ & $\widehat{\text{MP}}^{(O1)}$ & $\widehat{\text{MP}}^{(O2)}$ & $\widehat{\text{MP}}^{(R)}$ &
 $\widehat{\text{MP}}^{(U)}$ &  $\widehat{\text{MP}}^{(G)}$ & $\widehat{\text{MP}}^{(O1)}$ & $\widehat{\text{MP}}^{(O2)}$ & $\widehat{\text{MP}}^{(R)}$  \\ 
  \hline
   10 & 0.295 & 2.7 (25.7) & 2.0 (7.5) & -0.5 (23.7) & -0.3 (23.9) & -0.8 (24.1) & 93.2/97.8 & 95.6/95.9 & 92.3/97.3 & 92.3/97.2 & 91.7/97.6 \\ 
  15 & 0.294 & 2.8 (25.7) & 2.0 (7.5) & -0.5 (23.7) & -0.2 (23.9) & -0.7 (24.1) & 93.2/97.8 & 95.6/95.9 & 92.3/97.3 & 92.3/97.2 & 91.8/97.6 \\ 
  20 & 0.294 & 2.8 (25.7) & 2.1 (7.5) & -0.4 (23.7) & -0.1 (23.9) & -0.7 (24.1) & 93.3/97.8 & 95.6/95.8 & 92.3/97.3 & 92.3/97.2 & 91.8/97.6 \\ 
  25 & 0.294 & 2.8 (25.7) & 2.1 (7.5) & -0.4 (23.7) & -0.1 (23.9) & -0.7 (24.1) & 93.3/97.8 & 95.6/95.8 & 92.3/97.3 & 92.3/97.2 & 91.8/97.6 \\ 
  30 & 0.295 & 2.6 (25.7) & 1.8 (7.5) & -0.6 (23.7) & -0.4 (23.9) & -0.9 (24.1) & 93.1/97.8 & 95.6/96.0 & 92.2/97.2 & 92.3/97.2 & 91.7/97.6 \\ 
  35 & 0.295 & 2.7 (25.7) & 1.9 (7.5) & -0.6 (23.7) & -0.3 (23.9) & -0.8 (24.1) & 93.1/97.8 & 95.6/95.9 & 92.3/97.2 & 92.3/97.2 & 91.7/97.6 \\ 
  40 & 0.296 & 2.3 (25.7) & 1.5 (7.5) & -0.9 (23.7) & -0.7 (23.9) & -1.2 (24.1) & 93.0/97.8 & 95.4/95.8 & 92.0/97.2 & 92.2/97.2 & 91.6/97.6 \\ 
  45 & 0.296 & 2.0 (25.7) & 1.3 (7.5) & -1.2 (23.7) & -0.9 (23.9) & -1.4 (24.1) & 93.0/97.7 & 95.2/95.8 & 92.0/97.2 & 92.2/97.2 & 91.5/97.6 \\ 
 \hline
   \hline
\end{tabular}
\begin{tablenotes}
\item[1] The data generation process is given in Web Appendix G.2. In this simulation, we considered a moderate sample size of the validation study ($n_2=250$) and a moderate exposure measurement error ($\rho_{AA^*}=0.6$). The percent bias is defined as the ratio of bias to the true value over 2,000 replications, i.e., $mean(\frac{\widehat{\tau}-\tau}{\tau}) \times 100\%$. The standard error (SE) is defined as the square root of empirical variance of mediation effect measure estimates from the 2,000 replications. The empirical coverage rate v1 is obtained based on a Wald-type 95\% confidence interval using the derived asymptotic variance formula. The empirical coverage rate v2 is obtained based on a non-parametric bootstrap confidence interval.
\end{tablenotes}
\end{threeparttable}}
\end{table}

\begin{table}[htbp]
\caption{Simulation results when the error term in the mediator model follows a uniform distribution. For each mediation effect measure $\tau\in\{\text{NIE},\text{NDE},\text{MP}\}$, we use $\widehat{\tau}^{(U)}$, $\widehat{\tau}^{(G)}$, $\widehat{\tau}^{(O1)}$, $\widehat{\tau}^{(O2)}$, and $\widehat{\tau}^{(R)}$ to denote the estimator given by the uncorrected, gold-standard, ORC1, ORC2, and RRC approaches, respectively, where the RRC uses 4 split points to determine the validation study risk sets. Upper panel: performance of the NIE estimator; middle panel: performance of the NDE estimator; lower panel: performance of the MP estimator.}
\centering
\scalebox{0.70}[0.70]{
\begin{threeparttable}
\label{tab:sim_M_uniform}
\setlength\tabcolsep{4pt}
\begin{tabular}{cc ccccc ccccc }
  \hline
& & \multicolumn{5}{c}{Percent Bias (SE$\times$100)} &   \multicolumn{5}{c}{Empirical Coverage Rate v1/v2} \\
\cmidrule(lr){3-7} \cmidrule(lr){8-12}
$t$ &   NIE$(t)$  & 
 $\widehat{\text{NIE}}^{(U)}$ &  $\widehat{\text{NIE}}^{(G)}$ & $\widehat{\text{NIE}}^{(O1)}$ & $\widehat{\text{NIE}}^{(O2)}$ & $\widehat{\text{NIE}}^{(R)}$ &
 $\widehat{\text{NIE}}^{(U)}$ &  $\widehat{\text{NIE}}^{(G)}$ & $\widehat{\text{NIE}}^{(O1)}$ & $\widehat{\text{NIE}}^{(O2)}$ & $\widehat{\text{NIE}}^{(R)}$ \\ 
  \hline
10 & log(1.129) & -41.2 (1.5) & -0.8 (2.2) & 0.8 (3.6) & 1.0 (3.6) & 0.8 (3.6) & 12.8/20.4 & 94.0/94.5 & 95.0/95.0 & 94.7/95.3 & 94.7/95.7 \\ 
  15 & log(1.129) & -41.2 (1.5) & -0.8 (2.2) & 0.8 (3.6) & 1.0 (3.6) & 0.8 (3.6) & 12.8/20.4 & 94.0/94.5 & 95.0/95.0 & 94.7/95.3 & 94.7/95.7 \\ 
  20 & log(1.129) & -41.2 (1.5) & -0.8 (2.2) & 0.8 (3.6) & 1.0 (3.6) & 0.8 (3.6) & 13.0/20.4 & 94.0/94.5 & 94.9/95.0 & 94.7/95.3 & 94.7/95.7 \\ 
  25 & log(1.129) & -41.2 (1.5) & -0.8 (2.2) & 0.8 (3.6) & 1.0 (3.6) & 0.8 (3.6) & 13.0/20.4 & 94.0/94.5 & 94.9/95.0 & 94.7/95.3 & 94.7/95.7 \\ 
  30 & log(1.129) & -41.2 (1.5) & -0.7 (2.2) & 0.9 (3.6) & 1.1 (3.6) & 0.9 (3.6) & 13.1/20.4 & 93.8/94.6 & 94.9/95.2 & 94.7/95.3 & 94.7/95.7 \\ 
  35 & log(1.129) & -41.1 (1.5) & -0.6 (2.2) & 1.0 (3.6) & 1.2 (3.6) & 1.0 (3.6) & 13.1/20.8 & 93.8/94.6 & 95.0/95.2 & 94.7/95.4 & 94.7/95.7 \\ 
  40 & log(1.129) & -41.0 (1.5) & -0.3 (2.2) & 1.3 (3.6) & 1.5 (3.6) & 1.2 (3.6) & 13.3/21.1 & 93.9/94.6 & 95.0/95.2 & 94.6/95.3 & 94.7/95.7 \\ 
  45 & log(1.128) & -40.7 (1.5) & 0.1 (2.2) & 1.7 (3.6) & 1.9 (3.6) & 1.7 (3.6) & 14.1/21.8 & 94.0/94.7 & 95.0/95.2 & 94.8/95.5 & 94.8/95.8 \\ 
   \hline
   \hline
 & & \multicolumn{5}{c}{Percent Bias (SE$\times$100)} &   \multicolumn{5}{c}{Empirical Coverage Rate v1/v2} \\
\cmidrule(lr){3-7} \cmidrule(lr){8-12}
$t$ &   NDE$(t)$ & 
 $\widehat{\text{NDE}}^{(U)}$ &  $\widehat{\text{NDE}}^{(G)}$ & $\widehat{\text{NDE}}^{(O1)}$ & $\widehat{\text{NDE}}^{(O2)}$ & $\widehat{\text{NDE}}^{(R)}$ &
 $\widehat{\text{NDE}}^{(U)}$ &  $\widehat{\text{NDE}}^{(G)}$ & $\widehat{\text{NDE}}^{(O1)}$ & $\widehat{\text{NDE}}^{(O2)}$ & $\widehat{\text{NDE}}^{(R)}$  \\ 
  \hline
  10 & log(1.327) & -41.7 (7.5) & 0.4 (7.5) & 1.4 (13.4) & 1.4 (13.4) & 1.8 (13.6) & 66.6/71.3 & 95.8/95.6 & 95.8/95.8 & 95.7/95.7 & 95.7/95.8 \\ 
  15 & log(1.326) & -41.6 (7.5) & 0.6 (7.5) & 1.5 (13.4) & 1.5 (13.4) & 1.9 (13.6) & 66.8/71.5 & 95.8/95.6 & 95.7/95.8 & 95.7/95.7 & 95.8/95.8 \\ 
  20 & log(1.326) & -41.6 (7.5) & 0.6 (7.5) & 1.5 (13.4) & 1.5 (13.4) & 1.9 (13.6) & 66.8/71.5 & 95.8/95.6 & 95.7/95.8 & 95.7/95.7 & 95.8/95.7 \\ 
  25 & log(1.327) & -41.7 (7.5) & 0.5 (7.5) & 1.4 (13.4) & 1.4 (13.4) & 1.8 (13.6) & 66.6/71.3 & 95.8/95.6 & 95.8/95.8 & 95.7/95.7 & 95.7/95.8 \\ 
  30 & log(1.326) & -41.6 (7.5) & 0.6 (7.5) & 1.5 (13.4) & 1.6 (13.4) & 2.0 (13.6) & 66.8/71.5 & 95.8/95.6 & 95.7/95.8 & 95.6/95.7 & 95.8/95.7 \\ 
  35 & log(1.325) & -41.4 (7.5) & 0.9 (7.5) & 1.9 (13.4) & 1.9 (13.4) & 2.3 (13.6) & 67.2/71.8 & 95.8/95.7 & 95.7/95.8 & 95.5/95.7 & 95.8/95.7 \\ 
  40 & log(1.325) & -41.4 (7.5) & 1.0 (7.5) & 1.9 (13.4) & 2.0 (13.4) & 2.4 (13.6) & 67.2/71.9 & 95.8/95.7 & 95.7/95.8 & 95.5/95.7 & 95.8/95.7 \\ 
  45 & log(1.321) & -40.8 (7.5) & 2.0 (7.5) & 2.9 (13.4) & 3.0 (13.4) & 3.3 (13.6) & 68.2/72.8 & 95.7/95.3 & 95.5/95.8 & 95.5/95.5 & 95.8/95.6 \\ 
   \hline
   \hline
    & & \multicolumn{5}{c}{Percent Bias (SE$\times$100)} &   \multicolumn{5}{c}{Empirical Coverage Rate v1/v2} \\
\cmidrule(lr){3-7} \cmidrule(lr){8-12}
$t$ &   MP$(t)$  & 
 $\widehat{\text{MP}}^{(U)}$ &  $\widehat{\text{MP}}^{(G)}$ & $\widehat{\text{MP}}^{(O1)}$ & $\widehat{\text{MP}}^{(O2)}$ & $\widehat{\text{MP}}^{(R)}$ &
 $\widehat{\text{MP}}^{(U)}$ &  $\widehat{\text{MP}}^{(G)}$ & $\widehat{\text{MP}}^{(O1)}$ & $\widehat{\text{MP}}^{(O2)}$ & $\widehat{\text{MP}}^{(R)}$  \\ 
  \hline
  10 & 0.301 & 2.3 (35.3) & -1.2 (7.6) & 1.8 (33.1) & 2.0 (33.5) & 2.5 (33.0) & 93.0/97.3 & 94.3/95.2 & 93.4/97.0 & 93.4/96.9 & 93.7/97.2 \\ 
  15 & 0.301 & 2.2 (35.3) & -1.2 (7.6) & 1.7 (33.1) & 1.9 (33.5) & 2.4 (33.0) & 92.9/97.3 & 94.3/95.2 & 93.4/97.0 & 93.4/96.9 & 93.7/97.2 \\ 
  20 & 0.301 & 2.2 (35.3) & -1.2 (7.6) & 1.7 (33.1) & 1.9 (33.5) & 2.4 (33.0) & 92.9/97.3 & 94.3/95.2 & 93.4/97.0 & 93.4/96.9 & 93.7/97.2 \\ 
  25 & 0.301 & 2.3 (35.3) & -1.2 (7.6) & 1.8 (33.1) & 2.0 (33.5) & 2.5 (33.0) & 93.0/97.3 & 94.3/95.2 & 93.4/97.0 & 93.4/96.9 & 93.7/97.2 \\ 
  30 & 0.301 & 2.2 (35.3) & -1.2 (7.6) & 1.7 (33.1) & 1.9 (33.5) & 2.4 (33.0) & 92.9/97.3 & 94.3/95.2 & 93.4/97.0 & 93.4/96.9 & 93.7/97.2 \\ 
  35 & 0.301 & 2.1 (35.3) & -1.4 (7.6) & 1.6 (33.1) & 1.7 (33.5) & 2.3 (33.0) & 92.7/97.3 & 94.3/95.3 & 93.4/97.0 & 93.3/96.9 & 93.6/97.0 \\ 
  40 & 0.301 & 2.2 (35.3) & -1.2 (7.6) & 1.7 (33.1) & 1.9 (33.5) & 2.4 (33.0) & 92.9/97.3 & 94.3/95.2 & 93.4/97.0 & 93.4/96.9 & 93.7/97.2 \\ 
  45 & 0.302 & 1.9 (35.3) & -1.6 (7.6) & 1.4 (33.1) & 1.5 (33.5) & 2.1 (33.0) & 92.6/97.2 & 94.0/95.2 & 93.2/97.0 & 93.2/96.9 & 93.4/97.0 \\ 
 \hline
   \hline
\end{tabular}
\begin{tablenotes}
\item[1] The data generation process is given in Web Appendix G.2. In this simulation, we considered a moderate sample size of the validation study ($n_2=250$) and a moderate exposure measurement error ($\rho_{AA^*}=0.6$). The percent bias is defined as the ratio of bias to the true value over 2,000 replications, i.e., $mean(\frac{\widehat{\tau}-\tau}{\tau}) \times 100\%$. The standard error (SE) is defined as the square root of empirical variance of mediation effect measure estimates from the 2,000 replications. The empirical coverage rate v1 is obtained based on a Wald-type 95\% confidence interval using the derived asymptotic variance formula. The empirical coverage rate v2 is obtained based on a non-parametric bootstrap confidence interval.
\end{tablenotes}
\end{threeparttable}}
\end{table}

\begin{table}[htbp]
\caption{Simulation results when the outcome follows a proportional odds model. For each mediation effect measure $\tau\in\{\text{NIE},\text{NDE},\text{MP}\}$, we use $\widehat{\tau}^{(U)}$, $\widehat{\tau}^{(G)}$, $\widehat{\tau}^{(O1)}$, $\widehat{\tau}^{(O2)}$, and $\widehat{\tau}^{(R)}$ to denote the estimator given by the uncorrected, gold-standard, ORC1, ORC2, and RRC approaches, respectively, where the RRC uses 4 split points to determine the validation study risk sets. Upper panel: performance of the NIE estimator; middle panel: performance of the NDE estimator; lower panel: performance of the MP estimator.}
\centering
\scalebox{0.68}[0.68]{
\begin{threeparttable}
\label{tab:sim_proportional_odds}
\setlength\tabcolsep{4pt}
\begin{tabular}{cc ccccc ccccc }
  \hline
& & \multicolumn{5}{c}{Percent Bias (SE$\times$100)} &   \multicolumn{5}{c}{Empirical Coverage Rate v1/v2} \\
\cmidrule(lr){3-7} \cmidrule(lr){8-12}
$t$ &   NIE$(t)$  & 
 $\widehat{\text{NIE}}^{(U)}$ &  $\widehat{\text{NIE}}^{(G)}$ & $\widehat{\text{NIE}}^{(O1)}$ & $\widehat{\text{NIE}}^{(O2)}$ & $\widehat{\text{NIE}}^{(R)}$ &
 $\widehat{\text{NIE}}^{(U)}$ &  $\widehat{\text{NIE}}^{(G)}$ & $\widehat{\text{NIE}}^{(O1)}$ & $\widehat{\text{NIE}}^{(O2)}$ & $\widehat{\text{NIE}}^{(R)}$ \\ 
  \hline
10 & log(1.188) & -42.5 (2.2) & -3.5 (3.3) & -4.5 (5.1) & -4.4 (5.0) & -4.2 (5.1) & 14.0/20.1 & 93.0/94.2 & 93.9/95.0 & 93.8/95.2 & 94.0/95.2 \\ 
  15 & log(1.188) & -42.5 (2.2) & -3.5 (3.3) & -4.5 (5.1) & -4.4 (5.0) & -4.2 (5.1) & 14.0/20.1 & 93.0/94.2 & 93.9/95.0 & 93.8/95.2 & 94.0/95.2 \\ 
  20 & log(1.188) & -42.5 (2.2) & -3.5 (3.3) & -4.5 (5.1) & -4.4 (5.0) & -4.2 (5.1) & 14.0/20.1 & 93.0/94.2 & 93.9/95.0 & 93.8/95.2 & 94.0/95.2 \\ 
  25 & log(1.188) & -42.4 (2.2) & -3.5 (3.3) & -4.5 (5.1) & -4.4 (5.0) & -4.2 (5.1) & 14.0/20.1 & 93.0/94.2 & 93.9/95.0 & 93.8/95.2 & 94.0/95.2 \\ 
  30 & log(1.188) & -42.4 (2.2) & -3.4 (3.3) & -4.4 (5.1) & -4.3 (5.0) & -4.1 (5.1) & 14.2/20.4 & 93.1/94.2 & 93.9/95.0 & 93.8/95.3 & 94.0/95.3 \\ 
  35 & log(1.187) & -42.3 (2.2) & -3.2 (3.3) & -4.2 (5.1) & -4.1 (5.0) & -3.9 (5.1) & 14.4/20.8 & 93.2/94.2 & 94.0/95.0 & 93.8/95.4 & 94.0/95.2 \\ 
  40 & log(1.186) & -41.9 (2.2) & -2.6 (3.3) & -3.6 (5.1) & -3.5 (5.0) & -3.3 (5.1) & 15.2/21.9 & 93.4/94.5 & 94.0/95.0 & 94.0/95.5 & 94.1/95.4 \\ 
  45 & log(1.183) & -41.1 (2.2) & -1.3 (3.3) & -2.3 (5.1) & -2.2 (5.0) & -2.0 (5.1) & 16.9/24.4 & 93.6/94.5 & 94.2/95.3 & 94.0/95.4 & 94.2/95.8 \\ 
   \hline
   \hline
 & & \multicolumn{5}{c}{Percent Bias (SE$\times$100)} &   \multicolumn{5}{c}{Empirical Coverage Rate v1/v2} \\
\cmidrule(lr){3-7} \cmidrule(lr){8-12}
$t$ &   NDE$(t)$ & 
 $\widehat{\text{NDE}}^{(U)}$ &  $\widehat{\text{NDE}}^{(G)}$ & $\widehat{\text{NDE}}^{(O1)}$ & $\widehat{\text{NDE}}^{(O2)}$ & $\widehat{\text{NDE}}^{(R)}$ &
 $\widehat{\text{NDE}}^{(U)}$ &  $\widehat{\text{NDE}}^{(G)}$ & $\widehat{\text{NDE}}^{(O1)}$ & $\widehat{\text{NDE}}^{(O2)}$ & $\widehat{\text{NDE}}^{(R)}$  \\ 
  \hline
 10 & log(1.522) & -42.1 (13.9) & -0.8 (13.9) & 1.9 (24.5) & 1.6 (24.5) & 1.6 (24.8) & 73.1/77.2 & 94.8/95.0 & 93.8/95.0 & 94.0/95.0 & 93.8/95.3 \\ 
  15 & log(1.521) & -42.0 (13.9) & -0.7 (13.9) & 2.0 (24.5) & 1.7 (24.5) & 1.7 (24.8) & 73.3/77.2 & 94.7/95.0 & 93.8/95.0 & 94.0/95.1 & 93.8/95.3 \\ 
  20 & log(1.520) & -41.9 (13.9) & -0.6 (13.9) & 2.1 (24.5) & 1.8 (24.5) & 1.8 (24.8) & 73.5/77.2 & 94.7/95.0 & 93.8/95.0 & 94.0/95.1 & 93.8/95.3 \\ 
  25 & log(1.520) & -41.9 (13.9) & -0.6 (13.9) & 2.1 (24.5) & 1.8 (24.5) & 1.8 (24.8) & 73.5/77.2 & 94.7/95.0 & 93.8/95.0 & 94.0/95.1 & 93.8/95.3 \\ 
  30 & log(1.521) & -42.0 (13.9) & -0.7 (13.9) & 2.0 (24.5) & 1.7 (24.5) & 1.6 (24.8) & 73.2/77.2 & 94.8/95.0 & 93.8/95.0 & 94.0/95.1 & 93.8/95.3 \\ 
  35 & log(1.521) & -42.0 (13.9) & -0.7 (13.9) & 2.0 (24.5) & 1.7 (24.5) & 1.7 (24.8) & 73.3/77.2 & 94.7/95.0 & 93.8/95.0 & 94.0/95.1 & 93.8/95.3 \\ 
  40 & log(1.516) & -41.5 (13.9) & 0.1 (13.9) & 2.9 (24.5) & 2.6 (24.5) & 2.5 (24.8) & 74.0/77.8 & 94.7/95.0 & 93.8/95.0 & 94.0/95.0 & 93.9/95.2 \\ 
  45 & log(1.505) & -40.5 (13.9) & 1.9 (13.9) & 4.7 (24.5) & 4.3 (24.5) & 4.3 (24.8) & 75.1/79.3 & 94.7/94.8 & 93.9/95.0 & 94.0/95.0 & 94.0/95.0 \\ 
   \hline
   \hline
    & & \multicolumn{5}{c}{Percent Bias (SE$\times$100)} &   \multicolumn{5}{c}{Empirical Coverage Rate v1/v2} \\
\cmidrule(lr){3-7} \cmidrule(lr){8-12}
$t$ &   MP$(t)$  & 
 $\widehat{\text{MP}}^{(U)}$ &  $\widehat{\text{MP}}^{(G)}$ & $\widehat{\text{MP}}^{(O1)}$ & $\widehat{\text{MP}}^{(O2)}$ & $\widehat{\text{MP}}^{(R)}$ &
 $\widehat{\text{MP}}^{(U)}$ &  $\widehat{\text{MP}}^{(G)}$ & $\widehat{\text{MP}}^{(O1)}$ & $\widehat{\text{MP}}^{(O2)}$ & $\widehat{\text{MP}}^{(R)}$  \\ 
  \hline
 10 & 0.291 & -1.3 (90.2) & -2.1 (10.8) & -4.5 (191.0) & -5.0 (3667.7) & -4.9 (441.8) & 90.8/97.2 & 92.5/95.5 & 90.7/96.8 & 90.5/96.6 & 90.5/96.8 \\ 
  15 & 0.291 & -1.4 (90.2) & -2.2 (10.8) & -4.6 (191.0) & -5.1 (3667.7) & -5.0 (441.8) & 90.7/97.2 & 92.5/95.5 & 90.7/96.8 & 90.5/96.5 & 90.5/96.8 \\ 
  20 & 0.291 & -1.5 (90.2) & -2.3 (10.8) & -4.6 (191.0) & -5.2 (3667.7) & -5.0 (441.8) & 90.7/97.2 & 92.4/95.5 & 90.7/96.8 & 90.3/96.5 & 90.4/96.8 \\ 
  25 & 0.291 & -1.4 (90.2) & -2.2 (10.8) & -4.6 (191.0) & -5.1 (3667.7) & -5.0 (441.8) & 90.7/97.2 & 92.4/95.5 & 90.7/96.8 & 90.3/96.5 & 90.4/96.8 \\ 
  30 & 0.291 & -1.3 (90.2) & -2.1 (10.8) & -4.5 (191.0) & -5.0 (3667.7) & -4.9 (441.8) & 90.8/97.2 & 92.5/95.5 & 90.7/96.8 & 90.5/96.6 & 90.5/96.8 \\ 
  35 & 0.290 & -1.2 (90.2) & -2.0 (10.8) & -4.4 (191.0) & -4.9 (3667.7) & -4.8 (441.8) & 90.8/97.2 & 92.5/95.5 & 90.8/96.8 & 90.5/96.6 & 90.5/96.8 \\ 
  40 & 0.291 & -1.3 (90.2) & -2.1 (10.8) & -4.5 (191.0) & -5.0 (3667.7) & -4.9 (441.8) & 90.7/97.2 & 92.5/95.5 & 90.7/96.8 & 90.5/96.6 & 90.5/96.8 \\ 
  45 & 0.291 & -1.6 (90.2) & -2.4 (10.8) & -4.7 (191.0) & -5.2 (3667.7) & -5.1 (441.8) & 90.5/97.2 & 92.4/95.5 & 90.6/96.7 & 90.3/96.5 & 90.3/96.8 \\  
 \hline
   \hline
\end{tabular}
\begin{tablenotes}
\item[1] The data generation process is given in Web Appendix G.2. In this simulation, we considered a moderate sample size of the validation study ($n_2=250$) and a moderate exposure measurement error ($\rho_{AA^*}=0.6$). The percent bias is defined as the ratio of bias to the true value over 2,000 replications, i.e., $mean(\frac{\widehat{\tau}-\tau}{\tau}) \times 100\%$. The standard error (SE) is defined as the square root of empirical variance of mediation effect measure estimates from the 2,000 replications. The empirical coverage rate v1 is obtained based on a Wald-type 95\% confidence interval using the derived asymptotic variance formula. The empirical coverage rate v2 is obtained based on a non-parametric bootstrap confidence interval.
\end{tablenotes}
\end{threeparttable}}
\end{table}

\begin{table}[ht]
\caption{RRC estimates on the NIE, NDE, and MP under the Cox proportional hazard model (2), HPFS, 1986--2016. We provided the results with and without considering the exposure-mediator interaction in the Cox outcome model (2), indicated by the columns labeled ``with interaction" and ``no interaction" respectively.}
\centering
\scalebox{0.8}[0.8]{
\begin{threeparttable}
\begin{tabular}{r cc cc cc}
  \hline
Age & \multicolumn{2}{c}{NIE$^\dagger$} & \multicolumn{2}{c}{NDE} & \multicolumn{2}{c}{MP} \\
\cmidrule(lr){2-3} \cmidrule(lr){4-5}\cmidrule(lr){6-7}
& with interaction & no interaction & with interaction & no interaction & with interaction & no interaction \\
  \hline
45 & -0.043 & -0.040 & -0.068 & -0.067 & 0.388 & 0.369 \\ 
    & (-0.059,-0.029) & (-0.053,-0.031) & (-0.136,-0.012) & (-0.137,-0.012) & (0.220,0.768) & (0.229,0.787) \\ 
  50 & -0.043 & -0.040 & -0.068 & -0.068 & 0.388 & 0.369 \\ 
    & (-0.059,-0.029) & (-0.053,-0.031) & (-0.136,-0.012) & (-0.136,-0.012) & (0.220,0.768) & (0.230,0.786) \\ 
  55 & -0.043 & -0.040 & -0.068 & -0.068 & 0.388 & 0.369 \\ 
    & (-0.059,-0.029) & (-0.053,-0.031) & (-0.136,-0.012) & (-0.136,-0.012) & (0.220,0.769) & (0.230,0.786) \\ 
  60 & -0.043 & -0.040 & -0.068 & -0.068 & 0.388 & 0.369 \\ 
    & (-0.059,-0.029) & (-0.053,-0.031) & (-0.136,-0.011) & (-0.136,-0.012) & (0.220,0.770) & (0.230,0.786) \\ 
  65 & -0.043 & -0.040 & -0.068 & -0.068 & 0.387 & 0.369 \\ 
    & (-0.059,-0.029) & (-0.053,-0.031) & (-0.136,-0.011) & (-0.136,-0.012) & (0.220,0.771) & (0.230,0.786) \\ 
  70 & -0.043 & -0.040 & -0.068 & -0.068 & 0.387 & 0.369 \\ 
    & (-0.059,-0.029) & (-0.053,-0.031) & (-0.136,-0.011) & (-0.136,-0.012) & (0.221,0.774) & (0.230,0.786) \\ 
  75 & -0.043 & -0.040 & -0.068 & -0.068 & 0.387 & 0.369 \\ 
    & (-0.059,-0.029) & (-0.053,-0.031) & (-0.136,-0.011) & (-0.136,-0.012) & (0.221,0.777) & (0.230,0.786) \\ 
  80 & -0.043 & -0.040 & -0.068 & -0.068 & 0.386 & 0.369 \\ 
    & (-0.059,-0.029) & (-0.053,-0.031) & (-0.136,-0.011) & (-0.136,-0.012) & (0.221,0.781) & (0.230,0.786) \\ 
  85 & -0.043 & -0.040 & -0.068 & -0.068 & 0.386 & 0.369 \\ 
    & (-0.059,-0.029) & (-0.053,-0.031) & (-0.136,-0.011) & (-0.136,-0.012) & (0.221,0.781) & (0.230,0.786) \\
   \hline
\end{tabular}
\begin{tablenotes}
\item[$\dagger$] The point and interval estimates of NIE vary across different age; they happen to be identical in the first three decimal places.
\end{tablenotes}
\end{threeparttable}}
\end{table}

\begin{table}[ht]
\caption{Sensitivity analysis for the mediation analysis allowing departures from the Cox model (2) from the proportional hazards assumption, HPFS, 1986--2016. We present the RRC estimates on the NIE, NDE, and MP for a Cox outcome model with piecewise constant coefficients applied to regressors for which the Grambsch-Therneau test indicated significant departures from proportional hazards. We provided the results with and without considering the exposure-mediator interaction in the Cox outcome model (2), indicated by the columns labeled ``with interaction" and ``no interaction" respectively.}
\centering
\scalebox{0.8}[0.8]{
\begin{threeparttable}
\begin{tabular}{r cc cc cc}
  \hline
Age & \multicolumn{2}{c}{NIE} & \multicolumn{2}{c}{NDE} & \multicolumn{2}{c}{MP} \\
\cmidrule(lr){2-3} \cmidrule(lr){4-5}\cmidrule(lr){6-7}
& with interaction & no interaction & with interaction & no interaction & with interaction & no interaction \\
  \hline
45 & -0.074 & -0.073 & -0.065 & -0.065 & 0.534 & 0.523 \\ 
    & (-0.109,-0.049) & (-0.106,-0.051) & (-0.135,-0.014) & (-0.134,-0.011) & (0.365,0.860) & (0.367,0.863) \\ 
  50 & -0.074 & -0.073 & -0.065 & -0.066 & 0.534 & 0.523 \\ 
    & (-0.109,-0.049) & (-0.106,-0.051) & (-0.134,-0.014) & (-0.133,-0.012) & (0.367,0.859) & (0.367,0.863) \\ 
  55 & -0.054 & -0.052 & -0.065 & -0.066 & 0.452 & 0.441 \\ 
    & (-0.075,-0.039) & (-0.067,-0.040) & (-0.133,-0.013) & (-0.133,-0.012) & (0.267,0.804) & (0.280,0.796) \\ 
  60 & -0.054 & -0.052 & -0.065 & -0.066 & 0.452 & 0.441 \\ 
    & (-0.075,-0.039) & (-0.067,-0.040) & (-0.133,-0.013) & (-0.133,-0.012) & (0.267,0.805) & (0.280,0.796) \\ 
  65 & -0.054 & -0.052 & -0.065 & -0.066 & 0.452 & 0.441 \\ 
    & (-0.075,-0.039) & (-0.067,-0.040) & (-0.133,-0.013) & (-0.133,-0.012) & (0.267,0.805) & (0.280,0.796) \\ 
  70 & -0.038 & -0.036 & -0.066 & -0.066 & 0.365 & 0.353 \\ 
    & (-0.057,-0.023) & (-0.051,-0.025) & (-0.133,-0.011) & (-0.133,-0.012) & (0.195,0.739) & (0.203,0.730) \\ 
  75 & -0.038 & -0.036 & -0.066 & -0.066 & 0.365 & 0.353 \\ 
    & (-0.057,-0.023) & (-0.051,-0.025) & (-0.133,-0.012) & (-0.133,-0.012) & (0.195,0.740) & (0.203,0.730) \\ 
  80 & -0.029 & -0.027 & -0.066 & -0.066 & 0.305 & 0.292 \\ 
    & (-0.047,-0.013) & (-0.042,-0.013) & (-0.133,-0.014) & (-0.132,-0.012) & (0.128,0.732) & (0.142,0.667) \\ 
  85 & -0.029 & -0.027 & -0.066 & -0.066 & 0.305 & 0.292 \\ 
    & (-0.047,-0.013) & (-0.042,-0.013) & (-0.133,-0.014) & (-0.132,-0.012) & (0.128,0.733) & (0.142,0.667) \\ 
   \hline
\end{tabular}
\end{threeparttable}}
\end{table}

\newpage

\end{document}